\documentclass[usenatbib]{mn2e}

\usepackage{natbibmnfix, graphicx, times, subfig}
\pdfoutput=1
\usepackage{epstopdf}
\usepackage{float}

\newcommand{\rev}{}

\newcommand{\msun}{M$_{\odot}$}
\newcommand{\msunyr}{M$_{\odot}$ yr$^{-1}$}
\newcommand{\lsun}{L$_{\odot}$}
\newcommand{\kms}{km s$^{-1}$}

\newcommand{\vesc}{$v_{\rm esc}$}
\newcommand{\tmrg}{$t_{\rm mrg}$}

\newcommand{\fgas}{$f_{\rm gas}$}
\newcommand{\tdnl}{$t_{\rm dNL}$}
\newcommand{\inv}{$^{-1}$}
\newcommand{\gadget}{{\footnotesize GADGET}}

\newcommand{\gadthree}{{\footnotesize GADGET-3}}

\newcommand{\hbeta}{H{$\beta$}}
\newcommand{\oiii}{[O{\rm\,III}]}
\newcommand{\lhbeta}{$L_{\rm H\beta}$}
\newcommand{\lbol}{$L_{\rm bol}$}

\title[Double-peaked Narrow-Line Signatures of Dual Supermassive Black Holes]{Double-peaked Narrow-Line Signatures of Dual Supermassive Black Holes in Galaxy Merger Simulations}

\author[Blecha et al.]{Laura Blecha$^1$\thanks{Currently an Einstein \& JSI Fellow at the University of Maryland -- College Park. (Email: lblecha@astro.umd.edu.)}, Abraham Loeb$^1$, \& Ramesh Narayan$^1$\\ $^1$ Harvard University, Department of Astronomy, 60 Garden St., Cambridge, MA 02138, USA}

\begin{document}
\maketitle

\begin{abstract}
We present a first attempt to model the narrow-line (NL) region of active galactic nuclei (AGN) in hydrodynamic simulations of galaxy mergers, using a novel physical prescription. This model is used to determine the origin of double-peaked NL AGN in merging galaxies and their connection to supermassive black hole (SMBH) pairs, motivated by recent observations of such objects. We find that double-peaked NLs induced by the relative motion of dual SMBHs are a generic but short-lived feature of gaseous major mergers. Double-peaked NL AGN should often be observed in late-stage galaxy mergers, during the kiloparsec-scale phase of SMBH inspiral or soon after the nuclear coalescence and subsequent SMBH merger. However, even within the kiloparsec-scale phase, only a {\em minority} of double-peaked NLs are directly induced by the relative motion of binary SMBHs; their lifetimes are typically a few Myr. The majority of double-peaked NLs result from gas kinematics near the SMBH, although prior to the SMBH merger up to $\sim 80$\% of all double-peaked NL profiles may be {\em influenced} by SMBH motion via altered peak ratios or overall velocity offsets. The total lifetimes of double-peaked NL AGN depend strongly on viewing angle and on properties of the merging galaxies; gas-rich, nearly-equal-mass mergers have more NL AGN activity but may also be more obscured. Furthermore, in a typical merger,  at least 10 - 40\% of the double-peaked NLs induced by SMBH motion have small projected separations, $\sim$ 0.1 - 1 kpc, making it difficult to clearly identify dual peaks of stellar surface brightness. Diffuse tidal features can indicate a late-stage merger, although they do not distinguish an SMBH pair from a merged SMBH. We demonstrate that double-peaked NL AGN spectra with large peak velocity splittings ($\ga 500$ \kms) or with discernible overall velocity shifts are often associated with inspiraling SMBH pairs. Our results support the notion that selection of double-peaked NL AGN is a promising method for identifying dual SMBH candidates, but demonstrate the critical importance of high-resolution, multi-wavelength follow-up observations, and the use of multiple lines of evidence, for confirming the dual nature of candidate SMBH pairs.
\end{abstract}


\section{Introduction}
\label{sec:intro}
Although supermassive black hole (SMBH) pairs are a natural result of major galaxy mergers, until recently, evidence for their existence has been scarce. On large scales, when the SMBHs simply follow the motion of their host galaxies, some constraints are obtained from quasar clustering surveys. About 0.1\% of quasars are known to be in pairs on scales of $\la 1$ Mpc, and evidence for excess quasar clustering on small scales suggests that some of these are induced by galaxy interactions \citep{hennaw06}. 

In later stages of galaxy merging, when the SMBH separation is $\sim 1 - 10$ kpc, dynamical friction drives the evolution of the SMBH pair toward the center of the merger remnant. This ``kiloparsec-scale" phase has seen by far the most recent progress in identification of candidate SMBH pairs. A few spatially-resolved AGN pairs on these scales have been found {\rev via X-ray imaging} \citep{komoss03, bianchi08, green10,koss11a}, {\rev including, notably, a candidate AGN pair with $\sim 150$ pc separation \citep{fabbia11}}. However, {\em spectroscopic} surveys of AGN have found consistently that about 1\% of all AGN have double-peaked narrow \oiii\ lines, a possible signature of SMBH orbital motion on approximately kiloparsec scales \citep{comerf09a, smith10, liu10a}. This finding has increased the number of candidate SMBH pairs from a small handful to several hundred. Additionally, \citet{ge12} recently conducted a search of SDSS AGN spectra with more lenient selection criteria (examining asymmetric as well as double-peaked profiles and AGN+SF composite galaxies) and find that $\sim$ 1\% of {\em all} emission-line galaxies are double-peaked, of which 40\% (1,318 objects) are AGN or composite galaxies.

While only a fraction of these double-peaked narrow-line (dNL) AGN are expected to actually contain SMBH pairs, follow-up observations have already revealed strong evidence that some of these are in fact dual SMBHs. Observations of dNL AGN that combine ground-based, high-resolution optical and near-infrared imaging with long-slit spectroscopy \citep{liu10b,shen11} or integral field unit (IFU) data \citep{fu12,mcgurk11} have revealed numerous systems with two resolved stellar components in images that are spatially coincident with the two emission components in their spectra. These are among the most compelling dual SMBH candidates. \citet{shen11} estimate that 10-50\% of dNL AGN host dual SMBHs, and that, when corrected for sample completeness,  up to 2.5\% of Type 2 AGN at $z < 0.3$ are active, kpc-scale SMBH pairs. 

Additionally, \citet{fu11a} and \citet{rosari11} have used adaptive optics (AO) imaging to constrain the environment and host galaxy morphology of some dNL AGN; their estimates of the dual SMBH fraction are in broad agreement with those of \citet{shen11}. \citet{comerf12} have conducted analysis of 81 slit spectra obtained for dNL AGN; they find that {\em all} of these systems show spatially-separated emission components on approximately kpc scales, and they identify 14 of these as strong dual SMBH candidates. Still more compelling evidence for dual SMBHs comes from the detection of a dual compact X-ray \citep{comerf11} and radio \citep{fu11b} source, respectively, in two dNL AGN. In contrast, \citet{rosari10} combined slit spectroscopy with VLA imaging to demonstrate that the dual SMBH scenario is disfavored in two systems where radio jets are likely responsible for the double-peaked spectral features. Thus, follow-up observations of dNL AGN have already shown great promise for confirming the nature of these candidate dual SMBHs.

On smaller scales ($\la 10$ pc), SMBH pairs evolve to form a tightly bound binary SMBH. These present a formidable challenge for observers, partly because they are difficult if not impossible to resolve with current telescopes. Only one confirmed example of a tightly bound SMBH binary is known, with a separation of 7 parsec \citep[][]{rodrig06}. Numerous sub-parsec (spectroscopic) binary candidates have been proposed, but confirming their binary nature will likely require a better understanding of broad line physics \citep[][]{dotti09a,bogdan09,borlau09,tsalma11,eracle12}. Therefore, the kiloparsec-scale phase of SMBH pair evolution currently appears to be the most promising avenue for studying dual SMBHs.

Theoretical studies of SMBH mergers face their own challenges, largely owing to the vast range of physical scales involved.  Significant progress has been made on the smallest scales; simulations of BH mergers using full general relativity are now possible and can generate precise waveforms of GW emission as well as the remnant BH properties \citep[e.g.,][]{pretor05,campan06,baker06}. However, there is still much uncertainty regarding the timescale required for the binary to evolve from scales of a few parsec down to milliparsec scales where gravitational-wave (GW) emission dominates the orbital decay.  In highly symmetric, spheroidal galaxies with little to no gas, SMBH binaries may ``stall" at $\sim 1$ parsec for more than a Hubble time \citep[e.g.,][]{begelm80, milmer01, yu02}.  If the galaxy is gas-rich or triaxial, however, the SMBHs may merge on a much shorter timescale \citep[$\sim 10^6 - 10^7$ yr from the hard binary stage; e.g.,][]{gerbin85, yu02, armnat02, berczi06, escala04, goumir97}.  

Galaxy mergers and SMBH pair formation have been studied extensively with hydrodynamic simulations \citep[e.g.,][]{dimatt05,robert06b,hopkin06,calleg09,coldot11}. \citet{vanwas12} have studied the triggering of AGN pairs in merging galaxies, finding that luminous dual AGN occur most frequently in the late phases of merging, for pair separations $\la$ 10 kpc, and that much of the activity in the two nuclei is not simultaneous. However, despite the current observational focus on NL signatures of dual SMBHs, galaxy merger simulations have not considered the NL region (NLR), and detailed photoionization models have not been applied to the rapidly-varying environment of a late-stage merger. Here, we make a first attempt to model the NLR during galaxy mergers using hydrodynamic simulations, with special attention to the kiloparsec-scale phase that may produce double-peaked NL AGN.

This paper is organized as follows. In \S~\ref{ssec:simulations} \&~\ref{ssec:models}, we describe our simulations and galaxy merger models. Our semianalytic prescription for the NL gas is detailed in \S~\ref{ssec:nlgas} - \S~\ref{ssec:lhbeta}. Our results are presented in \S~\ref{sec:results}. In \S~\ref{ssec:general} \& \S~\ref{ssec:params}, we describe the evolution of the NLR throughout a major merger and discuss dependence on merger parameters. We describe the morphological properties of the NLRs in \S~\ref{ssec:nlmorph}, and in \S~\ref{ssec:dnlkin} we explore the observable signatures of kiloparsec-scale double-NL AGN. The lifetimes of double-NL AGN are discussed in \S~\ref{ssec:tdnl}. Finally, we summarize and discuss our results in \S~\ref{sec:disc}. Throughout the paper, we assume a flat $\Lambda$CDM cosmology with $H_0 = 71$ km s$^{-1}$ Mpc$^{-1}$, $\Omega_{\rm m} = 0.27$, and $\Omega_{\Lambda} = 0.73$.

\section{Methods}
\label{sec:methods}

\subsection{Simulations}
\label{ssec:simulations}

For our simulations of SMBH pairs in galaxy mergers, we use \gadget, a smoothed-particle hydrodynamics (SPH) and N-body code that conserves both energy and entropy \citep{spring05a}. The version we use (\gadthree) includes radiative cooling as well as a sub-resolution model for a multi-phase interstellar medium \citep[ISM, ][]{sprher03} that accounts for star formation and supernova feedback.  In addition, the code models SMBHs as gravitational ``sink" particles that contain a SMBH seed and a gas reservoir.  The reservoir is replenished by stochastic accretion of neighboring gas particles, but the actual accretion rate onto the SMBH is calculated smoothly using the Bondi-Hoyle formula \citep{bonhoy44} with locally-averaged values for the density and sound speed.  The accretion rate is modified by a multiplicative factor to account for the increase in density and sound speed that should occur on sub-resolution scales near the SMBH accretion disk; we adopt a standard value of 100. The Bondi-Hoyle model is of course a simplification of the accretion process, as some assumptions must be made in any large-scale simulations that include BHs. The exact nature of gas accretion from large scales down to the BH is not well understood, and the Bondi model has some physical motivation in that the accretion should scale with the local gas density and sound speed, at least in an average sense.

We include a model for thermal feedback from the SMBH by assuming that 5\% of the luminous output from the SMBH is coupled to the surrounding gas particle as thermal energy. We note that this model does not allow for the possibility of AGN outflows, and thus we are unable to model this possible mechanism for producing some double-peaked spectral features in AGN. Angular momentum is conserved during accretion of gas particles, but because this is a stochastic process we also introduce an accretion drag force calculated from the Bondi accretion rate.  These prescriptions are described in more detail in \citet{spring05b}.

\begin{table}
\begin{center}
\begin{tabular}{r r r r r r r r r}
$q$ & \fgas\ & B/T & $M_{\rm seed}$ & $q_{\rm EOS}$ & orbit & $f_{\rm res}$ \\
 &  & & [$10^6$ \msun] & & & \\
\hline 
1.0  & 0.3 & 0.0 & 0.14 & 0.25 & a & 1.0 \\
1.0  & 0.1 & 0.0 & 0.14 & 0.25 & a & 1.0 \\
1.0  & 0.04 & 0.0 & 0.14 & 0.25 & a & 1.0 \\
0.5  & 0.3 & 0.0 & 0.14 & 0.25 & a & 1.0 \\
0.5  & 0.1 & 0.0 & 0.14 & 0.25 & a & 1.0 \\
0.5  & 0.04 & 0.0 & 0.14 & 0.25 & a & 1.0 \\
0.333  & 0.3 & 0.0 & 0.14 & 0.25 & a & 1.0 \\
0.333  & 0.1 & 0.0 & 0.14 & 0.25 & a & 1.0 \\
\hline
1.0  & 0.1 & 0.0 & 0.14 & 0.25 & b & 1.0 \\
1.0  & 0.1 & 0.0 & 0.14 & 0.25 & c & 1.0 \\
1.0  & 0.1 & 0.0 & 0.14 & 0.25 & d & 1.0 \\
\hline
0.5 & 0.3 & 0.2 & 13 & 0.25 & a & 1.0 \\
0.5 & 0.3 & 0.3 & 24 & 0.25 & a & 1.0 \\
0.5 & 0.1 & 0.2 & 13 & 0.25 & a & 1.0 \\
0.5 & 0.1 & 0.3 & 24 & 0.25 & a & 1.0 \\
\hline
1.0  & 0.04 & 0.0 & 0.14 & 0.5 & a & 1.0 \\
1.0  & 0.04 & 0.0 & 0.14 & 0.05 & a & 1.0 \\
0.5  & 0.1 & 0.0 & 0.14 & 0.5 & a & 1.0 \\
0.5  & 0.1 & 0.0 & 0.14 & 0.05 & a & 1.0 \\
0.333  & 0.3 & 0.0 & 0.14 & 0.5 & a & 1.0 \\
0.333  & 0.3 & 0.0 & 0.14 & 0.05 & a & 1.0 \\
\hline
0.5  & 0.1 & 0.0 & 0.14 & 0.25 & a & 0.5 \\
0.5  & 0.1 & 0.0 & 0.14 & 0.25 & a & b0.5dm2.5 \\
0.5  & 0.1 & 0.0 & 0.14 & 0.25 & a & b1.0dm5.0 \\
0.5  & 0.1 & 0.0 & 0.14 & 0.25 & a & 2.0 \\
\hline 
\hline
1.0 & 0.3 & 0.3 & 32 & 0.25 & a & 0.1 \\
1.0 & 0.3 & 0.2 & 17 & 0.25 & a & 0.1 \\
1.0 & 0.1 & 0.3 & 32 & 0.25 & a & 0.1 \\
0.5 & 0.5 & 0.2 & 13 & 0.25 & a & 0.1 \\
0.5 & 0.3 & 0.2 & 13 & 0.25 & a & 0.1 \\
0.5 & 0.1 & 0.3 & 24 & 0.25 & a & 0.1 \\
0.333 & 0.3 & 0.3 & 21 & 0.25 & a & 0.1 \\
\end{tabular}
\end{center}
\caption{Parameters of merger simulations used. In all models, the primary galaxy has a total mass (baryons + DM) of $1.4 \times 10^{12}$ \msun; $q$ indicates the mass ratio of the secondary galaxy. \fgas\ indicates the fraction by mass of gas in each disk. {\rev B/T gives the bulge-to-total baryonic mass ratio, and} $M_{\rm seed}$ is the initial seed mass of the SMBH in each galaxy. $q_{\rm EOS}$ is the softening factor for the gas equation of state, as described in the text. Each merger orbit is assigned a label and its parameters are detailed in Table~\ref{table:orbits}. Finally, $f_{\rm res}$ is the factor by which the mass resolution is varied with respect to the fiducial resolution, either for all particles or for baryonic ('b') or dark matter ('dm') separately. When the mass resolution is varied by $f_{\rm res}$, the gravitational softening length is varied by $f_{\rm res}^{1/3}$. Note that the last section of models (separated by a double line) is a low-resolution set of simulations designed to better explore the parameter space of galaxies with initial stellar bulges. The NL model is not applied to these simulations owing to their lower resolution. \label{table:models}}
\end{table}

\subsection{Galaxy and SMBH Merger Models}
\label{ssec:models}

The progenitor galaxies for our merger simulations consist of a dark matter (DM) halo, a disk of gas and stars, and a central SMBH sink particle as described above.  We also include a stellar \citet{hernqu90} bulge component in some models.  In each case the total mass of the primary galaxy is $M_{\rm tot} = 1.4 \times 10^{12}$ \msun, and 4.1\% of this mass is in a gas and stellar disk. In the models that also have a stellar bulge, the total baryonic fraction is 5.1 - 5.9\%. The angular momentum of each DM halo is determined as in \citep{spring05b}, with a dimensionless spin parameter $\lambda = 0.033$. The disk angular momentum is assigned to be a fraction $M_{\rm disk}/M_{\rm tot}$ of the total angular momentum, where $M_{\rm disk}$ is the disk mass. We simulate major galaxy mergers with mass ratios of $q = $1, 0.5, \& 0.333, and the initial gas fraction by mass of the disk is varied between 4 and 30\%. We use fairly high mass and spatial resolution in order to resolve as best as possible the NLR around each SMBH. In the fiducial simulations, the gravitational softening length adopted is $r_{\rm soft} = 37$ pc for baryons and $r_{\rm soft,DM} = 111$ pc for DM, and the particle masses for each particle type are $m_{\rm star} = 4.2\times 10^4$ \msun, $m_{\rm gas} = 2.8 \times 10^4$ \msun, and $m_{\rm DM} = 5.4 \times 10^5$ \msun. 

Each galaxy is given a central SMBH. In the fiducial simulations, the galaxies are initially bulgeless, so a small seed mass ($1.4\times10^5$ \msun) is assumed, motivated by the observed BH-bulge relations. In these simulations, the SMBH masses grow rapidly after the first close encounter of the galaxies, such that by the time the SMBHs merge, their masses are a few $\times 10^6 - 10^7$ \msun. The galaxy merger also causes a stellar bulge to form, such that the final merger remnant is in good agreement with the BH-bulge relations. The fact that these SMBHs grow by factors of 10 - 100 during the simulation (and ultimately lie on the BH-bulge relations) motivates the choice of small SMBH seed mass, as the initial mass is insignificant compared to the accreted mass. Nonetheless, the results of present study may be affected by the {\em timing} of this growth, for which both the initial SMBH mass and bulge mass may play a role. Thus, we consider some simulations in which the progenitor galaxies contain a stellar bulge initially. In these cases, the initial BH seed mass is chosen according to the $M_{\rm BH}$-$M_{\rm bulge}$ relation measured by \citet{harrix04}. {\rev The stellar bulge is initialized with a \citet{hernqu90} profile and a bulge-to-total baryonic mass ratio (B/T) of 0.2 - 0.3, where ``total" refers to the disk+bulge mass.}

We use the same SMBH merger prescription as in \citet{blecha11}, in that the SMBH merger time (\tmrg) is determined based on the SMBH separation ($a_{\rm sep}$) and relative velocity ($v_{\rm rel}$), and then the simulation is restarted at an earlier point to merge the SMBHs at that exact time. We define \tmrg\ as the time at which $a_{\rm sep} < r_{\rm soft}$ and $v_{\rm rel} < 0.5\, c_{\rm sound}$. Restarting the simulations is particularly important for our present study, because a detailed analysis of the short-lived kpc-scale phase of SMBH evolution requires good time resolution. In order to provide this resolution where needed without generating excessive data in the early merger phase, we run the initial simulations with a time resolution (snapshot output frequency) of 10 Myr, and restart the simulation 100 Myr before the kpc-scale phase (as defined below) with a time resolution of 1 Myr. 

Table~\ref{table:models} summarizes the parameters of the merger simulations conducted. The different sections of the table denote (i) the fiducial set of merger simulations, followed by simulations with (ii) varied merger orbits (the details of which are given in Table~\ref{table:orbits}) (iii) galaxies with initial bulges and correspondingly larger BH masses (iv) varying values of the ISM equation-of-state parameter (v) varying mass and spatial resolution. The final section (separated by a double line) denotes a set of simulations with initial galactic bulges and larger BH masses conducted at 10x lower resolution, in order to better explore this parameter space. Two of these simulations have identical initial conditions as in the fiducial-resolution set, to ensure that the lower mass resolution does not influence the results. Even so, we do not perform NL modeling with these low resolution simulations, but rather focus on the evolution of the SFR and BH accretion rate.

\setlength{\tabcolsep}{5pt}
\begin{table}
\begin{center}
\begin{tabular}{l r r r r r r l}

Orbit & $a_{\rm init}$ & $r_{\rm peri}$ & $\theta_1$ & $\phi_1$ & $\theta_2$ & $\phi_2$ & Description \\
 & [kpc] & [kpc] & [deg] & [deg] & [deg] & [deg] & \\
 \hline
 % orb0
 a & 143 & 7.1 & -30 & 45 & 30 & 60 & fiducial \\
% orb2
 b & 143 & 14.3 & -30 & 45 & 30 & 60 & large peri. \\
% orb6
 c & 143 & 7.1 & 0 & 0 & 90 & 0 & polar \\
% orb11
 d & 143 & 7.1 & -10 & 0 & 10 & 0 & near-coplanar\\

\end{tabular}
\end{center}
\caption{Initial orbital parameters for galaxy mergers simulated. All orbits are initially parabolic. $\theta$ and $\phi$ denote the angular momentum vector of each galaxy in the global coordinate system, such that $(\theta, \phi) = (0, 0)$ is oriented along the positive $z$-axis. The fiducial model is a ``generic" orbit, in that there are no symmetries in the initial disk orientations. In all orbits, the initial separation of the galaxies is set such that $a_{\rm init}/R_{200} = 0.625$. Note that the actual orbital parameters of the merger vary over time owing to dynamical friction. \label{table:orbits}}
\end{table}

\subsection{NLR Identification}
\label{ssec:nlgas}

Here we outline a basic procedure for determining the location, kinematics, and (\hbeta) luminosity of the narrow-line region around one or more AGN in a gaseous galaxy merger simulation with \gadthree. All of the calculations described here are done in post-processing, that is, after the \gadget\ simulation has finished.

\subsubsection{BH accretion \& luminosity}
As described in \S~\ref{ssec:simulations}, the accretion rate onto the BHs is calculated using the Bondi-Hoyle formula \citep{bonhoy44} with a multiplicative factor, capped at the Eddington limit:
\begin{eqnarray}
\dot M = {\rm min}( f_{\rm acc} \dot M_{\rm Bondi}, \dot M_{\rm Edd} ),  \\
\dot M_{\rm Bondi} = { 4 \pi (G M_{\rm BH})^2 \rho_{\rm g,\infty}  \over (v_{\rm rel}^2 + c_{\rm s,\infty}^2)^{3/2} } \\
{\dot M}_{\rm Edd} = {4\pi G M_{\rm BH} \over \eta\, \kappa_{\rm es\,} c }
\end{eqnarray}
where $v_{\rm rel}$, $c_{\rm s, \infty}$, and $\rho_{\rm g, \infty}$ are all computed as averages over the gas particles neighboring the SMBH, $\kappa_{\rm es}$ is the electron-scattering opacity, and $\eta = 0.1$ is the radiative efficiency. Using this accretion rate, we can calculate the bolometric luminosity
\begin{equation}
L_{\rm bol} = \eta \dot M \, c^2.
\end{equation}
Here, $\eta = 0.1$ unless $\dot M \ll \dot M_{\rm Edd}$, in which case the SMBH is assumed to be radiatively inefficient, with the following scaling for $\eta$ \citep{narmcc08}:
\begin{equation}
\eta = 0.1 \left ({ \dot M \over 0.01 \dot M_{\rm Edd}} \right),\:\: \dot M < 0.01 \dot M_{\rm Edd}.
\end{equation}

\subsubsection{Ionizing photon production rate}
To calculate properties of the NL region, we need to know the number of ionizing photons produced by the accreting SMBHs.  We begin by applying a reverse bolometric correction to $L_{\rm bol}$ to obtain the B-band luminosity, and then we assume a broken power-law form for the optical - UV spectrum.  Following the method of \citet{marcon04}, the bolometric correction is:
\begin{equation}
\log \left ( { L_{\rm bol} \over \nu_{\rm B} L_{\nu_{\rm B}} }\right ) = 0.8 - 0.067\mathcal{L} + 0.017\mathcal{L}^2 - 0.0023\mathcal{L}^3,
\end{equation}
where $\mathcal{L} = \log L_{\rm bol} - 12$ and $L_{\rm bol}$ is in units of $L_{\odot}$.
For the broken power-law, we use the following spectral indices from the AGN SED fits of \citet{marcon04}: 
\begin{eqnarray} 
\alpha_{\rm opt} = -0.44,\:\: 1\, \mu {\rm m} > \lambda > 1300 {\rm \AA} \\
\alpha_{\rm UV} = -1.76,\:\: 1200 {\rm \AA} > \lambda > 500 {\rm \AA}.  
\end{eqnarray}
Then we can get the normalization factor $A_{\rm opt}$ from the bolometric correction by taking $ \nu_{\rm B} L_{\nu_{\rm B}} = A_{\rm opt} \nu_{\rm B}^{\alpha_{\rm opt}+1}$.  To get the normalization for the UV range we can take $\nu L_{\nu}(1300 {\rm \AA}) = \nu L_{\nu} (1200 {\rm \AA})$, which gives
\begin{equation}
{A_{\rm UV} \over A_{\rm opt}} = 2.02 \times 10^{20}.
\end{equation}

The normalization $A_{\rm UV}$ allows us to calculate the number of ionizing photons emitted per unit time by the source:
\begin{equation}
Q = \int_{\nu_0}^{\infty} { L_{\nu} \over h \nu } d\nu,
\end{equation}
where $\nu_0 = 13.6$ eV.

\subsubsection{Selection criterion for cold-phase gas}
\label{sssec:coldgas}
In order to determine which gas in the galactic center will be ionized by the AGN and produce narrow-line emission, we must impose several criteria on the SPH particles. The multi-phase model for the ISM in \gadget\ assumes that the gas is comprised of a ``cold" phase and a ``hot" phase that can exchange mass via star formation, cooling, and cloud evaporation by supernovae. First, we select only those gas particles that have a nonzero fraction of mass $f_{\rm cold}$ in the ``cold" phase. We assume that the cold-phase gas has fragmented into discrete clouds on sub-resolution scales. The NL clouds are heated by photoionization and should be warmer than the cold clouds in the multiphase ISM ($\sim 10^4$ K versus $\sim 10^3$ K) and thus less dense (assuming pressure equilibrium between phases).  Therefore, we use the parameters of the multiphase model only for selection of SPH particles that contain cold gas, and we instead calculate a ``cloud density" for each of these particles: $\rho_{\rm cl} =  \rho_{\rm sph} ( T_{\rm sph} / 10^4$ K). (Here, the subscript ``sph" denotes a mass-weighted average between the hot and cold phases, and the subscript ``cl" denotes the quantities for our NL cloud model.)

\subsubsection{Selection criterion for gas particle covering fraction}
\label{sssec:fomega}
We also impose a criterion on the SPH particles such that the solid angle subtended by NL clouds does not exceed $4\pi$. We cannot determine exactly which clouds will have unobscured sight lines to the AGN, as their size is below our resolution limit and our simulations do not include radiative transfer. However, we account for the problem of particle self-shielding in an average sense, as follows. An SPH particle that subtends a solid angle $\Omega_{\rm sph}$ has a covering fraction 
\begin{eqnarray}
f_{\Omega_{\rm sph}} = {\Omega_{\rm sph} \over 4\pi} = {r_{\rm sph}^2 \over 4\, r_{\rm BH}^2}, \\
r_{\rm sph} = \left( {m_{\rm sph} \over {4\pi\over3} \left < \rho \right > } \right )^{1/3},
\end{eqnarray}
where $m_{\rm sph}$ is the mass of the SPH particle, $r_{\rm sph}$ is the effective size of the particle and $r_{\rm BH}$ is the distance to the SMBH. The ``area filling factor" ($\epsilon_{\rm A}$) and volume filling factor ($\epsilon_{\rm V}$) are the fractions of the SPH particle's area and volume covered by the cold clouds, respectively:
\begin{eqnarray}
\epsilon_{\rm A} = \epsilon_{\rm V}^{2/3}\, N_{\rm cl}^{1/3}, \\[2pt]
\epsilon_{\rm V} = f_{\rm cold}\, {\rho_{\rm sph} \over \rho_{\rm cl}},
\end{eqnarray} 
where $N_{\rm cl} = f_{\rm cold}\,m_{\rm sph}  / m_{\rm cl}$ is the number of clouds within the particle. In this formulation, $m_{\rm cl}$ is a free parameter that serves mainly to set the surface area to volume ratio of the NL clouds.  For a typical area covering fraction of order unity on NLR scales of $\sim 0.1 - 1$ kpc (as inferred observationally), we are led to adopt  $m_{\rm cl} = 200 M_{\odot}$.

The covering fraction of the clouds in each particle is then
\begin{equation}
f_{\Omega} = {\epsilon_{\rm A}\, \Omega_{\rm sph} \over 4\pi}.
\end{equation}

In order to avoid allowing a photon to be absorbed by multiple clouds, we truncate the NLR beyond the radius where the total covering fraction of clouds reaches unity. While this method is not exact, it does allow for the correct rate of photoionization in an average sense. 

\subsubsection{Selection criteria for ionization parameter \& density}
\label{sssec:U_ndens}

For the selected gas particles, we calculate the ionization parameter for the cold clouds in a given SPH particle when ionized by a single SMBH:
\begin{equation}
U = {Q \over 4 \pi r_{\rm BH}^2 c\, n_{\rm H,cl}}, 
\end{equation}
where $n_{\rm H,cl} = \rho_{\rm cl} / \mu m_{\rm H}$ is the number density of the NL clouds and $r_{\rm BH}$ is the distance from the SMBH to the cloud. The ionization parameter quantifies the ratio of the ionizing photon density to the electron density at each cloud. In the case of a galaxy merger where two active SMBHs may be present, the ionization parameter becomes 
\begin{equation}
U = {1 \over 4 \pi c\, n_{\rm H,cl}} \left ( {Q_1 \over r_{\rm BH1}^2} + {Q_2 \over r_{\rm BH2}^2} \right ),
\end{equation}
where the subscripts ``1" and ``2" refer to the primary and secondary SMBHs. We impose an additional cut on the NL gas particles by selecting only those with $U$ in the range $10^{-4.5} - 10^{-1.5}$, under the assumption that the gas is transparent to ionizing radiation for AGN luminosities.

Finally, to ensure that the cloud densities are reasonable, we impose a final cut on the SPH particles such that only those with $n_{\rm H, cl}$  in the range $10^2 - 10^6$ cm$^{-3}$ are included.  The maximum of this range is more often the limiting criterion, and is chosen to be roughly equal to the critical density for \oiii\ emission. Above this density, collisional de-excitation begins to dominate over the forbidden-line emission. While we do not consider forbidden lines in our model, we know they are present in real NLRs, and further that NL AGN typically have \oiii/\hbeta\ $\ga 3$. Thus, gas above the maximum density will not contribute to the NL AGN profile in a conventional manner, and we exclude these particles from our NL model. 

Average densities in the simulated NL gas are typically $n_{\rm H, cl} \sim 10^3 - 10^4$ cm$^{-3}$, in agreement with many photoionization models of NL AGN \citep[e.g.,][]{koski78,groves04,villar08,kraeme09}. However, for short periods at the height of merger activity (when central gas density increases dramatically in response to the rapid inflow of cold gas), the NLRs may be dominated by gas with densities as high as $10^5 - 10^6$ cm$^{-3}$. Although these large average densities are atypical of NL-emitting gas, we stress that the environment of a major galaxy merger is itself atypical, and that standard photoionization models have not been applied to such extreme, rapidly-varying conditions. Given that gas densities are known to be higher during the peaks of merger activity, and that the NLR model yields reasonable densities throughout the rest of the simulations, we consider the NL densities in our models to be consistent with physical intuition.

We consider a NLR to be ``active" if at least 10 SPH particles meet all of the above criteria, but in practice the NLRs in our simulations typically contain hundreds to thousands of SPH particles.

\subsubsection{Identification of gas particles with each SMBH}
After the galaxies have undergone a close passage, and especially during their final coalescence, particles are easily exchanged between galaxies, and the initial identifications of which particles are in which galaxy are no longer relevant. Nonetheless, we wish to know which NL particles are associated with each SMBH; this is helpful for understanding the NLR kinematics even when the SMBHs are near coalescence. Accordingly, we assign particles to each SMBH based on their proximity to and degree of photoionization from each SMBH. Specifically, we switch gas particles from their initial galaxy identification if (a) they are closer to the SMBH in the other galaxy and (b) the quantity $Q/r_{\rm BH}^2$ is substantially larger for the SMBH in the other galaxy (we use $U_{1,2}/U_{2,1} > 4$). 

\subsection{\hbeta\ Luminosity and Velocity Profiles}
\label{ssec:lhbeta}

\subsubsection{\hbeta\  luminosity}
Once we have selected the NL particles according to the above procedure, we may estimate the \hbeta\ luminosity, \lhbeta, of each particle. While \oiii\ is generally the strongest narrow emission line, and thus has been the focus of most searches for dNL AGN, for our models the \hbeta\ line offers the advantage that its recombination coefficient depends only weakly on temperature, and thus the line strength is much less sensitive to the exact conditions in the ISM. Because AGN typically have \oiii/\hbeta\ $\ga$ 3 - 10, our results for the lifetime of observable NL emission, calculated from \lhbeta, may be considered lower limits in this sense. \lhbeta\ may be written as
\begin{equation}
L_{\rm H\beta} = h\, \nu_{\rm H\beta}\, {\alpha^{\rm eff}_{\rm H\beta}(T) \over \alpha^{\rm eff}_{\rm  B}(T)}\, f_{\Omega}\, Q\,  ,
\end{equation}
where $Q$ is the rate of ionizing photon production as before, $f_{\Omega}$ is the cloud covering fraction discussed above, and $\alpha^{\rm eff}_{\rm H\beta}(T) / \alpha^{\rm eff}_{\rm B}(T) \approx 1/8.5$ is the number of \hbeta\ photons produced per hydrogen recombination for $T = 10^4$ K \citep{ostfer06}.
Thus, when two ionizing sources (two SMBHs) are present, the total \hbeta\ luminosity is
\begin{equation}
L_{\rm H\beta} = {h\, \nu_{\rm H\beta} \over 8.5 } { \epsilon_{\rm A} \over 4 \pi }  \left ( {\Omega_{1,{\rm sph}}\, Q_1} + {\Omega_{2,{\rm sph}}\, Q_2} \right ).
\end{equation}

\subsubsection{\hbeta\ velocity profiles}
\label{sssec:hbeta_profs}
In order to understand the kinematics of the NLR as they relate to observations, we construct and analyze velocity profiles for each NLR. From the simulations we have the 3-D velocities for each NL-emitting SPH particle, measured with respect to the stellar center of mass. After projecting these along a given sight line, we assume that within each particle, the NL clouds have a Gaussian internal velocity dispersion with a full-width at half maximum (FWHM) equal to 0.5 $c_{\rm sound}$. The resulting total velocity profile is then convolved with another Gaussian to degrade it to the desired resolution, chosen to correspond to the spectral resolution of SDSS or DEIMOS at typical redshifts for dNL AGN (we use a fiducial value of 65 \kms). In order to understand the dependence on viewing angle, this procedure is repeated for 40 random sight lines for each snapshot.

We also apply criteria to determine whether a given NL profile should appear double-peaked. In order for the two peaks to be distinct, the ratio of peak luminosities must be greater than 0.05, and the velocity separation of the two peaks much be greater than their FWHM. We additionally require that the peak luminosity be $> 5 \times 10^4$ \lsun, corresponding to a minimum observable line flux of $\sim 10^{-17}$ erg s$^{-1}$ cm$^{-2}$ for objects at the mean redshift of the SDSS sample \citep[$z \sim 0.1$; limit based on][and Y. Shen, private communication]{liu10a}. As AGN typically have \oiii/\hbeta\ flux ratios $\ga$ 3-10,  \oiii\ should be easily detected. even when \lhbeta\ is near this detectability limit.

\section{Results}
\label{sec:results}

\subsection{General Properties of Narrow-Line AGN in Mergers}
\label{ssec:general}

The maximum angular SMBH separation for which double-peaked profiles resulting from distinct, orbiting NLRs could be seen on the same spectrum is set by the size of the spectral slit or fiber of the instrument used. The size of the spectral slit on the DEIMOS spectrograph is $0.75"$, corresponding to 5.36 kpc at $z = 0.7$.  Thus, only a dual SMBH with separation $a_{\rm max} \la 5.36$ kpc could be observed in a single DEIMOS spectrum.  The diameter of the SDSS spectroscopic fiber is 3", which projects to 5.47 kpc and 21.4 kpc at $z = 0.1$ \& 0.7, respectively.  Thus, for the mean redshifts of the AGN samples studied by \citet{comerf09a, smith10, liu10a}, the dNL AGN diagnostic is sensitive to dual SMBHs with projected separation $\la 5.5$ kpc. Most of our results assume $a_{\rm max} = 5.5$ kpc, but we also consider $a_{\rm max} = 21$ kpc, corresponding to higher-redshift systems.

For our analysis of NLRs in galaxy merger simulations, we divide the merger evolution into phases based on these limits for observing double NLs.  We refer to the early merger stage as Phase I, when the NLRs are well-separated and could not be observed in a single spectrum ($a_{\rm sep} > a_{\rm max}$).  Phase II refers to the ``kiloparsec-scale phase", which occurs when the following criteria are met: (i) $a_{\rm sep} < a_{\rm max}$, (ii) at least one SMBH has an active NLR, and (iii) the SMBHs have not yet merged.  We define the post-BH-merger phase as Phase III.  If for any time between Phases I \& III the criteria (i) \& (ii) are not met, we define this period as Phase IIb. 

\begin{figure}
\resizebox{\hsize}{!}{\includegraphics{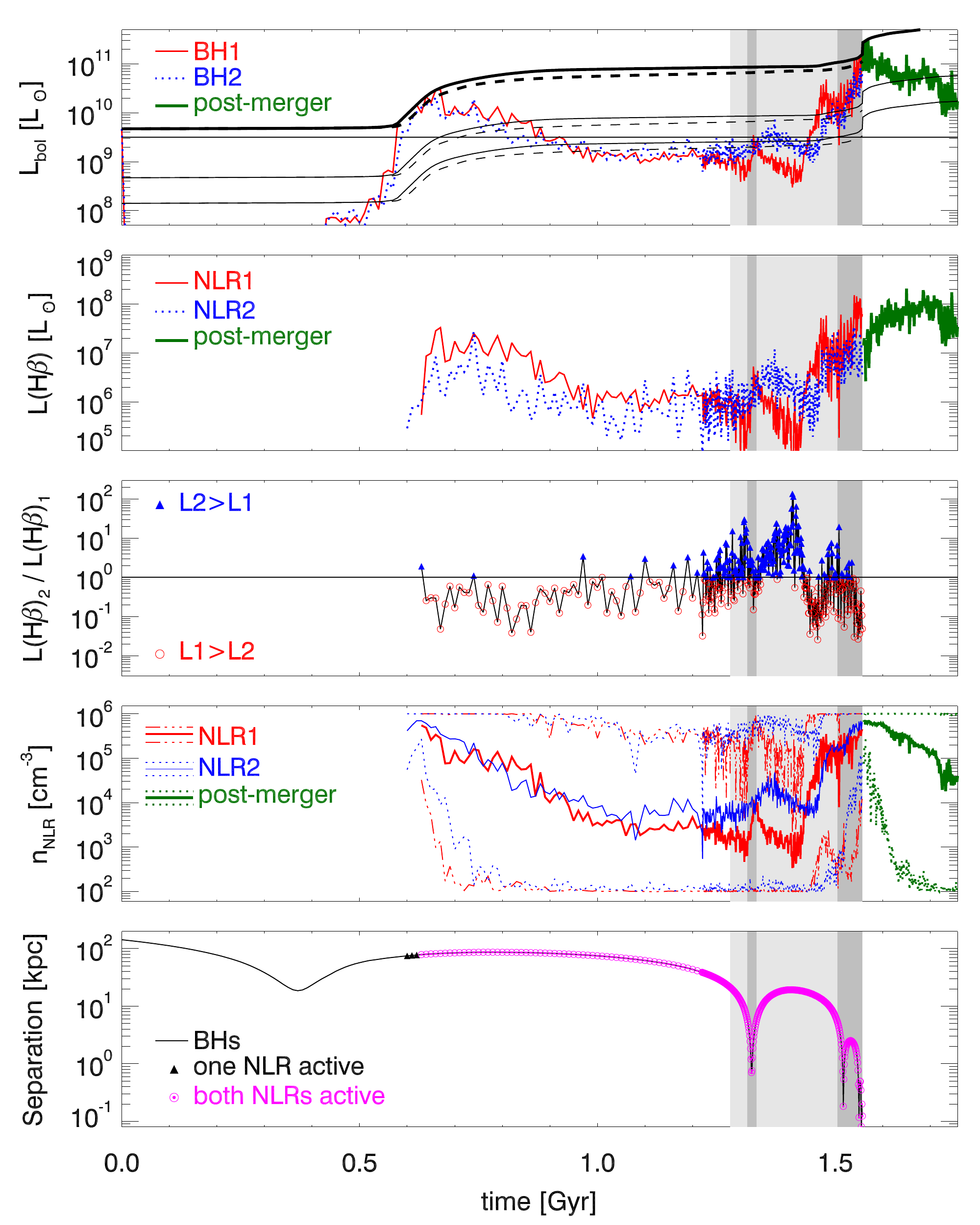}}
\caption[]{Time evolution of relevant quantities for an equal-mass merger containing 10\% gas initially. In each panel, the gray-shaded regions denote Phase II, i.e., the kpc-scale phase. The dark-gray shade denotes (3-D) SMBH separations of $< 5.5$ kpc, which corresponds to the size of the SDSS spectroscopic fiber for objects at $z \sim 0.1$, or the size of the DEIMOS spectral slit for objects at $z \sim 0.7$. The light-gray shade denotes SMBH separations of $< 21$ kpc, corresponding to the size of the SDSS fiber for objects at $z \sim 0.7$. Top plot: bolometric luminosity ($L_{\rm bol}$) versus time. Blue {\rev dotted} and red {\rev solid} curves denote each SMBH's luminosity prior to the SMBH merger, and the {\rev thick} green curve denotes $L_{\rm bol}$ after the merger. The thick solid and dashed {\rev black} lines (uppermost) denote the Eddington limit for each SMBH, and the thin solid and {\rev dashed black} lines denote three different definitions of an AGN (3\% \& 10\% $L_{\rm Edd}$ for each SMBH, and a constant-luminosity definition of $3\times10^9$ L$_{\odot}$). Upper-middle plot: \hbeta\ luminosity, $L$(\hbeta). Same color scheme as in top plot. Middle plot: Ratio of \lhbeta\ for each SMBH. {\rev Solid blue triangles denote points where $L_{\rm H\beta 2} > L_{\rm H\beta 1}$}, and {\rev open red circles denote where $L_{\rm H\beta 1} > L_{\rm H\beta 2}$}. Lower-middle plot: Number density of NLR gas {\rev particles. Thick red and thin blue solid lines denote the mean number density for all gas particles in NLR1 and NLR2, respectively, while triple-dot-dashed red and dotted blue lines denote the maximum and minimum densities for each respective NLR. Thick solid and dotted green lines similarly denote the post-merger NLR densities.} Bottom plot: SMBH separation vs time. The magenta {\rev circles} denote the NLR center-of-mass separation versus time for the snapshots in which both NLRs are simultaneously {\rev active, while black triangles denote where only one NLR is active.} 
\label{fig:lc}}
\end{figure}

In Fig.~\ref{fig:lc}, we show the evolution of AGN and NL activity throughout a major merger. The \hbeta\ luminosity traces the bolometric luminosity; both curves have Eddington-limited peaks after the first pericentric passage of the two galaxies and a larger peak during the final coalescence. Note that the NLR is not active at all for the first 600 Myr of the merger simulation, owing to the low SMBH luminosity. Only after the galaxies undergo a close pericentric passage do the gas particles meet our minimum criteria for an active NLR, after which they remain active for the rest of the simulation. The \hbeta\ luminosities of the two NLRs are within a factor of ten throughout most, but not all, of the simulation. 

Figure \ref{fig:lc} also demonstrates a key feature of AGN triggered by major, gaseous mergers: the peak AGN and NL activity often occurs during the final coalescence of the two galaxies. Under the assumption of efficient SMBH mergers, this means that peak activity should also occur near the time of the SMBH merger. Thus, merger-triggered AGN are typically brightest between the kiloparsec-scale phase of SMBH inspiral and the post-BH-merger phase. This simple fact enhances the probability of observing dNL AGN in the kpc-scale phase versus earlier stages in the merger, as is discussed further in \S~\ref{ssec:tdnl}. This statement does have some dependence on the initial morphology and BH mass of the progenitors; in particular, relatively gas-poor, disk-dominated galaxies with BHs and stellar bulges that lie on the $M_{\rm BH}$-$M_{\rm bulge}$ relation prior to merger may have more AGN activity in the early merger phase than at final coalescence. The dependence of NL AGN activity on galaxy morphology is discussed in detail in \S~\ref{sssec:morphology}, and it is shown despite this caveat, our conclusion that observed NL AGN activity in major mergers is dominated by the late merger stages agrees with observations, at least for double-peaked NL AGN.

Another interesting feature in Fig.~\ref{fig:lc} is a slight dip in \lhbeta\ that occurs near the time of SMBH merger and peak $L_{\rm bol}$ (note the much sharper peak in $L_{\rm bol}$ versus \lhbeta). This occurs when the central gas density in the merger remnant reaches its peak and some of the gas briefly exceeds our maximum density criterion for NLRs, as shown in the NL gas density panel Fig.~\ref{fig:lc}. As described in \S~\ref{sssec:U_ndens}, in such conditions the emission from collisional de-excitation begins to outweigh forbidden-line emission, so this gas cannot be considered part of the narrow-line region as it is typically defined for an AGN (with, e.g., \oiii/\hbeta\ $\ga 3$). For most of the simulation, NL gas densities span the full allowed range of $n_{\rm NLR} = 10^2 - 10^6$ cm$^{-3}$, but for short periods after the first passage and at coalescence, the NLRs are dominated by high-density ($n_{\rm NLR} > 10^5$ cm$^{-3}$) gas. The slight dip in NL emission occurs only in the mergers that attain the highest peak densities, i.e., those with nearly equal mass and moderate-to-high gas fractions, but it is nonetheless an interesting feature of NL emission in a dense, rapidly-varying environment. We note that because these dips in emission are brief, they do not affect our results qualitatively and  have only a small effect on our quantitative results. Specifically, the dNL AGN lifetimes increase by $< 50\%$ for $n_{\rm max}$ as high as $10^8$ cm$^{-3}$, which is much less than the variation in these lifetimes for different viewing angles. 

In order to compare the simulated \hbeta\ luminosities with those of observed dNL AGN, we can infer \lhbeta\ from the observed total \oiii\ luminosities and typical \oiii/\hbeta\ line ratios. The \oiii\ luminosities range from log \lsun\ $\sim$ 6.5 - 9.5 \citep[e.g.,][]{fu11a,liu10b}, and \oiii/\hbeta\ ratios are typically $\sim 10$, but range from $\sim$ 3-30 \citep[e.g.,][]{liu10b,rosari10,shen11}. Thus, we can infer that the typical observed \lhbeta\ in dNL AGN range from log \lsun\ $\sim$ 5.5 - 8.5, with some possibly extending $\sim$ 0.5 dex beyond this range. In the simulations, the total \hbeta\ luminosities for ``observable" profiles (as defined in \S~\ref{sssec:hbeta_profs}) range from log \lsun\ $\sim$ 6.6 - 9.2, and most simulations have a maximum \lhbeta\ $\sim$ 7.7 - 8.7 log \lsun. Thus, our simulated \hbeta\ luminosities agree well with the range of observed values.

\begin{figure}
\resizebox{\hsize}{!}{\includegraphics{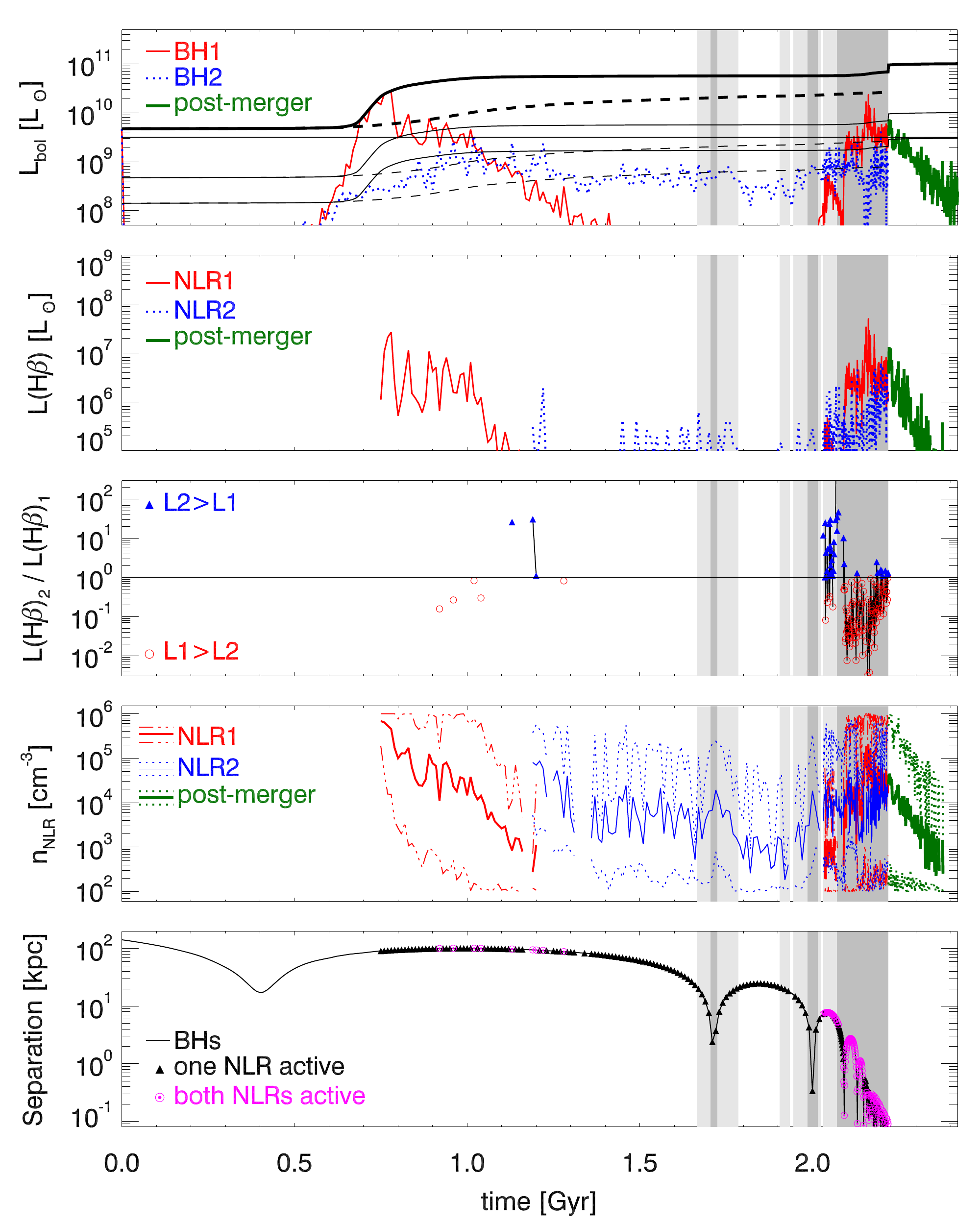}}
\caption[]{Same quantities as in Fig.~\ref{fig:lc}, but for an unequal-mass merger (mass ratio $q = 0.5$) with a lower initial gas fraction ($f_{\rm gas} = 0.04$). 
\label{fig:lc_q50fg04}}
\end{figure}

We can also make a very rough estimate of the typical bolometric AGN luminosities by further extrapolating from the \oiii\ luminosity according to the correction $L_{\rm bol} = 3500 L_{\rm \oiii}$ \citep{zakams03,heckma04}. {\rev Note that this relation assumes there is no substantial reddening of the \oiii\ line \citep[c.f.][]{lamast09}; the same assumption is made for both the observed luminosities cited above and the simulated NLR luminosities. We can infer a} typical bolometric luminosity range of $\sim 10^9 - 10^{12}$ \lsun\ for double-peaked NL AGN. As is readily apparent from Fig.~\ref{fig:lc}, typical simulated bolometric luminosities lie in this range for most of the merger, and even the weaker merger event shown in Fig.~\ref{fig:lc_q50fg04} has $L_{\rm bol} > 10^{10}$ \lsun\ during peak activity. Mergers with higher gas content (\fgas\ $= 0.3$) have peak $L_{\rm bol} \ga 10^{11.5}$ \lsun. Thus, the bolometric AGN luminosities are consistent between the simulations and observations.

For the merger shown in Fig.~\ref{fig:lc}, the total merger time from the start of the simulation to the time of SMBH merger is 1.6 Gyr, but Phase II, shaded in dark (light) gray for $a_{\rm max} = 5.5$ (21) kpc, has a duration of 72 (278) Myr. In general, efficient dynamical friction and gas drag in the dense merger remnant potential ensure that Phase II is always a small fraction of the total merger timescale.
 
Initially, 4 - 30\% of the disk mass of the progenitors is gas. This gas is depleted substantially via star formation during the course of the merger; at the onset of Phase II, the gas content is typically about half of its initial value. However, major mergers are efficient at rapidly fueling gas to the central regions of galaxies, such that by Phase II the remaining gas and newly-formed stars dominate the central region of each merging galaxy.

\subsection{Dependence on Merging Galaxy Parameters}
\label{ssec:params}

\subsubsection{Galaxy Mass Ratio and Gas Content}
\label{sssec:qfgas}

We find that the amount of NL activity in galaxy mergers is strongly influenced by the conditions in the host galaxies. Figure~\ref{fig:lc_q50fg04} shows the same quantities as Fig.~\ref{fig:lc}, but for an unequal-mass merger ($q = 0.5$) with a low initial gas fraction of $f_{\rm gas} = 0.04$. The SMBH in the primary galaxy has an Eddington-limited burst of accretion following the first close passage, triggering strong feedback that heats the surrounding gas and limits further accretion until final coalescence. The secondary SMBH never reaches its Eddington limit but maintains a fairly constant $L_{\rm bol}$ following the pericentric passage until final coalescence. While at least one of the NLRs is active for most of the simulation, they are simultaneously active virtually only during the kpc-scale phase, and even then the \lhbeta\ ratio fluctuates significantly. 

Unlike the example in Fig.~\ref{fig:lc}, in this example the peak of AGN and NL activity occurs several tens of Myr prior to the SMBH merger, and thus most of the NL activity coincides with the kpc-scale phase. Nonetheless, we see that substantially less NL activity occurs in this low-\fgas, unequal-mass merger than in the previous example. This holds true of our merger models in general; mergers with lower mass ratios and gas fractions have less AGN and NL activity, and thus are less likely to produce observable dNL AGN. (See \S~\ref{ssec:tdnl} for a comparison of dNL lifetimes.) This is as expected, because equal-mass mergers induce the strongest perturbations in the merging galaxies, causing more gas to lose angular momentum and flow rapidly to the galaxy centers. Galaxies with substantial gas reservoirs will likewise provide more fuel to the central SMBHs than those that are gas-poor. However, some additional effects described below cause the NL activity in Phases II and III to be especially sensitive to the merging galaxy parameters.

As mentioned in \S~\ref{ssec:general}, the peak SMBH accretion rates for major, gaseous mergers are generally higher during final coalescence than following the first close passage. However, in mergers with lower $q$ and \fgas, accretion, star formation, and feedback early in the merger may deplete the central reservoir of cold gas. After the accretion and SF burst following the first close passage, the two galaxies combined have only $\sim 3 \times 10^8$ \msun\ of gas in the cold phase of the multiphase ISM model (the equal-mass, higher-\fgas\ merger has three times this amount of gas at a similar phase). Fig.~\ref{fig:lc_q50fg04} illustrates that in such cases, the AGN luminosity may actually be {\em lower} during coalescence than after the first passage, and that there is much less NL activity overall. Thus, we expect that gas-poor, unequal-mass mergers will contribute little to the population of dNL AGN.

A related effect is that in nearly-equal-mass, gaseous mergers, the inflow of gas that fuels SMBH accretion can also cause rapid changes to the central potential of the merger remnant. Specifically, the central escape speed (\vesc) may increase substantially during final coalescence. This is the same effect noted by \citet{blecha11} in the context of its importance for the timing of GW recoil kicks. As the central potential deepens, the SMBH and central gas velocities increase. This causes longer-lived and more-pronounced double-peaked NLs via SMBH motion and via rotating gas disks. We find that this effect generally leads to a larger fraction of dNL AGN in Phase III (see \S~\ref{sssec:diagnostics} \& \ref{ssec:tdnl}). However, this increase in \vesc\ occurs only for nearly-equal-mass, relatively gas-rich mergers. For the merger models presented here, only those with \fgas\ $= 0.3$ show a rapid \vesc\ increase at coalescence, and only in the $q=1$, \fgas\ $= 0.3$ model does \vesc\ increase by more than $\sim 200$ \kms.

We note that the same processes that fuel AGN can also trigger rapid star formation, which may produce large amounts of obscuring dust. The problem is especially complex for the NLR, which exists at large enough radii to be intermingled with the newly-forming central cusp of stars. Because the present work, as a first attempt to model the NLR in galaxy mergers, does not account for the potential effects of obscuration or reprocessing of emission by dust, we must consider this an important caveat to interpretations of our simulated NLRs as ``observable". 

In order to avoid as much as possible a strong-starburst regime, we do not consider initial gas fractions above 30\%, and much of the analysis presented here concerns galaxies with \fgas\ $\le 0.1$ initially. Thus, our simulations produce low- to moderate luminosity AGN, rather than luminous quasars. This choice is justified because most searches of double-peaked NLs have also focused on moderate luminosity AGN. As detailed in \S~\ref{ssec:general}, the NL luminosities calculated from our simulations agree well with the observed luminosities.
 
Furthermore, while the peak star formation rate (SFR) following the first pericentric passage can be in excess of 100 \msunyr\ for our higher-$q$ and \fgas\ simulations, the peak SFR during final coalescence (i.e., during Phase II) is typically much lower, $\la 1 - 10$ \msunyr\ for these same simulations.  For lower-$q$, \fgas\ mergers, the peak SFR is much lower than this;  the example in Fig.~\ref{fig:lc_q50fg04} has a peak SFR at coalescence of only $0.4$ \msunyr. At these SFRs, the cumulative UV emission from massive stars should be much weaker than the AGN emission, such that we may neglect its contribution to photoionization of circumnuclear gas.

\subsubsection{Progenitor Galaxy Morphology}
\label{sssec:morphology}

As mentioned in \S~\ref{ssec:general}, the morphology of the progenitor galaxies has an effect on the timing of the peaks of AGN (and NL) activity. Although the fiducial merger models have initially bulgeless galaxies, we have also conducted simulations including progenitors with stellar bulges, and with SMBH masses chosen following the $M_{\rm BH}$-$M_{\rm bulge}$ relation of \citet{harrix04}. Specifically, a \citet{hernqu90} bulge with {\rev B/T $=$ 0.2 - 0.3} is added to the progenitor galaxy models.

The simulations with and without bulges (and corresponding SMBH masses) both show good agreement with expected trends. Although the fractional increase in SMBH mass is much larger in the bulgeless simulations, the actual amount of gas accreted is similar in all cases. This reflects the fact that SMBH accretion is driven largely by the supply of cold gas provided to the BH during merger, and although the timing of gas inflow should vary with galactic morphology, major mergers should ultimately cause catastrophic collapse of the gas disk.

Additionally, in mergers with and without bulges, the SMBHs and final bulges lie on the BH-bulge relations by the end of the simulation. This is to be expected, because the efficiency of the BH feedback model is tuned to give this result, but given the systematic variation of the merger parameters and the fact that the merger remnants are not entirely relaxed by the end of the simulations, the agreement is quite good. Specifically, fitting a line to the final SMBH masses and stellar velocity dispersions for all simulations in Table~\ref{table:models} yields a correlation log $M_{\rm BH} = 7.84 + 4.46({\rm log}\: \sigma_{*,200})$, with a scatter of 0.32 dex ($\sigma_{*,200} = \sigma_*$/ 200 km s$^{-1}$).

It is well-known that stellar bulges can stabilize a galactic disk to perturbations during encounters, thereby delaying catastrophic loss of gas angular momentum until the final coalescence of the galaxies \citep[e.g.,][]{mihher96}. Star formation and SMBH growth should thus be similarly delayed to some extent, and indeed this behavior is seen in our simulations with bulges. In particular, a strong starburst typically occurs after the first close passage in the bulgeless simulations, while in simulations with bulges this initial burst is smaller (a factor of a few increase in SFR versus a factor of a few tens for bulgeless simulations). In some cases, the strongest starburst occurs at coalescence.

For SMBH accretion, however, there is a competing effect owing to the presence of a much larger (factor of $\sim$ 100 - 200) initial SMBH mass relative to the bulgeless simulations with initial $M_{\rm seed} = 1.4 \times 10^5$ \msun. The larger seed mass allows for more significant AGN activity in the early merger phase. The net result is that simulations with larger initial bulges (B/T = 0.3) have less SF and SMBH growth in the early merger phase than at final coalescence, but those with smaller initial bulges (B/T = 0.2) may display the opposite trend. Recall that in most cases, the bulgeless simulations lie in the former category; their AGN activity typically peaks at final coalescence. However, Fig.~\ref{fig:lc_q50fg04} shows that even in bulgeless galaxies, if the gas fraction is low, the fuel supply may substantially depleted at final coalescence, such that AGN activity in the early merger phase is dominant.

When a larger SMBH is present initially {\em and} the galaxies are unequal-mass and relatively gas-poor, this effect is amplified. In the simulation with $q=0.5$, \fgas\ $= 0.1$, and B/T = 0.2, the peak SMBH accretion rate after the first passage is 10 times larger than at coalescence. Increasing B/T to 0.3 reverses this trend, however; the larger bulge mass is able to suppress much of the early AGN phase, such that the peak $\dot M_{\rm BH}$ is $\sim 4$ times larger at coalescence than at first passage. Furthermore, the simulation with $q=0.5$, \fgas\ $= 0.1$, and B/T $= 0.2$ is the {\em only} merger in which AGN activity in the early merger phase is strongly dominant. In the other eight unique simulations with initial bulges and corresponding SMBH masses, three have comparable peak $\dot M_{\rm BH}$ at first passage and final coalescence (peak $\dot M_{\rm BH,late}$/$\dot M_{\rm BH,early}$ = 0.8 - 2), and the remaining five simulations achieve much higher $\dot M_{\rm BH}$ at coalescence, by factors of 4 - 127.

It is also instructive to compare the total amount of SMBH growth in the early versus late-merger phases. Because the low-resolution simulations with bulges do not have NL analysis, we simply use the time when the BH separation first falls below 5.5 kpc as the definition of the late-merger phase. For the $q=0.5$, \fgas\ $= 0.1$, and B/T $= 0.2$ simulation described above, the ratio $\Delta M_{\rm late}/\Delta M_{\rm early}$ is only 0.1. However, for all other simulations with bulges, this ratio lies in the range 0.4 - 7, in very good agreement with the range of $\Delta M_{\rm late}/\Delta M_{\rm early} = $ 0.6 - 6 measured for the fiducial of simulations. In other words, the timing of SMBH accretion is similar in the mergers with and without stellar bulges. In both cases, at most marginally more gas is accreted onto the SMBHs in the early merger phase than in the late phase, despite the fact that the former phase is typically $\sim 10$ times longer. Thus, our simulations indicate that AGN in the early phase of galaxy merging should not strongly dominate the population of major-merger-triggered AGN.

This claim is also supported by empirical evidence. \citet{ge12} have recently analyzed the spectra of SDSS emission-line galaxies using more lenient selection criteria than previous studies, yielding a much larger sample of double-peaked objects (including SF+AGN composite galaxies as well as AGN). They find 1318 dNL AGN and dNL AGN+SF galaxies, and visual inspection yielded a subsample of 40 that have visible companions beyond the 3" spectral fiber of SDSS. Some of these are presumably AGN triggered in the early phases of merging, where one AGN has double-peaked NLs produced by gas kinematics. (The production of double-peaked NLs by gas kinematics versus dual SMBH motion is discussed in detail in \S~\ref{ssec:nlmorph} \& \S~\ref{sssec:kinematics}.)  If we assume, following \citet{shen11}, that $\ga 10\%$ of the dNL AGN are likely to contain dual SMBHs (most of which will not be resolvable with SDSS imaging), then the ratio of ``wide separation" to ``small separation" dual AGN with dNLs can be roughly estimated as $\sim 40 / (0.1 * 1318)$, or about 0.3. This suggests that dNL AGN in late-stage mergers may outnumber those in early-stage mergers. This ratio could be even lower if the fraction of unresolved dual AGN is larger or if some of the ``companions" are fore/background objects, but it could increase if many of the dNL AGN are in isolated galaxies. ``Wide" and ``small" separations are loosely defined here, as they are based on an angular scale (3") rather than a physical size, but the redshift distribution of the sample is strongly peaked at $z \sim 0.1$.

Similar analysis can be undertaken with the results of \citet{liu10a}. They find 167 dNL AGN in their SDSS sample, 6 of which have resolved dual cores in the SDSS images. Additionally, 30 of these objects have companions beyond 3" (X. Liu, private communication). Follow-up observations of this sample found that $\sim 10\%$ of the dNL AGN are strong dual SMBH candidates \citep{shen11}, so the ratio of ``wide" to ``small-separation" dual SMBHs is $\sim 30 / (0.1 * 167)$, or about 1.8. That this value is larger than that inferred from the \citet{ge12} sample is not surprising, because their inclusion of composite galaxies in addition to true AGN likely populates the sample with more isolated galaxies that are not undergoing a merger. An additional 40\% of the \citet{shen11} dNL follow-up sample were indeterminate as to their dual nature; if all of these contained dual SMBHs (which seems unlikely), then the estimated ratio of wide to small-separation dual SMBHs would be $\sim 0.4$. Thus, we see that NL AGN observed in the early-merger phase should outnumber those observed in the late-merger phase by at most a factor of 1.8, and may in fact be less numerous. Accordingly, we conclude that our fiducial merger models with bulgeless galaxies and small initial SMBH masses produce results that are in general agreement with observations.

\subsubsection{Galaxy Merger Orbits}
\label{sssec:orbits}

In addition to the mass and composition of the progenitor galaxies, we consider the effect of varying the orbital parameters of their interaction and merger. The fiducial orbit used (orbit {\em a}) has an initial pericenter of 7.1 kpc, and the galactic angular momentum vectors are prograde but tilted relative to the orbital plane. (The actual orbital parameters evolve over time owing to dynamical friction.) For the merger model with $q = 1$ and \fgas\ $= 0.1$, we additionally consider an orbit with the same orientations but a larger pericenter (orbit {\em b}), a polar orbit (orbit {\em c}) and a nearly-coplanar orbit (orbit {\em d}). 

Previous work by, e.g., \citet{mihher96} and \citet{cox08} has shown that coplanar, prograde orbits tend to maximize the tidal forces on galactic disks, leading to more efficient bursts of star formation and BH accretion, while polar orbits are less efficient. Also, larger pericenters increase the orbital angular momentum, but because the merger takes longer, more gas is consumed throughout. Indeed, we find this to be the case; orbit {\em b} and orbit {\em d} produce the largest final BH masses and consume the most gas in star formation, while orbit {\em c} has the least cumulative BH growth and star formation. The total ranges are small, however: $M_{\rm BH,fin} = 1.2 - 2.9 \times 10^7$ \msun, and $M_{\rm *,new} = 7.7 - 8.7 \times 10^9$ \msun. 

We also find that the variation in the lifetimes of double-peaked NL AGN phases for different orbits is smaller, or at most comparable to, the variation with observer viewing angle. The dNL AGN lifetime in the kpc-scale phase (Phase II) has the least variation with orbital parameters; for the four orbits simulated, this lifetime ranges from 13 - 23 Myr, while the intrinsic line-of-sight variation is 2 - 4 times larger. Thus, we conclude the orbital parameters of a galaxy merger have a moderate effect on AGN and SF activity relative to galactic structure (e.g., mass ratio and gas content), and the latter is the focus of most subsequent analysis.

\subsubsection{Gas Equation of State}
\label{sssec:qeos}

The multiphase ISM model of \citet{sprher03} includes a parameter, $q_{\rm EOS}$, that determines the degree of ``softening" of the gas equation of state (EOS). When this parameter is set to zero, the EOS is purely isothermal, and when it is set to one, the full multiphase ISM model is used. Intermediate values of $q_{\rm EOS}$ interpolate between these two. We adopt a fiducial value of $q_{\rm EOS} = 0.25$, as in \citet{sprher03}, but because of the potential for the choice of EOS to affect our results, we also test values of 0.5 and 0.05 for select merger models (see Table~
\ref{table:models}). 

In general, the gas disk is more stable for the softer EOS ($q_{\rm EOS}$ = 0.5), while for $q_{\rm EOS} = 0.05$ the gas is much more prone to fragmentation and has a very clumpy distribution. The latter case results in higher SMBH accretion rates, including sustained higher accretion rates following the first close passage of the two galaxies. The final SMBH masses are up to an order of magnitude higher for $q_{\rm EOS} = 0.05$ versus 0.5. We consider the nearly-isothermal EOS ($q_{\rm EOS} = 0.05$) to be a fairly extreme model, but regardless, it is clear that the SMBH activity in these galaxy mergers does depend on the gas EOS. Specifically, gas distributions that are clumpier and closer to isothermal should result in more AGN and NL activity. 

The observable dNL AGN lifetimes (\tdnl) in these simulations vary as much or more with $q_{\rm EOS}$ as do the total AGN lifetimes. In Phase I, the $q_{\rm EOS} = 0.05$ models have line-of-sight averaged \tdnl\ ranging from 0 - 435 Myr, while the other models all have \tdnl\ $< 1$ Myr. In Phase III, the trend of increasing \tdnl\ with lower $q_{\rm EOS}$ is more linear, but the variation is still 1-2 orders of magnitude between $q_{\rm EOS} = 0.5 - 0.05$. \tdnl\ in Phase II (the kpc-scale phase) exhibits somewhat less variation with $q_{\rm EOS}$, but still increases by more than a factor of 10 for the factor of 10 decrease in $q_{\rm EOS}$.

Although the dNL AGN lifetimes exhibit this steep dependence on $q_{\rm EOS}$, we find that this results directly from the variation in {\em total} observable NL AGN lifetimes with $q_{\rm EOS}$. The latter are calculated using a minimum observable line flux as described in \S~\ref{sssec:hbeta_profs}, but without any restrictions on the profile shape. These NL AGN lifetimes exhibit the same magnitude of variation as the double-NL AGN lifetimes, which indicates that the gas EOS affects the AGN luminosity but {\em not} the gas kinematics that produce the NL profiles. Thus, our quantitative results for the (double) NL AGN lifetimes depend on the choice of $q_{\rm EOS}$, but the NL profile shapes and relative fraction of single and double-peaked profiles are robust.

\subsubsection{Mass and Spatial Resolution}
\label{sssec:resolution}

Finally, we have examined the dependence of our results on the mass and spatial resolution of the simulations, using a subset of four additional simulations with the $q=0.5$, \fgas\ $= 0.1$ merger model. Specifically, we test baryonic mass resolutions of a factor of two higher and lower than our fiducial value ($f_{\rm res}  = $0.5-2) and DM mass resolutions up to $f_{\rm res} = 5$ (see Table~\ref{table:models}). The gravitational softening lengths are scaled accordingly ($\propto M^{1/3}$).

We do find a trend toward higher central densities for higher resolutions;  this essentially reflects the need for the multiplicative factor for the accretion rate, described in \S~\ref{ssec:simulations}, which accounts for the higher densities not captured on sub-resolution scales.  Over the factor of 4 in mass resolution tested, the maximum central density (occurring shortly after the first pericentric passage) varies by a factor of 4 for the smaller BH, and by a factor of $> 40$ for the larger BH. The bigger disparity in the latter arises mainly from the lowest resolutions, where the larger gas particles are more sensitive to the amount of BH feedback.

Owing to the approximate nature of the sub-resolution SMBH accretion model, it is impossible to know the ``true" value to which the accretion rates should converge, and this must be considered a source of uncertainty in any numerical study utilizing semi-analytic SMBH models. However, the dependence of our results on the maximum central density is weak. First, we note that the maximum average density of the {\em narrow-line} gas varies by only 20\%. The maximum bolometric and \hbeta\ luminosities vary by factors of $\sim 2$ and $\sim 9$ respectively over the resolutions sampled, and more importantly, the {\em time-averaged} SMBH accretion rates vary by a factor of only 1.6. Further, these accretion rates and luminosities do not increase monotonically with mass resolution, indicating that this variation is within the level of random fluctuations. And finally, the line-of-sight averaged lifetimes of kpc-scale dNL AGN in this simulation subset have a spread $\Delta t_{\rm avg} / \left< t_{\rm avg} \right > = 0.9$ relative to their mean, which is less than half the variation owing to viewing angle effects and to the definition of ``kpc-scale" (5.5 vs. 21 kpc).

\begin{center}
\begin{figure*}
\subfloat[$\: \mathbf{t_{\rm mrg} - 707}$ Myr]{
\includegraphics[width=0.24\textwidth]{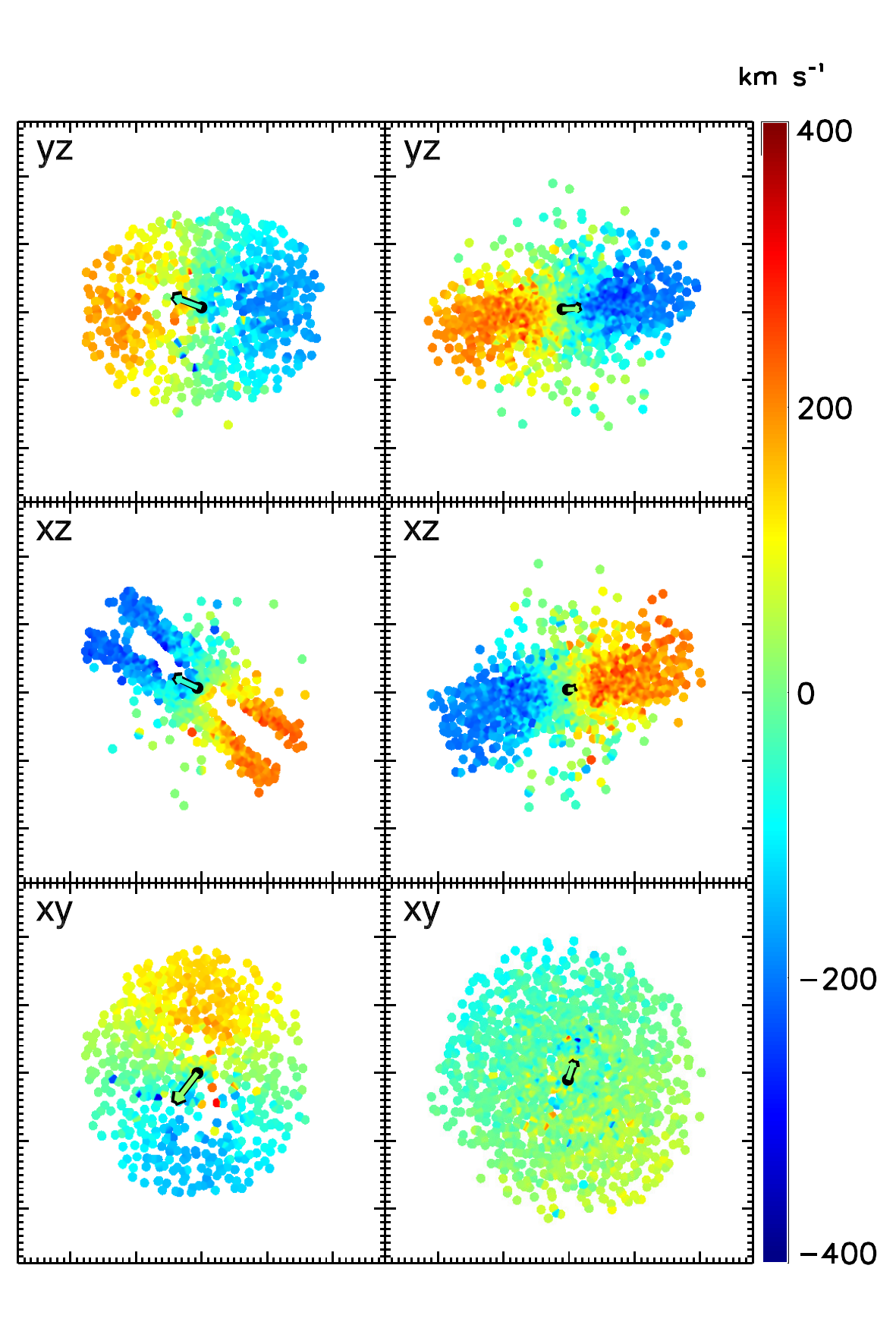}
\hspace{-3pt}
\includegraphics[width=0.24\textwidth]{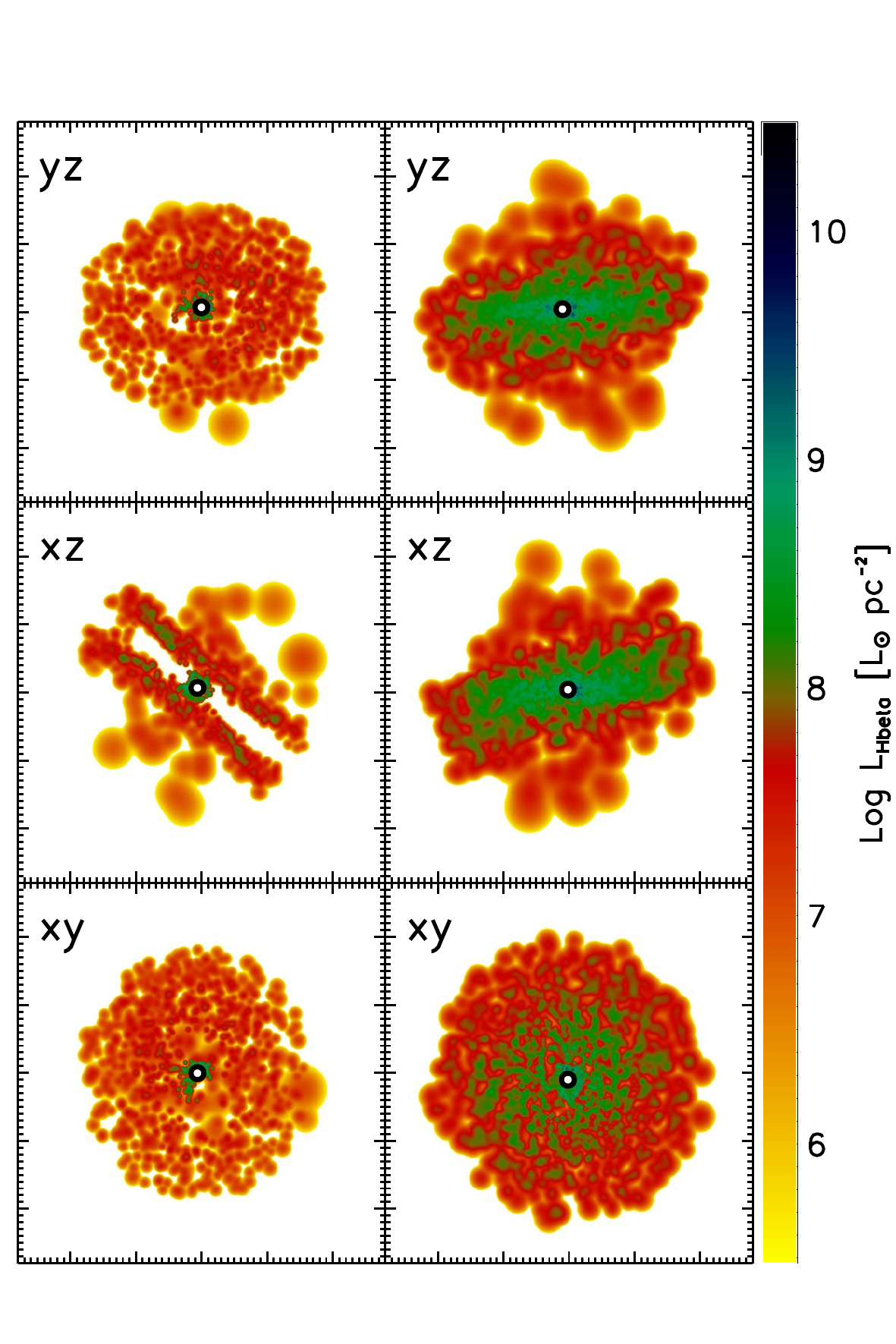}}
\hspace{12pt}
\subfloat[$\: \mathbf{t_{\rm mrg} - 397}$ Myr]{
\includegraphics[width=0.24\textwidth]{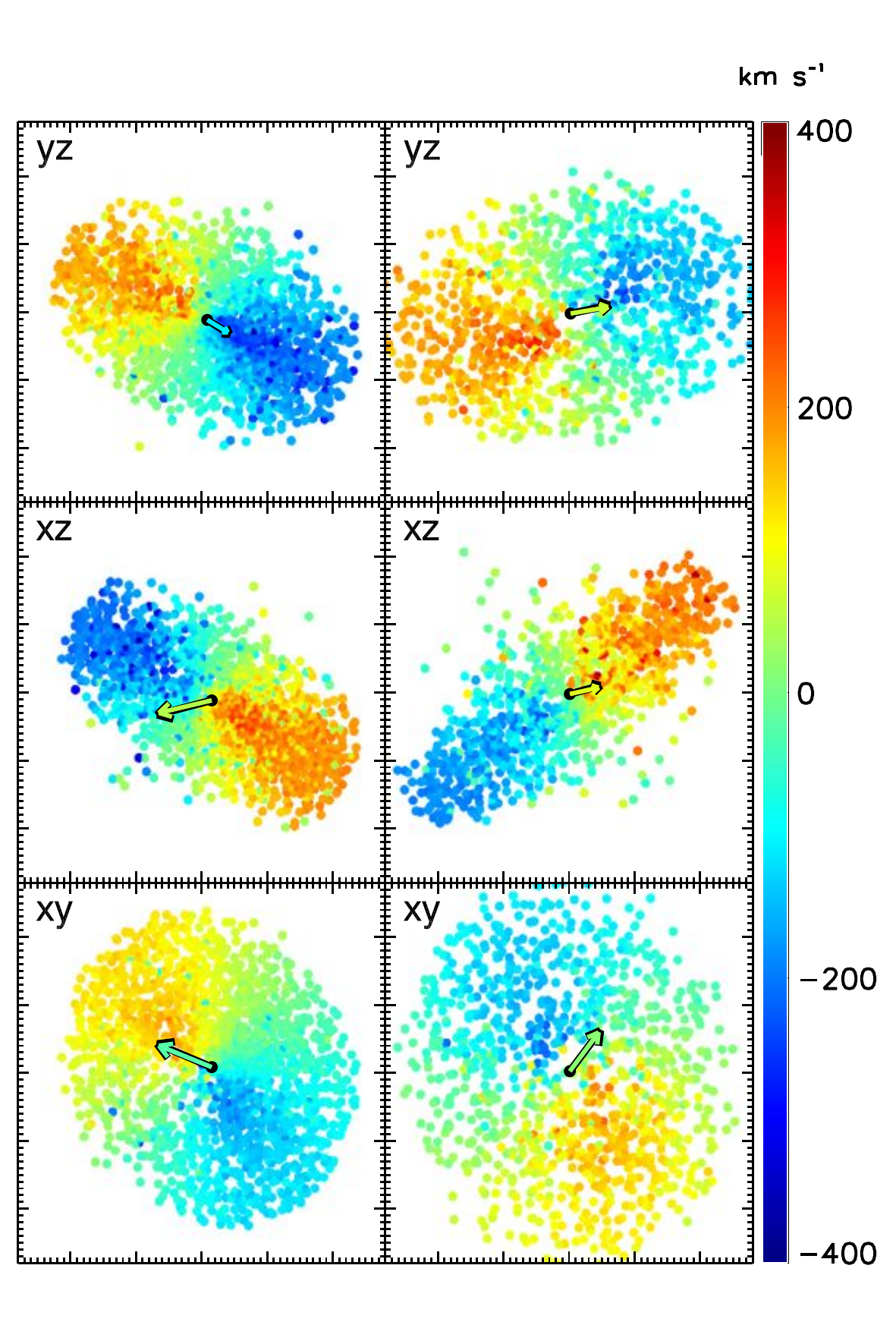}
\hspace{-3pt}
\includegraphics[width=0.24\textwidth]{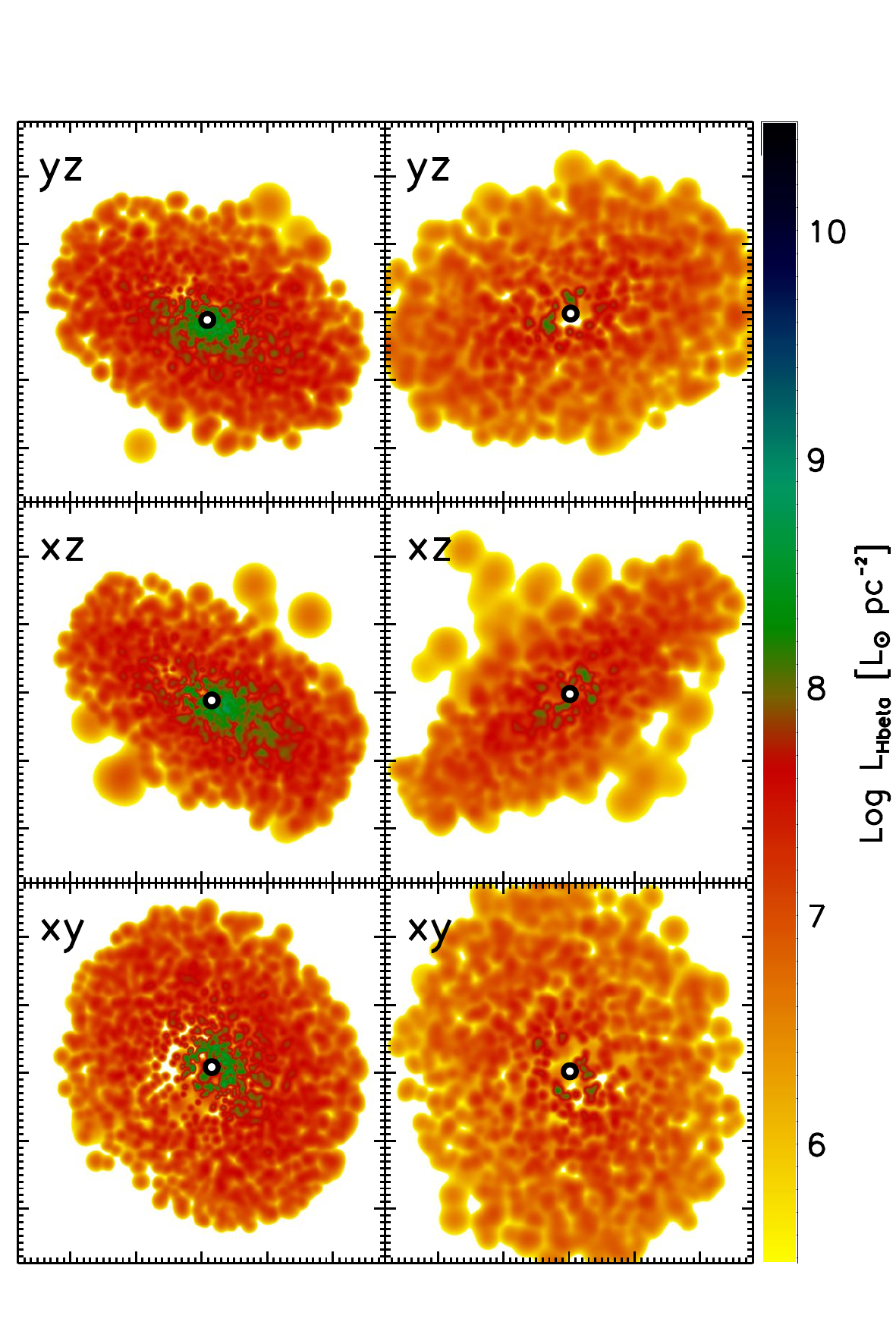}}

\subfloat[$\: \mathbf{t_{\rm mrg} - 232}$ Myr]{
\includegraphics[width=0.24\textwidth]{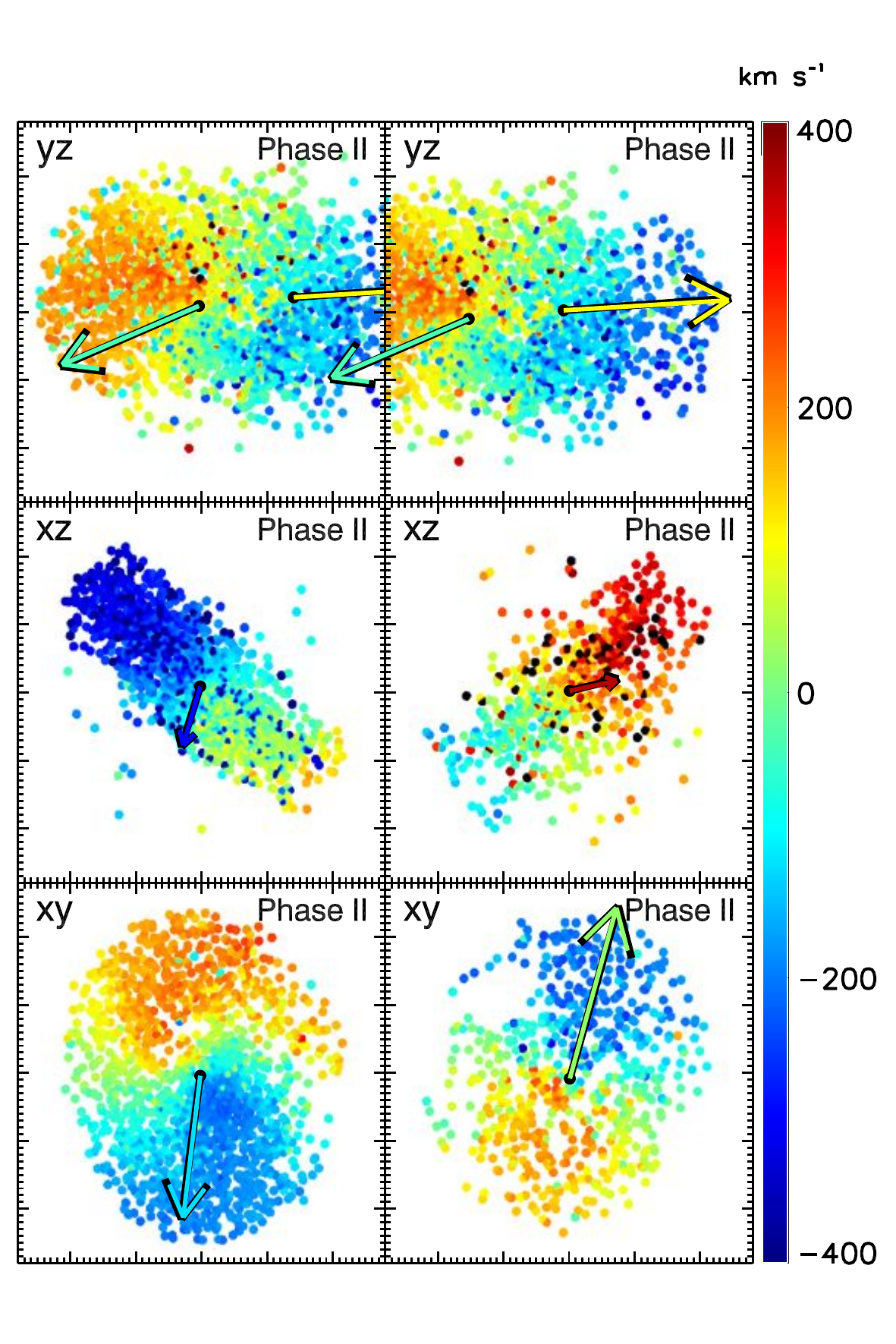}
\hspace{-3pt}
\includegraphics[width=0.24\textwidth]{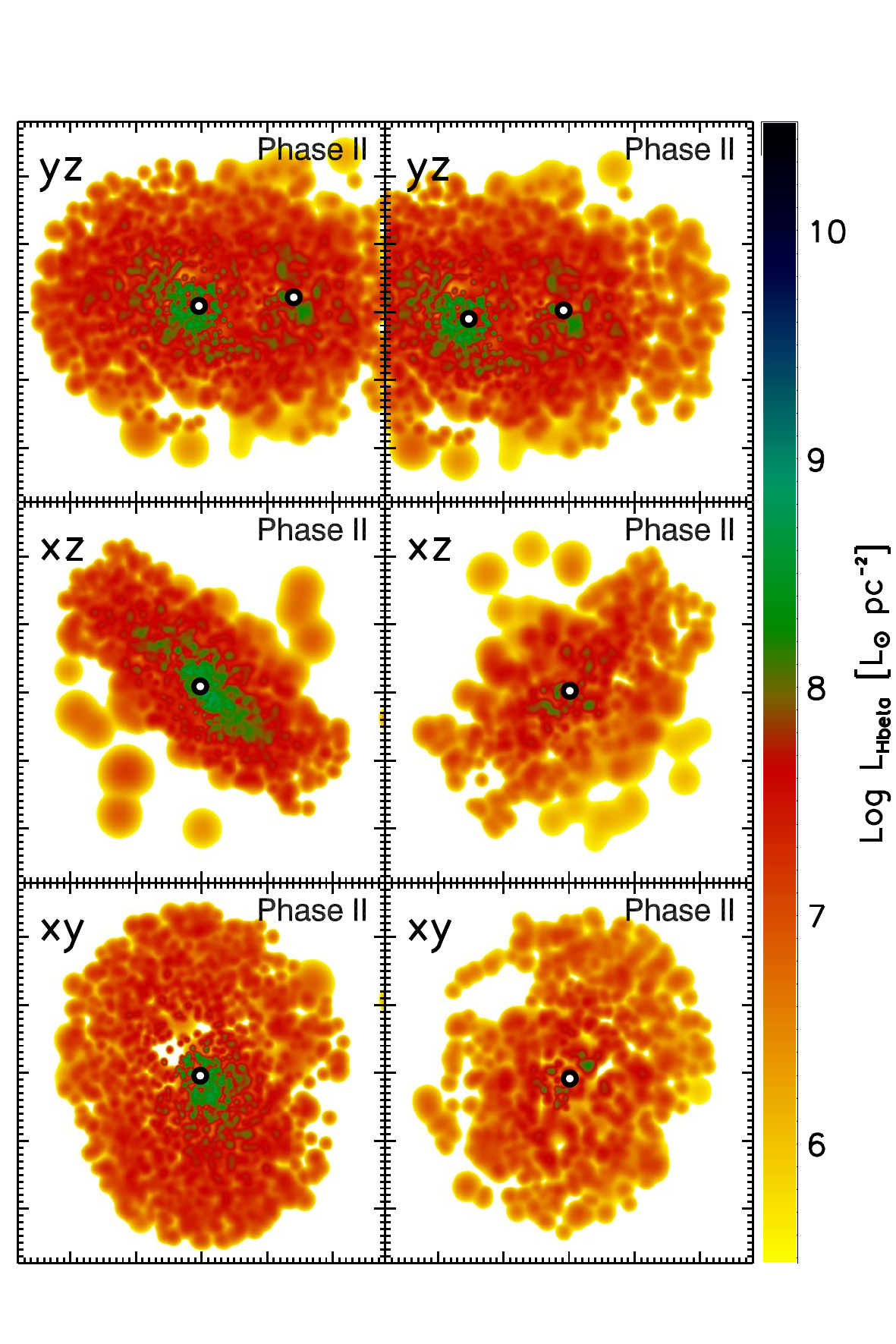}}
\hspace{12pt}
\subfloat[$\: \mathbf{t_{\rm mrg} - 81}$ Myr]{
\includegraphics[width=0.24\textwidth]{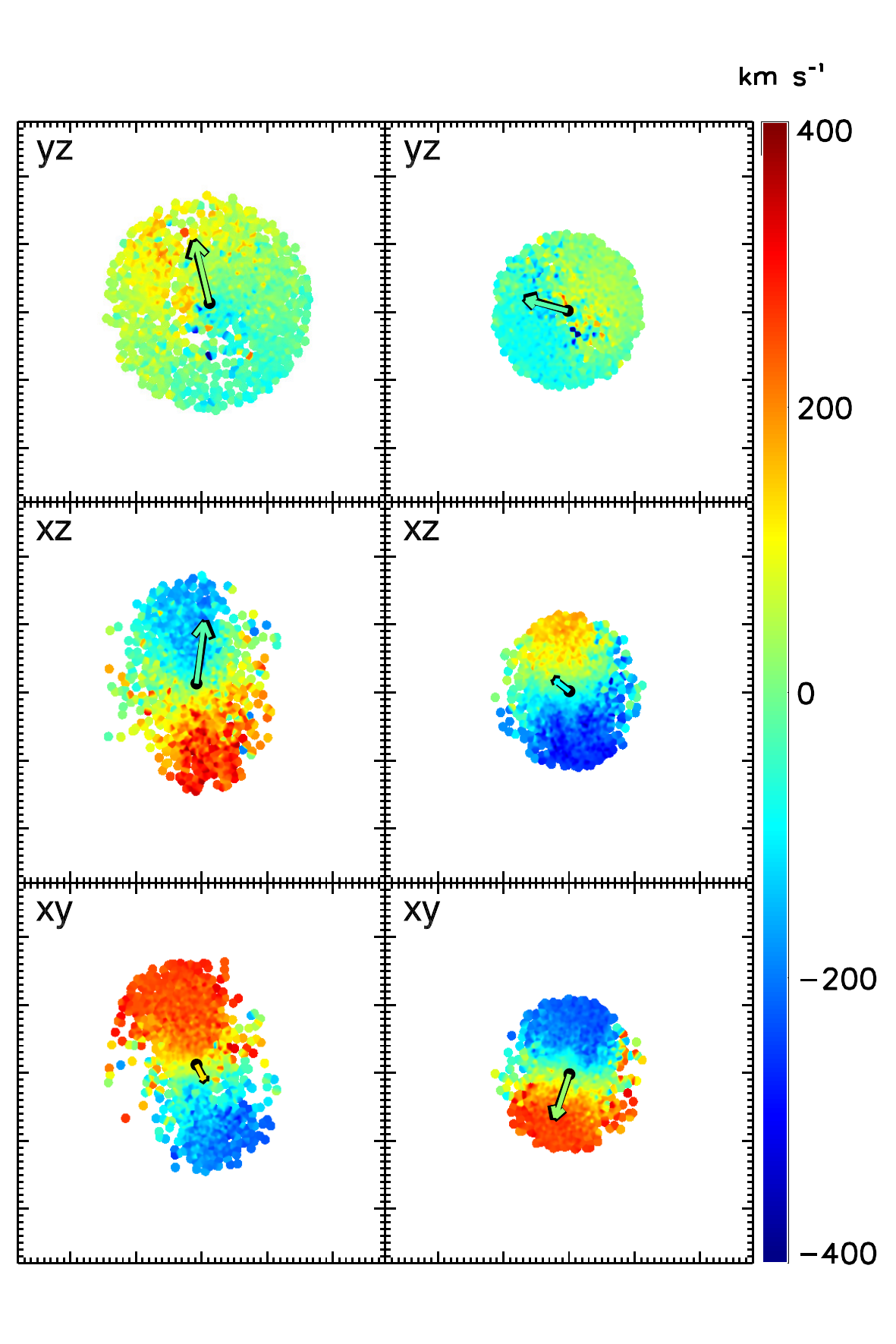}
\hspace{-3pt}
\includegraphics[width=0.24\textwidth]{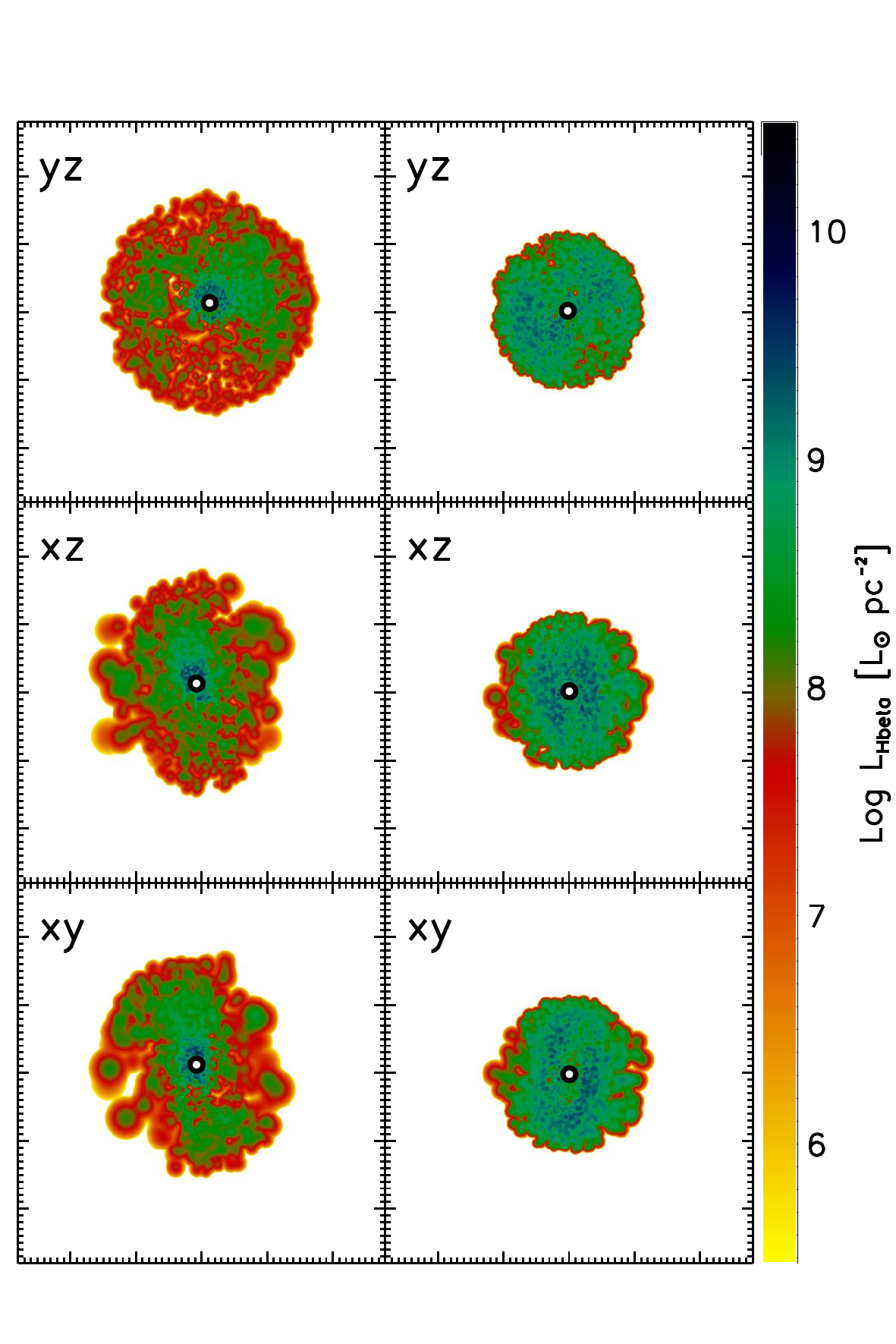}}

\subfloat[$\: \mathbf{t_{\rm mrg} - 41}$ Myr]{
\hspace{-6pt}
\includegraphics[width=0.24\textwidth]{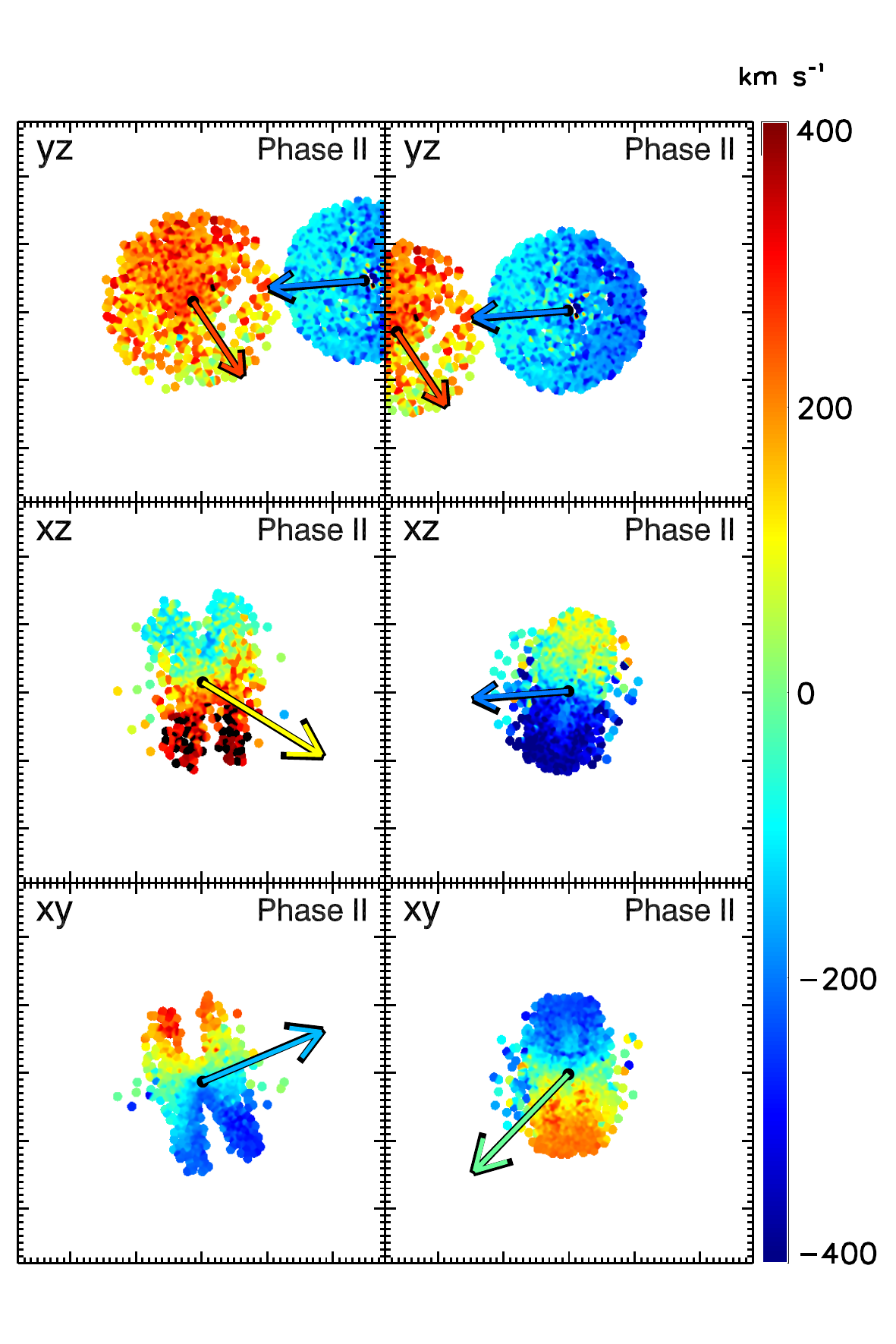}
\hspace{-3pt}
\includegraphics[width=0.24\textwidth]{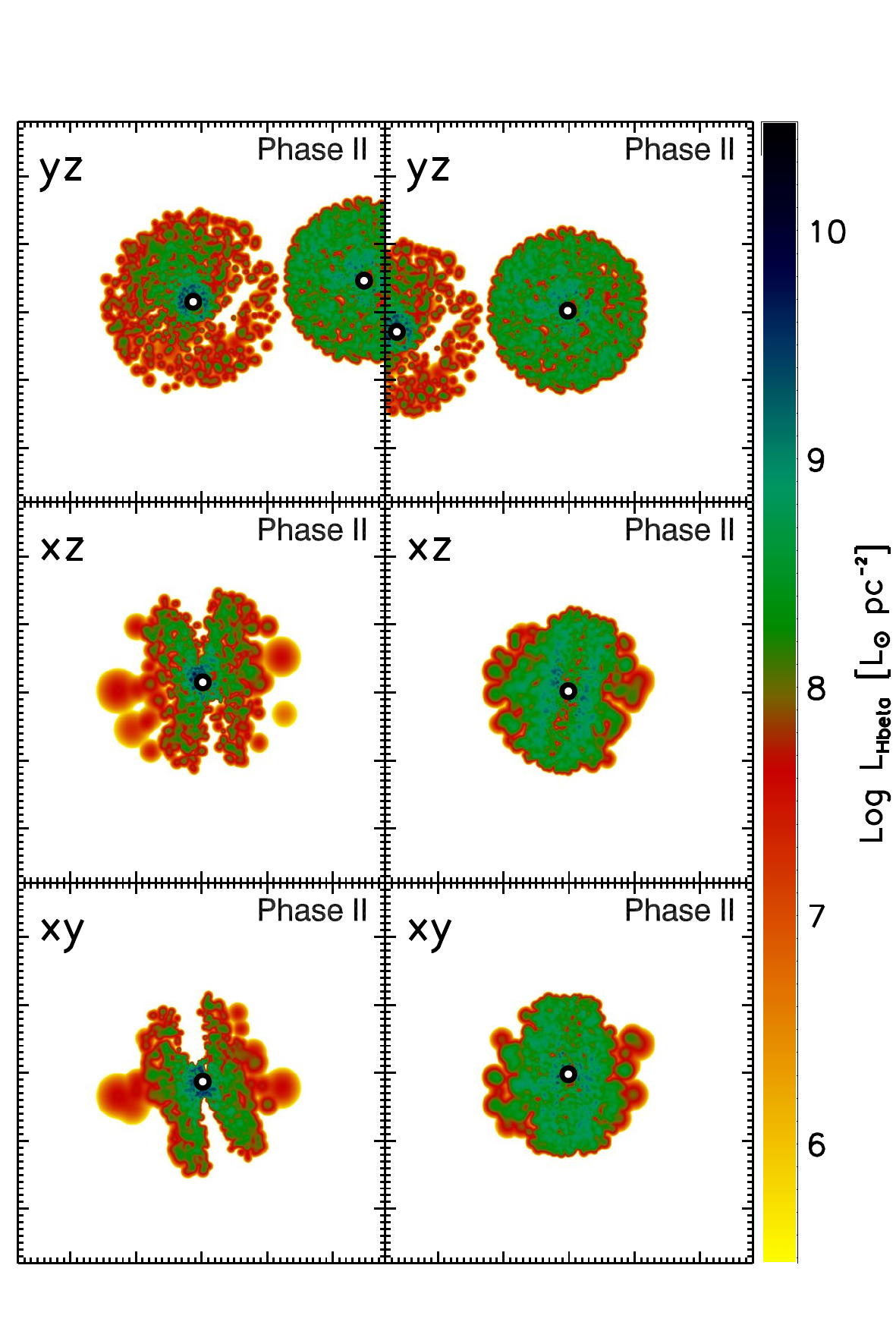}}
\hspace{60pt}
\subfloat[$\: \mathbf{t_{\rm mrg} + 200}$ Myr]{
\includegraphics[width=0.14\textwidth]{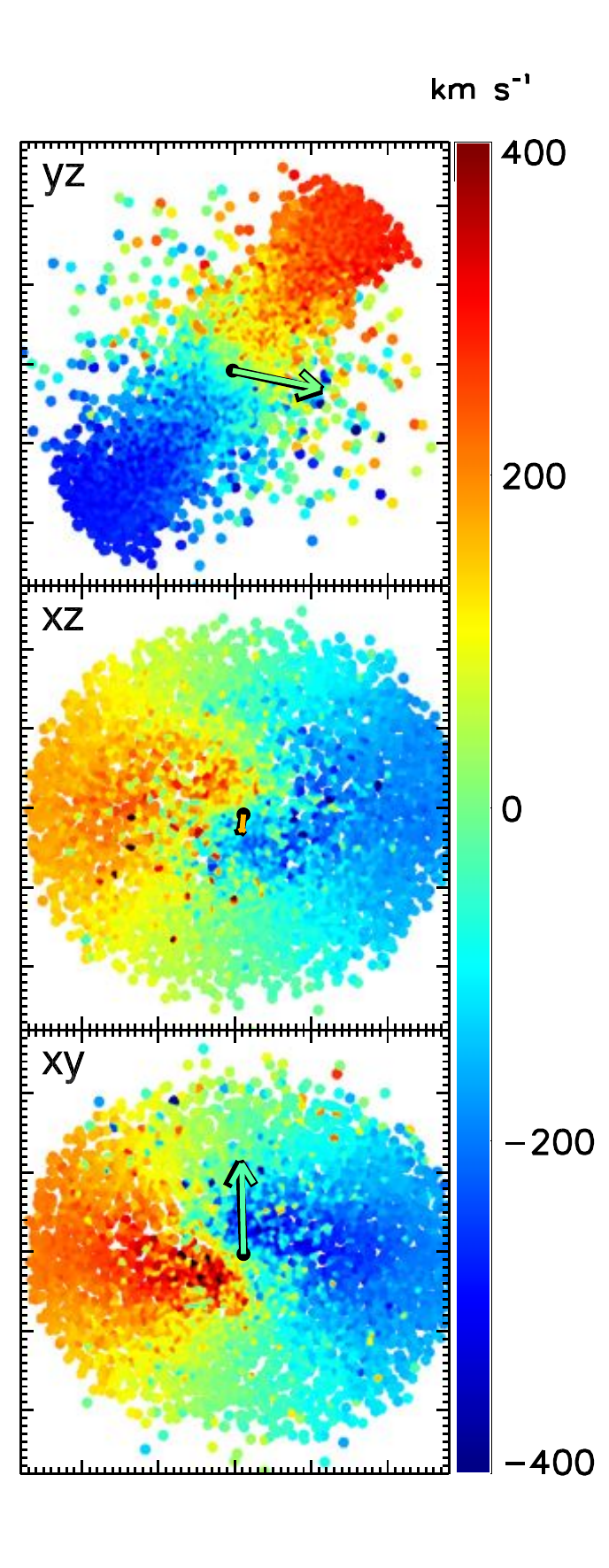}
\includegraphics[width=0.14\textwidth]{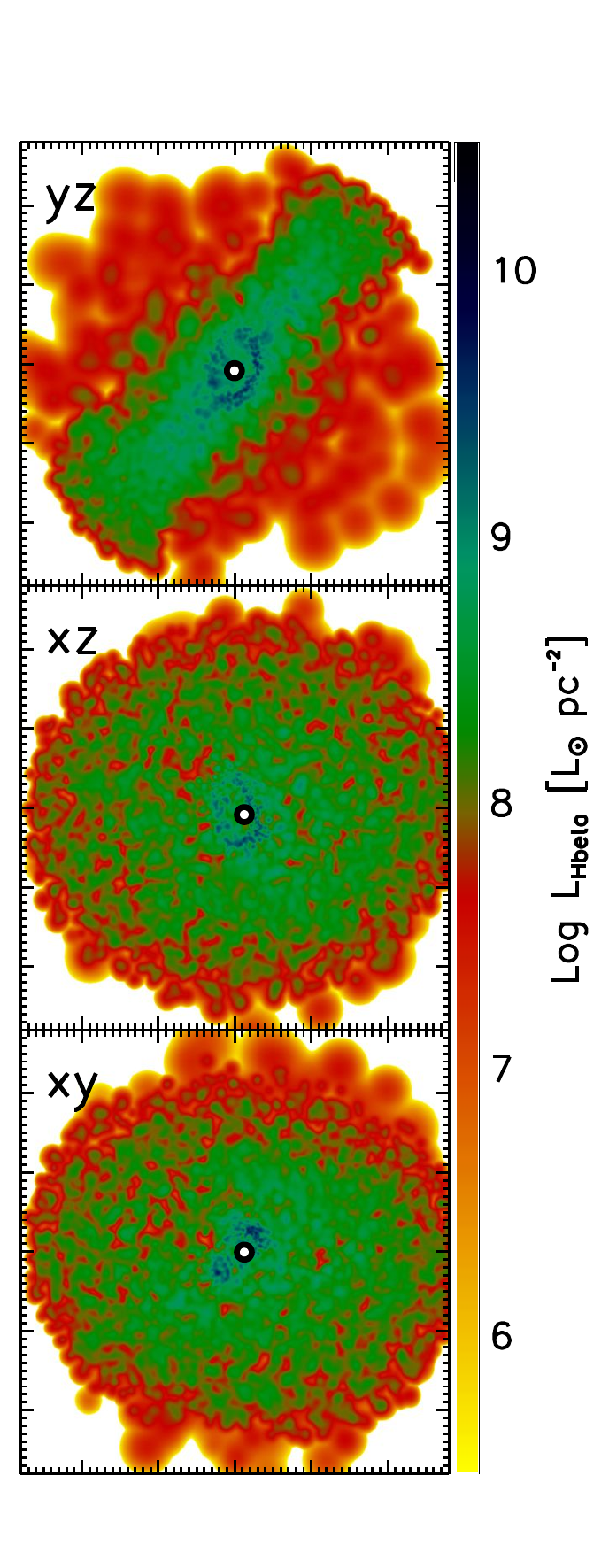}}
\hspace{45pt}
\caption[]{Line-of-sight (LOS) velocity maps and \lhbeta\ maps of NLRs at six different snapshots throughout a single galaxy merger with $q=1$ and \fgas\ $=0.1$.  In each subfigure {\bf(a)} - {\bf(f)}, the left plot is the velocity map and the right plot is the \lhbeta\ map. Within each 6-panel plot, the left column of shows three orthogonal projections centered on SMBH2, and the right column shows the same three projections centered on SMBH1. Fig.~\ref{fig:hbvel}{\rev f} is a post-BH-merger snapshot, so only one column of panels is shown. In the velocity maps, the magnitude of the colored arrows denotes the projected SMBH velocity, and their hue denotes the LOS SMBH velocity. Each panel is 800 pc on a side, and the velocity scale spans from -400 \kms\ (blue) to $+$400 \kms\ (red). This sign convention for LOS velocity is used throughout the paper. The panels in the \lhbeta\ maps are also 800 pc on a side. \label
{fig:hbvel}}
\end{figure*}
\end{center}

\subsection{NLR Morphology and Kinematics}
\label{ssec:nlmorph}
 
\begin{figure*}
\subfloat[]{\includegraphics[width=0.1517\textwidth]{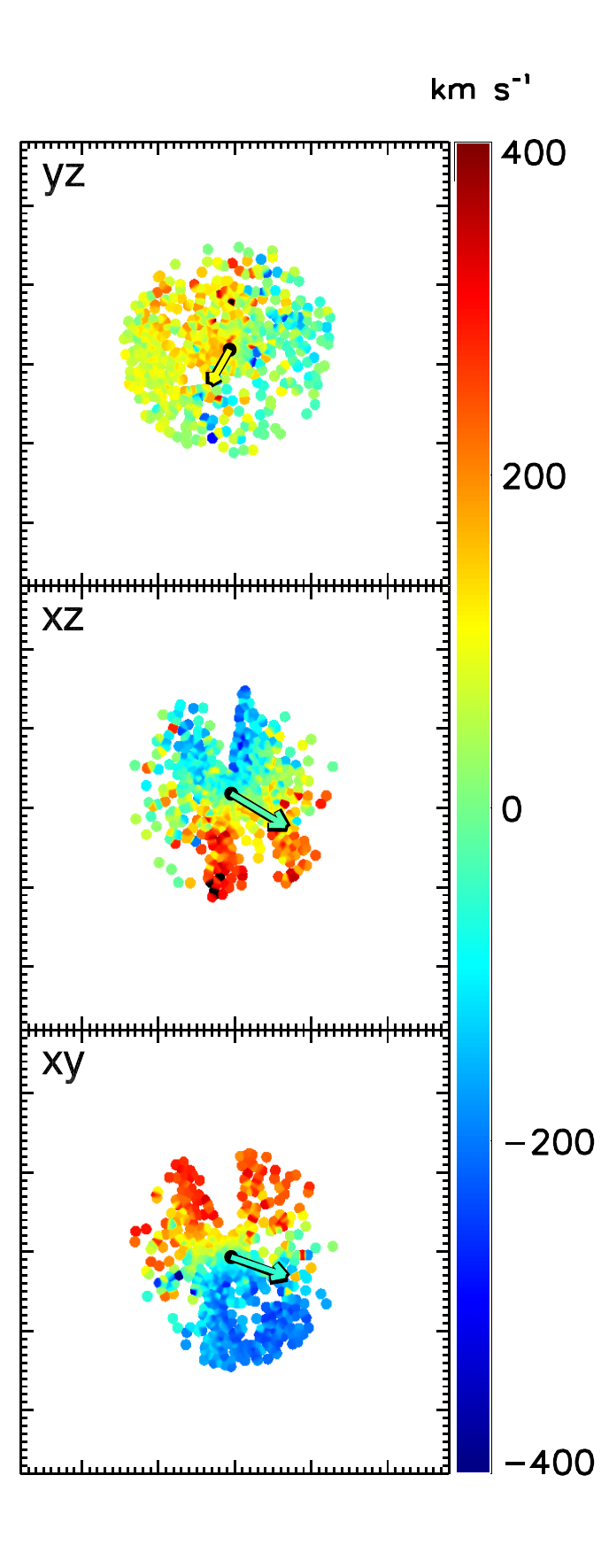}
\includegraphics[width=0.26\textwidth]{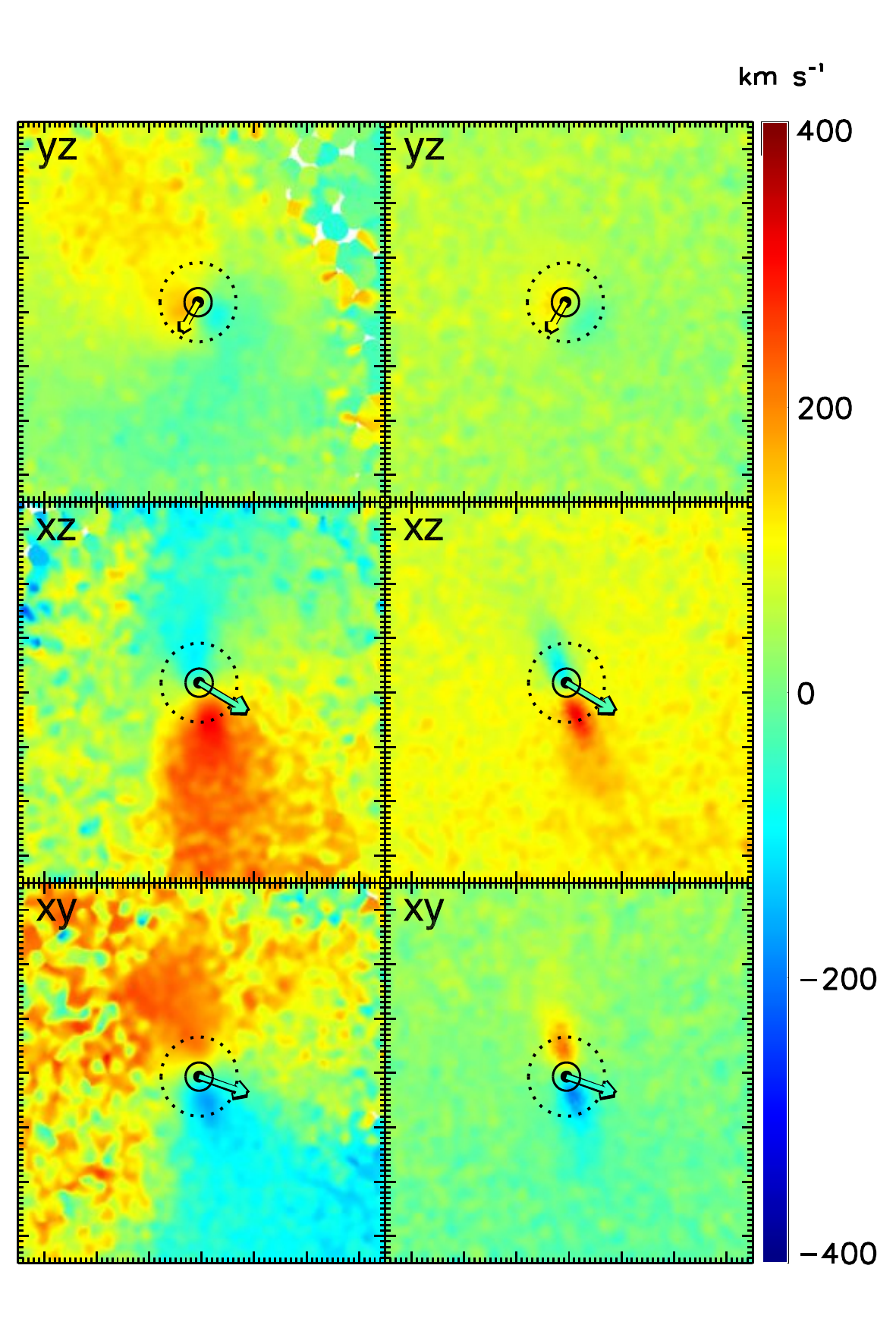}}
\hspace{10pt}
\subfloat[]{\includegraphics[width=0.26\textwidth]{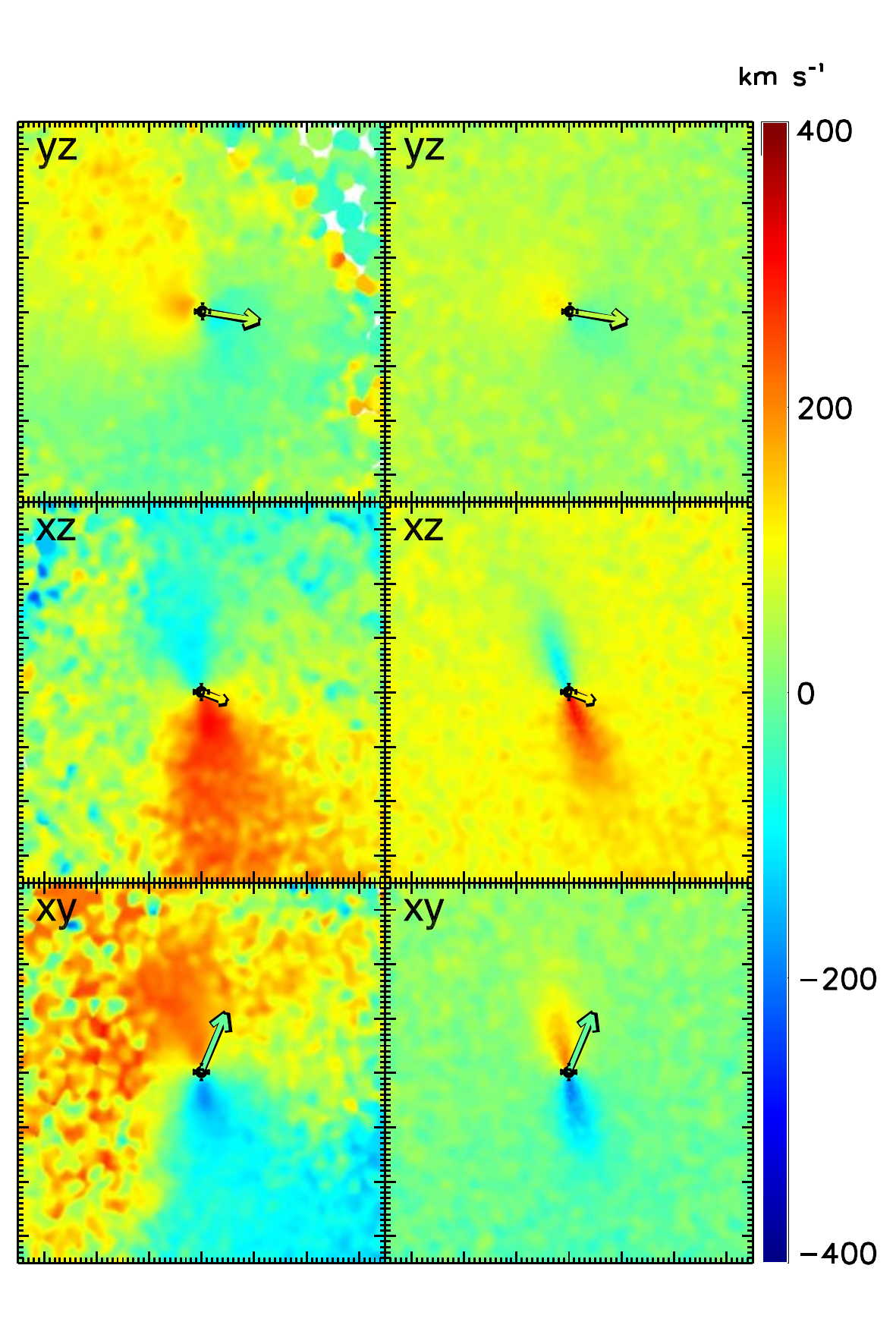}}
\hspace{10pt}
\subfloat[]{\includegraphics[width=0.26\textwidth]{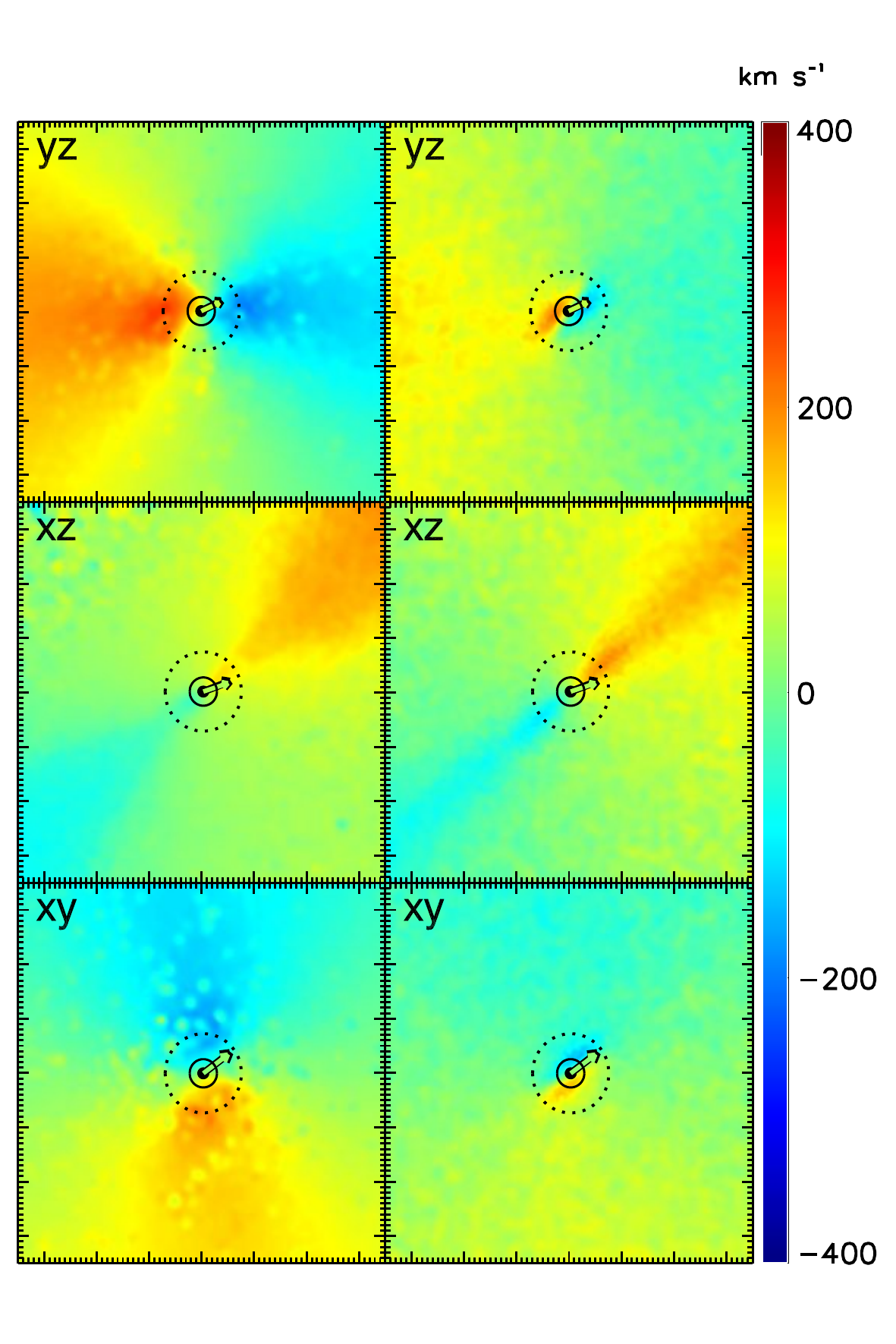}}
\caption{{\bf (a, left plot)}: \hbeta\ velocity map centered on one SMBH, in three projections, in a similar manner as in Fig.~\ref{fig:hbvel}. The SMBH velocity vectors are also plotted here as in Fig.~\ref{fig:hbvel}. The simulation shown has $q = 1$ and \fgas\ $= 0.1$ initially. {\bf (a, right plot):} For the same simulation, the velocity map is shown for the {\em total} gas distribution (rather than only the NL gas) in the left panels, and for the total {\em stellar} distribution in the right panels. The solid circle drawn around the SMBH denotes the gravitational softening radius ($r_{\rm soft}$), and the dashed circle denotes $2.8 r_{\rm soft}$, the point at which the gravitational force becomes fully Newtonian. {\bf (b)}: The same gas and stellar velocity maps are shown as in the previous plot, for the same merger model, but with the merger partially re-simulated with an $r_{\rm soft}$ that is 5 times smaller. The same circles are drawn here, but because $r_{\rm soft}$ is smaller, their size barely exceeds that of the dot that marks the SMBH position. {\bf (c)}: The same gas and stellar velocity maps are shown, but for a different merger model with $q = 1$ and \fgas\  $= 0.3$. The larger gas supply causes enhanced central star formation and a kinematic feature extending well beyond $r_{\rm soft}$. \label{fig:gsvel}}
\end{figure*}

Here we examine in detail the structure, kinematics, and luminosity of the NLRs in our simulations. Figure~\ref{fig:hbvel} shows velocity maps of the NLRs at various stages of a single merger simulation, along with corresponding \lhbeta\ maps. The simulation shown is an equal-mass merger with initial \fgas $=0.1$. We use it for illustration because its moderate gas fraction prevents the formation of a heavily-obscured starburst while still allowing substantial NLR activity throughout the simulation. Figure \ref{fig:hbvel}a shows a snapshot about 150 Myr after the onset of NL activity following the first close passage of the galaxies. We see that each NLR is $\ga 500$ pc  across, has a disky structure as seen in the $x z$ projection, and is in rotation. Additionally, in the first galaxy (left panels in each plot), a gap in the NLR is seen in the disk midplane. Because the gas density is highest along the midplane, the gas here may at times exceed our maximum-density criterion. 

Figure \ref{fig:hbvel}b shows a later snapshot when \lbol, \lhbeta, and the gas density are near their minima between the first passage and final coalescence of the galaxies. Accordingly, the NLRs are fainter and more diffuse. In the following snapshot (Fig.~\ref{fig:hbvel}c), the NLRs are shown just after a pericentric passage, $\sim 200$ Myr prior to SMBH merger. The SMBH separation here is only 0.76 kpc, and a large relative line-of-sight (LOS) velocity is apparent in the $x z$ projection. The NLRs are still quite diffuse and faint here, with \lhbeta\ $\sim$ few $\times 10^6$ \lsun.

Soon thereafter, the central gas density increases as the galaxies near their final coalescence, and the NLRs become brighter and more compact (Fig.~\ref{fig:hbvel}d). Their shape becomes more spherical, as their size is now limited mostly by the self-shielding criterion described in \S~\ref{sssec:fomega}, but a preferred axis for the brightest \hbeta\ emission can still be seen. 

The next snapshot (Fig.~\ref{fig:hbvel}e) occurs just before a close passage of the SMBHs. The red- and blueshifted NLRs are apparent as in Fig.~\ref{fig:hbvel}c, but here \lbol\ and \lhbeta\ are about 10 times higher. This is in fact an example of a double-peaked NL AGN induced by SMBH motion, as will be discussed in the next section. Finally, Fig.~\ref{fig:hbvel}f shows the last snapshot of the simulation, 200 Myr after the SMBH merger. The gas density, \lbol\ and \lhbeta\ have all declined, causing the NLR to become somewhat more diffuse. 

The persistence of flattened, rotating NLR morphologies throughout the merger is critical for the resulting velocity profiles; many double-peaked profiles arise simply from the rotation of these disk features, rather than from the relative SMBH motion. As such, these features merit closer examination. Because the NL gas particles are selected self-consistently from the total gas distribution, as outlined in \S\ \ref{sssec:coldgas}-\ref{sssec:U_ndens}, nothing is assumed {\em a priori} about the angular momentum of the NLRs themselves. Rather, an initial angular momentum is imparted to the galactic disks as described in \S~\ref{ssec:models}, such that the NL gas should rotate along with the total distribution of gas and stars.

If we define the SMBH radius of influence as the radius at which the baryonic mass surrounding the SMBH equals twice its mass, $r_{\rm infl} \equiv r(M_{\rm b}=2M_{\rm BH})$, then $r_{\rm infl}$ ranges from $\ga 100$ pc early in the merger, when the gas density is low, to only a few pc at late stages when the gas density is much higher. Thus, at all times, $r_{\rm NLR} \gg r_{\rm infl}$, so the gravitational potential of the SMBH alone is not responsible for keeping the NL clouds bound in coherent rotation. However, following the first close passage of the galaxies, a dense cusp of new stars begins to form and dominates the central region of each galaxy. Figure~\ref{fig:gsvel}a illustrates the net rotation of the central stellar and gas distribution; it is clear that the NL gas kinematics match that of the central gas distribution, which itself traces the kinematics of the stellar cusp. Thus, it is indeed the angular momentum of the central gas and newly-formed stars that underlies the flattened rotation features in the NLRs seen in Fig.~\ref{fig:hbvel}. 

A remaining question is whether the softened gravitational potential used in our simulations has a nonnegligible effect on the kinematic structure of the central region. The gravitational softening length used in our simulations is $r_{\rm soft} = 37$ pc, which is smaller than the size of our NLRs, but because of the softening kernel used, the gravitational forces are not strictly Newtonian until $2.8 r_{\rm soft}$, i.e., 104 pc. When the galaxies are near coalescence and the NLRs are most compact, their size can indeed be comparable to this value. We have drawn these two radii, $r_{\rm soft}$ and $2.8 r_{\rm soft}$, on the panels in Fig.~\ref{fig:gsvel}.  

In the first example (Fig.~\ref{fig:gsvel}a), the central stellar rotation feature is comparable in size to the extent of the softening kernel. However, we have resimulated part of this merger with a softening length five times smaller, and we show the result for the same snapshot in Fig.~\ref{fig:gsvel}b. Here, the circle drawn at $2.8 r_{\rm soft}$ is barely larger than the size of the dot that denotes the SMBH position, yet the kinematic stellar structure has the same spatial extent. In fact, the smaller softening length seems to allow the stellar rotation to persist down to smaller scales around the SMBH. This is evidence that if anything, the gravitational softening ``washes out" rotation features on scales of $< 10$ pc, and it is certainly not artificially inducing or supporting rotation on larger scales. In both cases the motion of the gas traces that of the stars, so we can have similar confidence in the rotation observed in our NLRs. 

Additionally, we find that the central, rotating cusp of stars does not appear until after the first burst of star formation following the close passage of the galaxies. This  argues against a numerical origin for the gas rotation features, as the gravitational softening remains constant throughout the simulation. Further evidence along these lines comes from Fig.~\ref{fig:gsvel}c, which shows a snapshot from an equal-mass merger with a higher gas fraction (\fgas\ $= 0.3$). In this simulation, the higher central SFR creates a more extended, rotating, disk-like stellar feature that can be seen in the $x z$ velocity map, well beyond the influence of the softened gravitational potential.

\subsection{Observable Signatures of kpc-scale Double-peaked NL AGN}
\label{ssec:dnlkin}

\begin{figure*}
\hspace{20pt}
\subfloat[]{\includegraphics[width=0.34\textwidth]{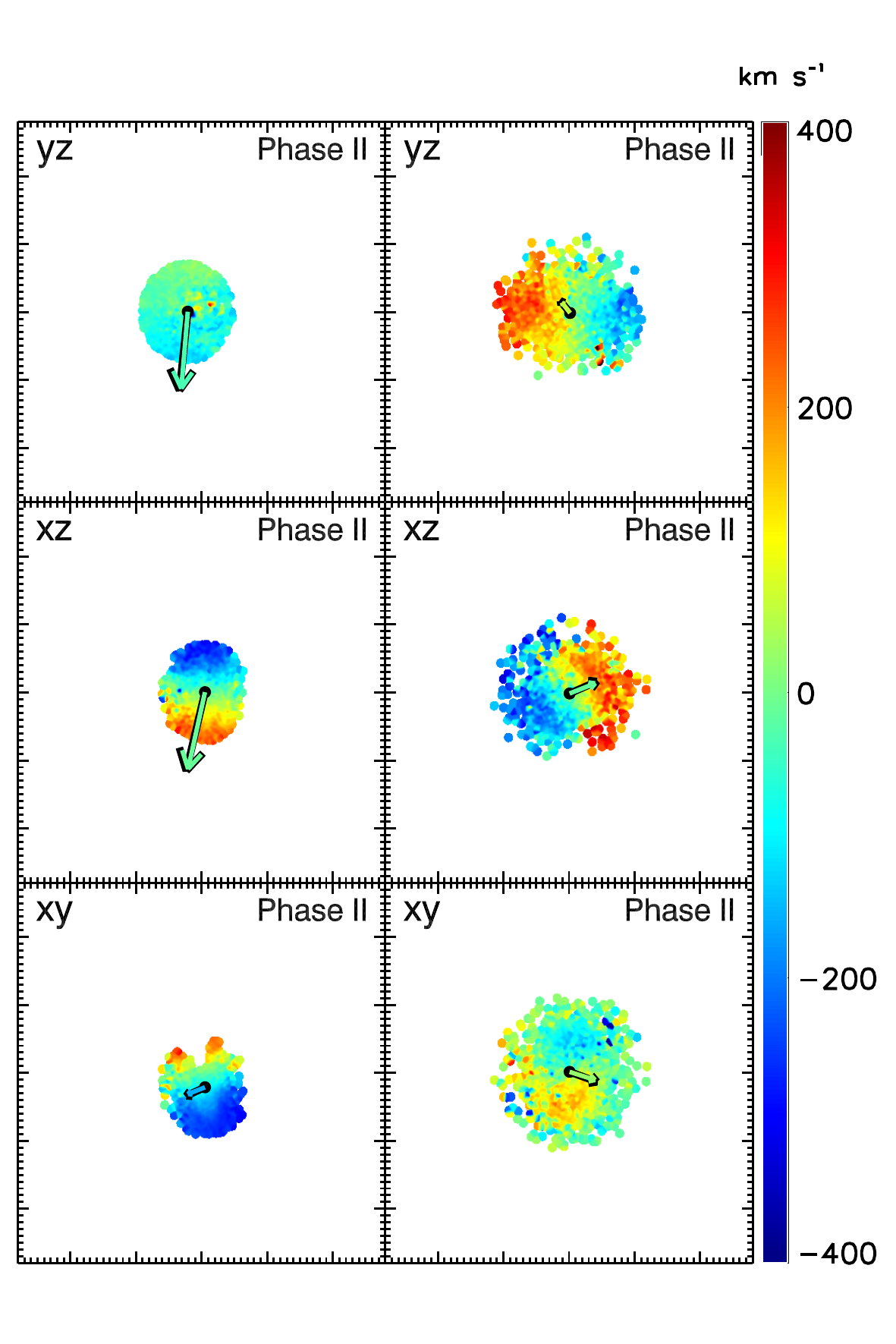}}
\hspace{5pt}
\subfloat[]{\includegraphics[width=0.34\textwidth]{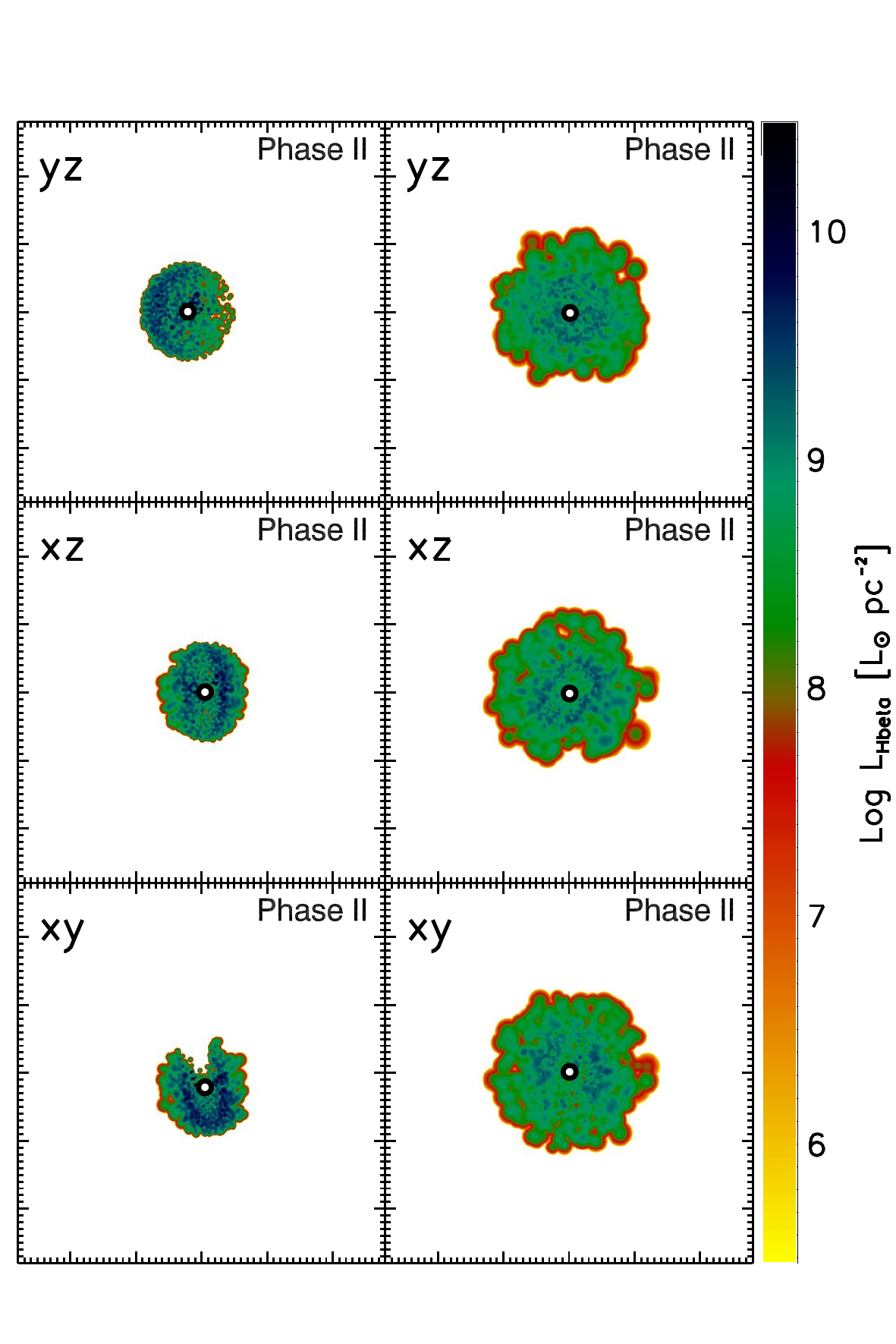}}
\hspace{15pt}
\subfloat[]{\includegraphics[width=0.1983\textwidth]{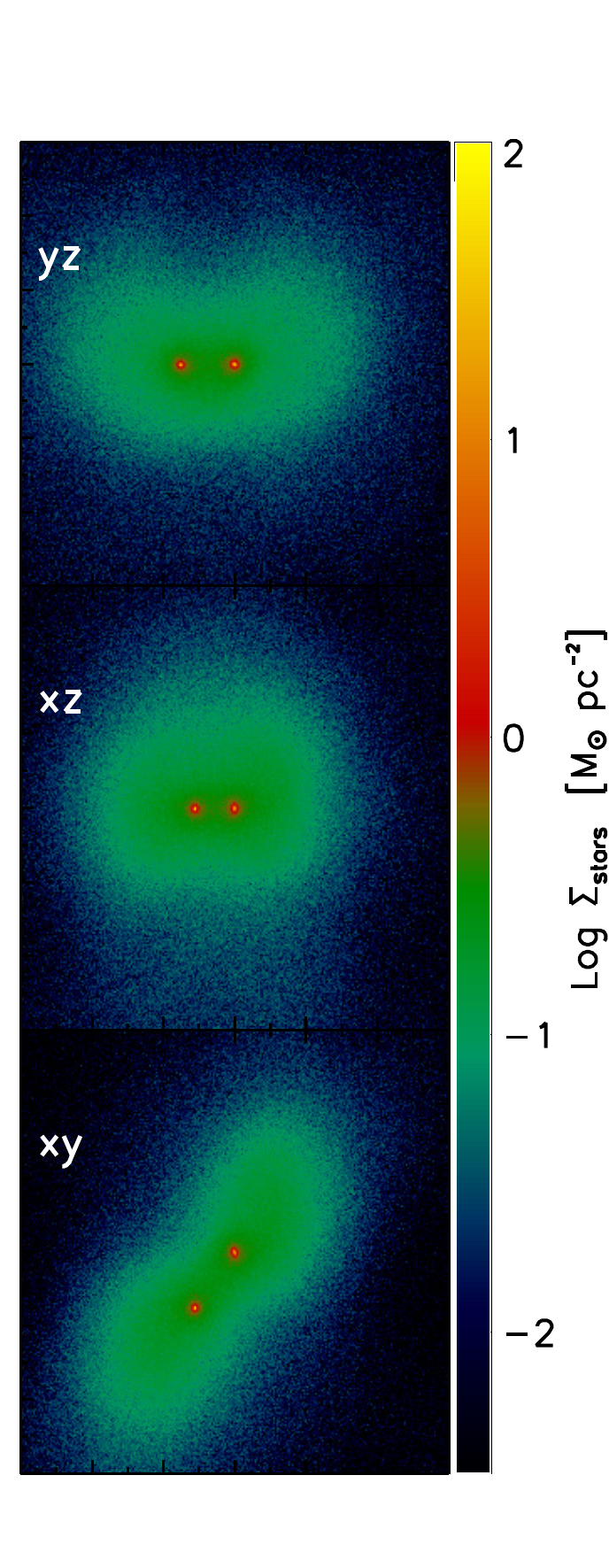}}

\subfloat[]{\includegraphics[width=0.7\textwidth]{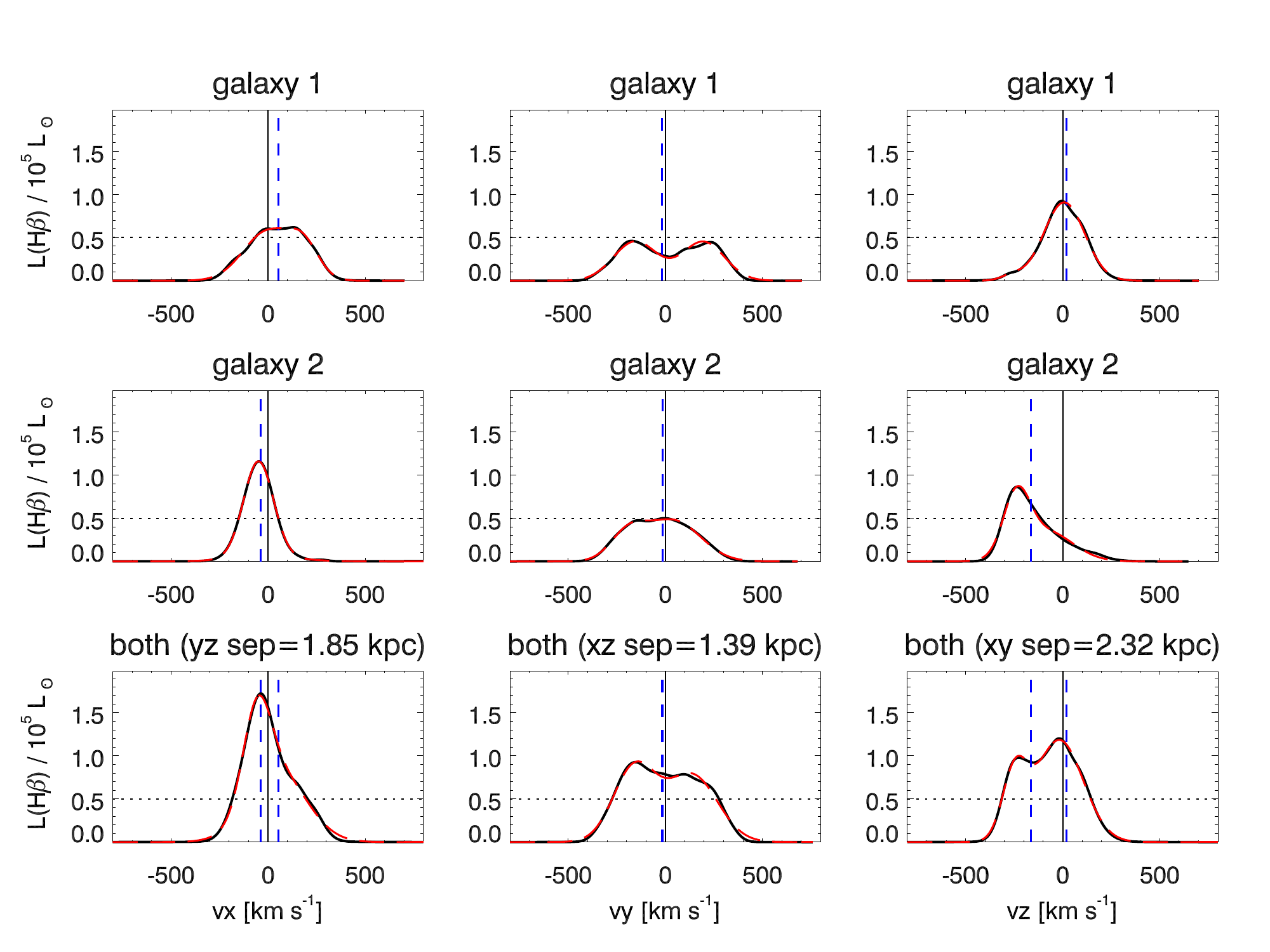}}
\hspace{32pt}
\subfloat[]{\includegraphics[width=0.1983\textwidth]{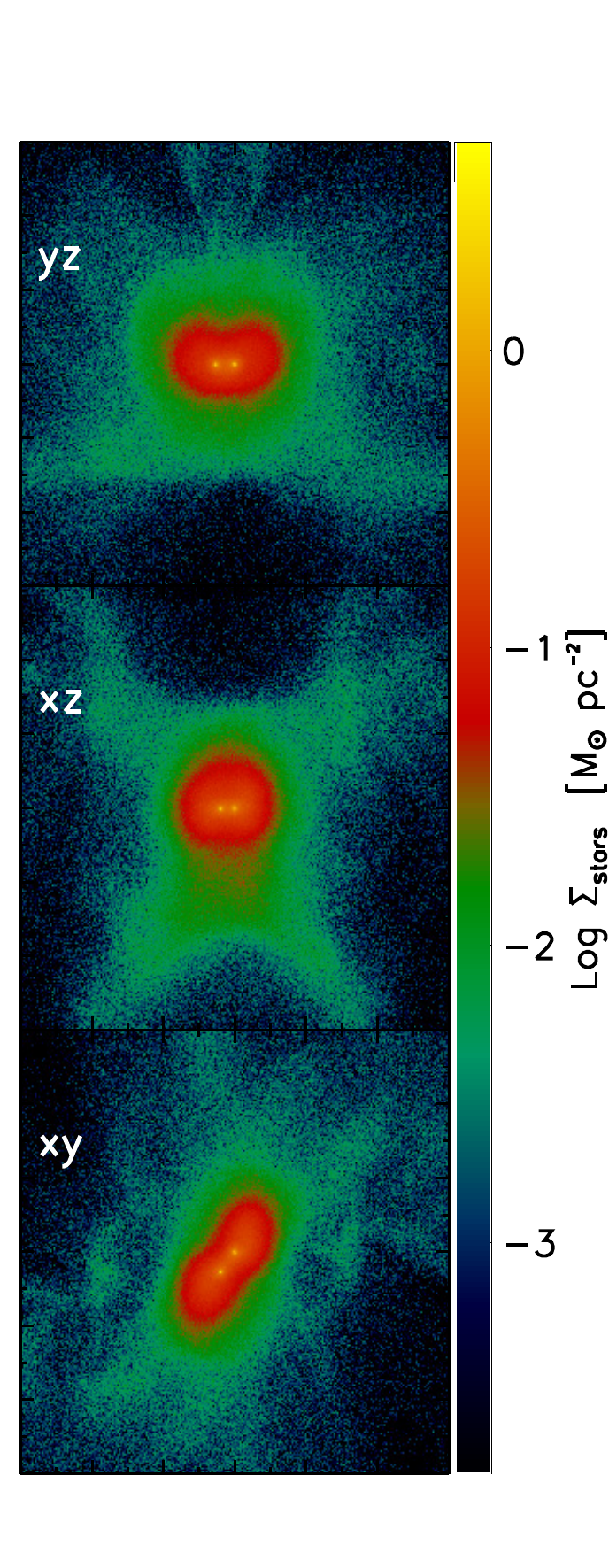}}

\caption[]{The NLR velocity maps, \hbeta\ luminosity, and 1-D velocity profiles, as well as the stellar density maps, are shown for three orthogonal projections of a single simulation snapshot during the kpc-scale phase (Phase II). The merger model for the simulation shown has $q=1$ and \fgas$ = 0.1$ initially. The snapshot shown occurs 20 Myr prior to the SMBH merger, when the SMBH separation is 2.3 kpc. \,\,{\bf (a):} LOS velocity maps for NLR particles, in the same manner as in Fig.~\ref{fig:hbvel}. As before, the left column of this plot shows three projections centered on SMBH2, the right column is centered on SMBH1, and the arrows denote the SMBH velocity.  \,\, {\bf (b):} map of \hbeta\ surface brightness. The orientation and scale of these panels is the same as in the velocity map (left plot). Here the SMBH positions are denoted by the black open circle. \,\,{\bf (c):} Projected stellar density shown from the same three orientations; each panel is 15 kpc on a side. \,\,{\bf (d):} 1-D velocity profiles for the NLRs shown. The three columns show $v_x$, $v_y$, \& $v_z$, corresponding to the $y z$, $x z$, and $x y$ projections, respectively. The top row shows the profiles for NLR particles associated with galaxy 1 (and SMBH1), and the middle row shows those associated with galaxy 2 (SMBH2). The bottom row shows the combined profile. Because this snapshot is within Phase II, the combined profile is by definition what would be seen by observers. In each panel, the black curve shows the profile calculated from simulation data, and the {\rev red long-dashed} curve is the best fit for a double-Gaussian profile. The horizontal black dotted line marks the minimum threshold we have set for observability of the \hbeta\ line: peak \lhbeta\ $= 5\times 10^4$ \msun. The vertical black line is the rest frame of the host galaxy (stellar center-of-mass velocity), and the blue dashed vertical lines denote the velocities of the SMBHs corresponding to each NLR; both are shown in the combined profile. The calculated profiles assume a Gaussian internal velocity dispersion with FWHM $= 0.5\, c_{\rm sound}$, and are degraded to a resolution of 65 \kms. 
\,\,{\bf (e):} projected stellar density from three orientations, but on a larger scale than above; each panel is 43 kpc on a side. \label{fig:dir_dnl_wide}}
\end{figure*}

\subsubsection{Double-Peaked NL Velocity Profiles}
\label{sssec:kinematics}

\begin{figure*}
\hspace{74pt}
\subfloat[]{\includegraphics[width=0.22\textwidth]{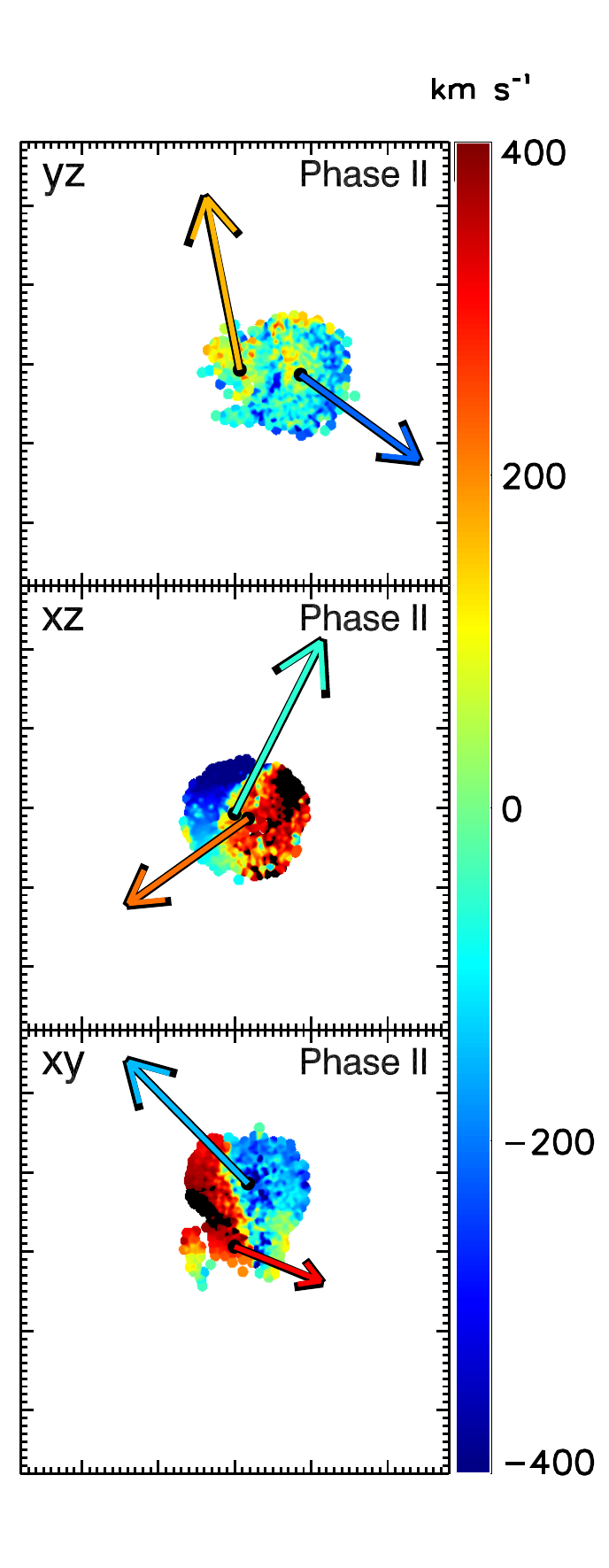}}
\hspace{10pt}
\subfloat[]{\includegraphics[width=0.22\textwidth]{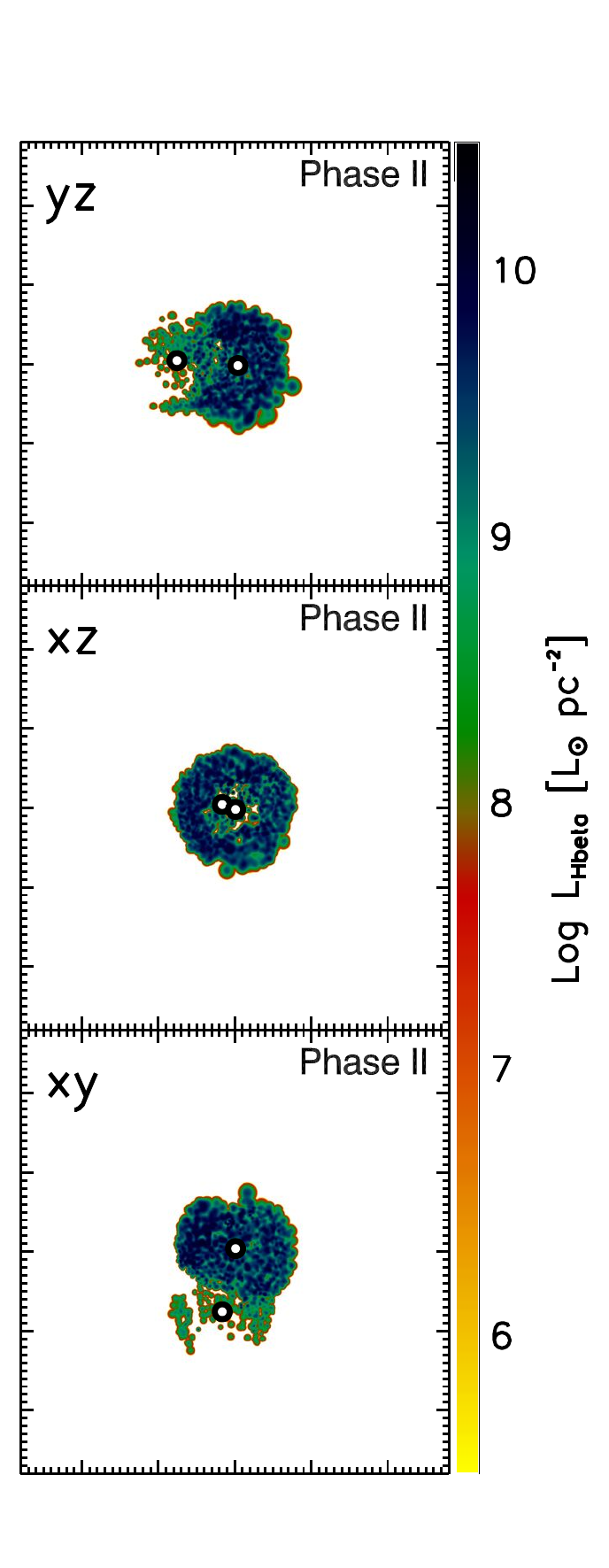}}
\hspace{52pt}
\subfloat[]{\includegraphics[width=0.22\textwidth]{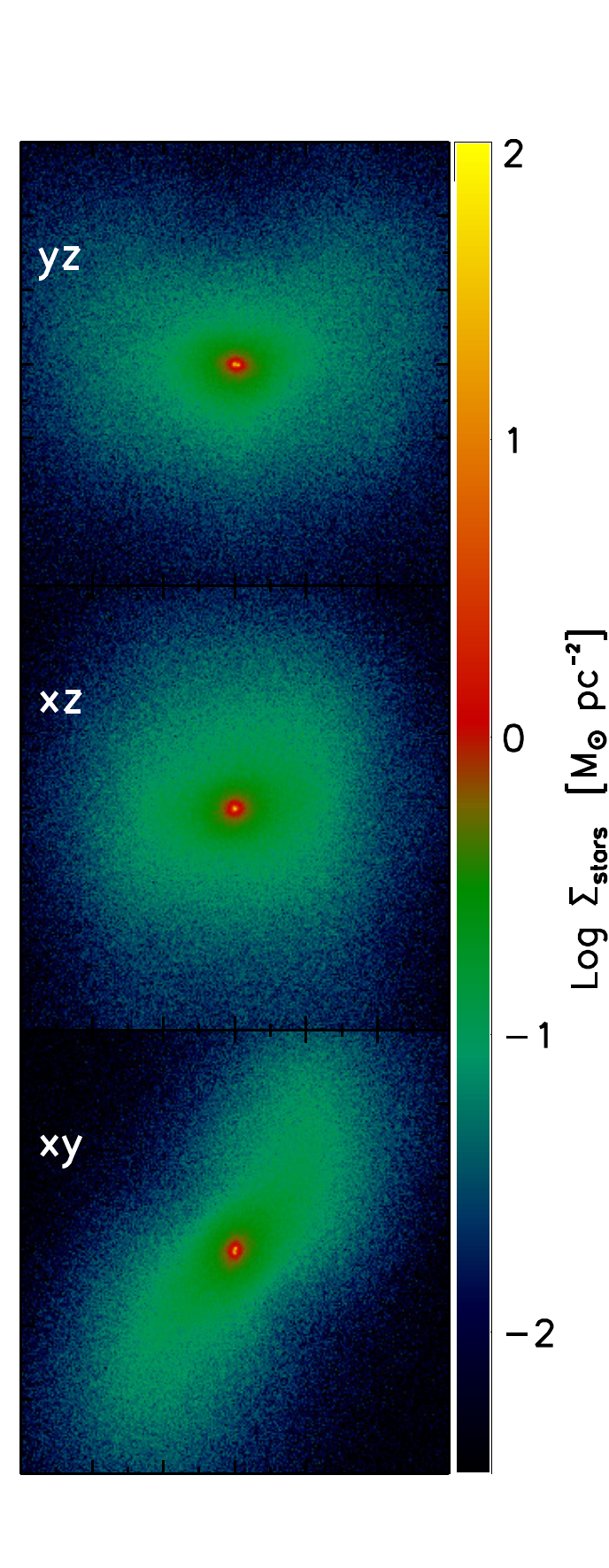}}

\subfloat[]{\includegraphics[width=0.7\textwidth]{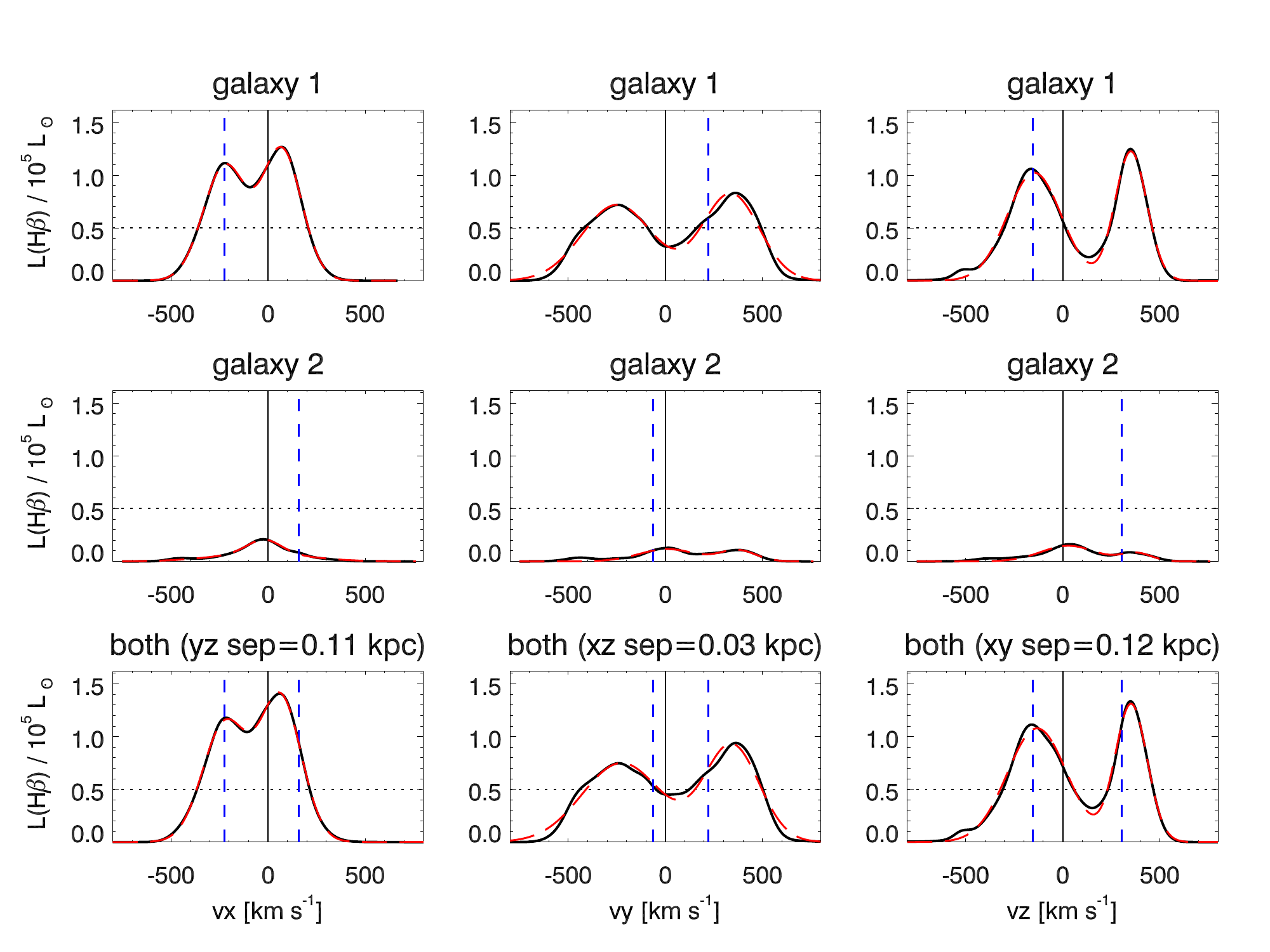}}
\hspace{10pt}
\subfloat[]{\includegraphics[width=0.22\textwidth]{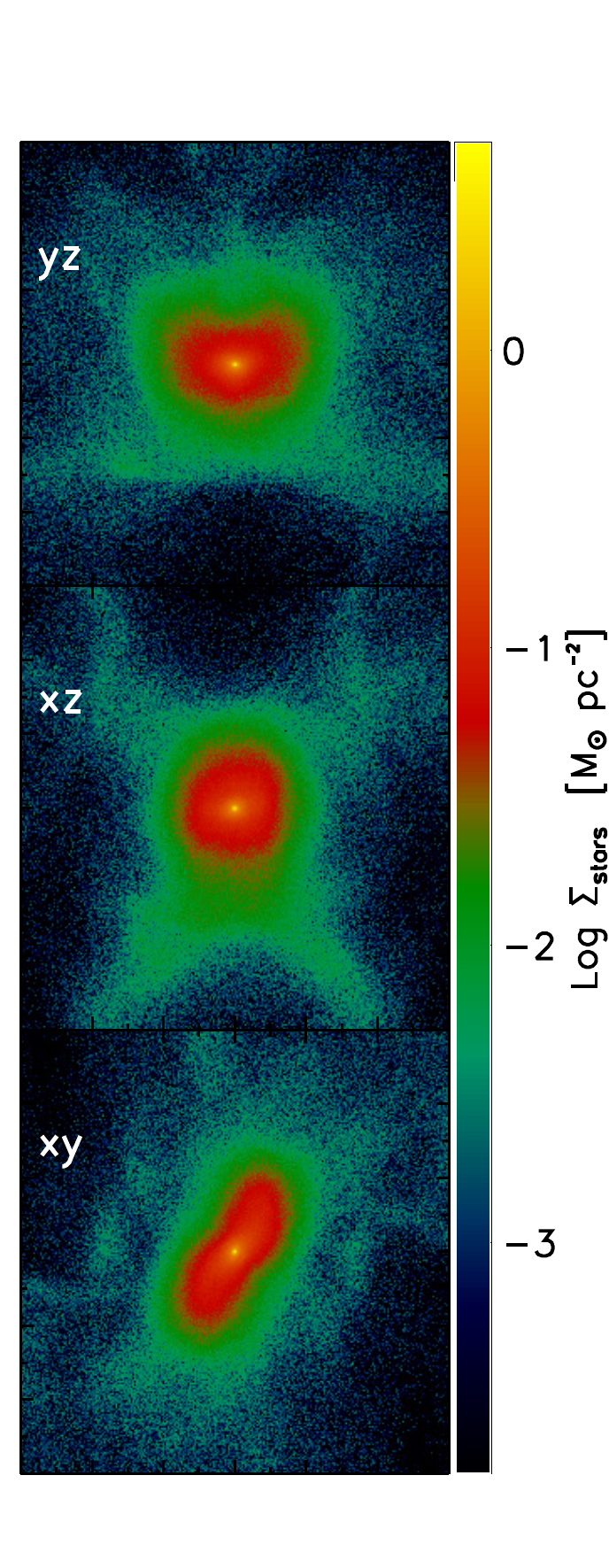}}
\caption[]{
NLR velocity maps, \hbeta\ luminosity maps, and 1-D velocity profiles, as well as stellar density maps, are shown in the same manner as Fig.~\ref{fig:dir_dnl_wide}, for a Phase II snapshot in a simulation with $q=1$ and \fgas $= 0.1$ initially. This snapshot occurs 8 Myr prior to SMBH merger, when the SMBH separation is only 0.17 kpc. \label{fig:dir_dnl_tight}}
\end{figure*}

\begin{figure*}
\hspace{20pt}
\subfloat[]{\includegraphics[width=0.34\textwidth]{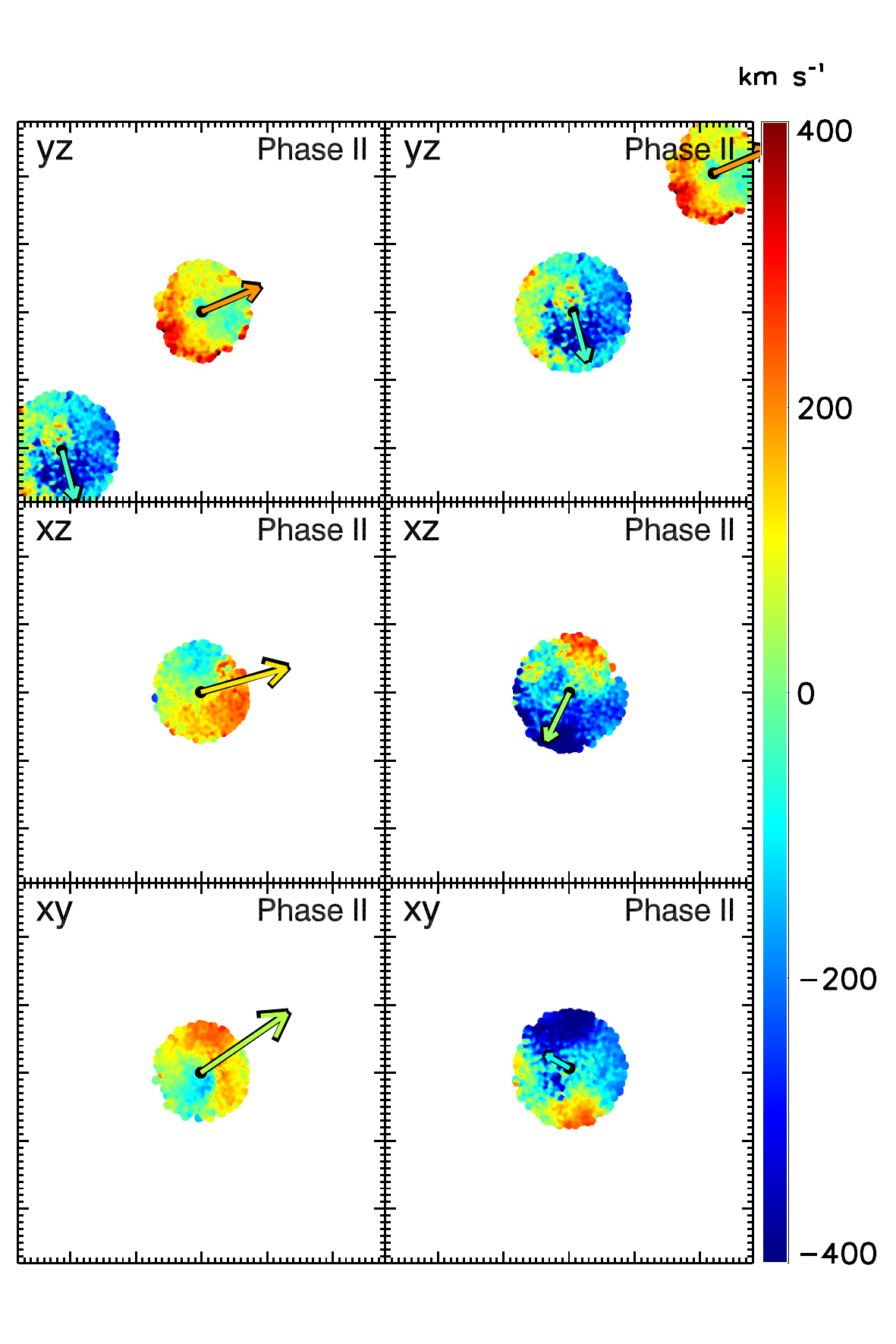}}
\hspace{5pt}
\subfloat[]{\includegraphics[width=0.34\textwidth]{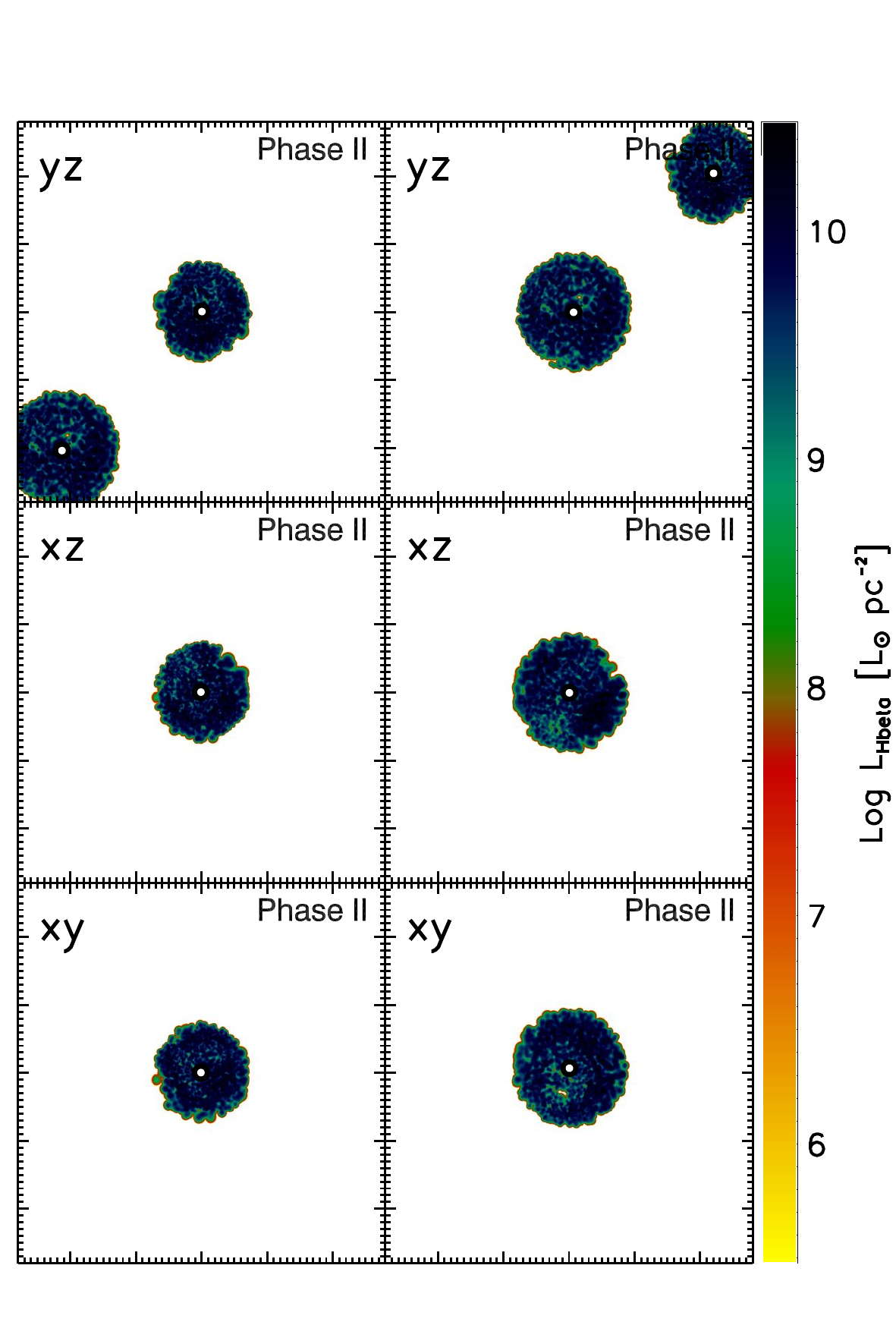}}
\hspace{15pt}
\subfloat[]{\includegraphics[width=0.1983\textwidth]{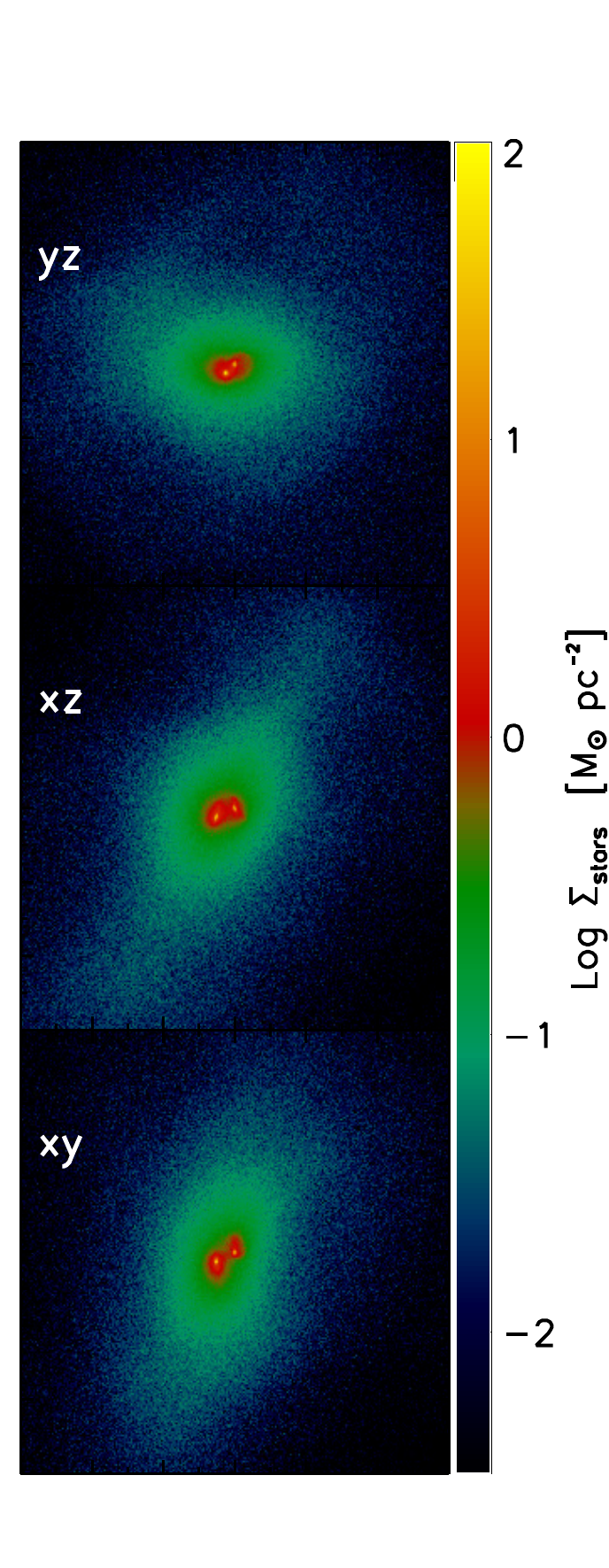}}

\subfloat[]{\includegraphics[width=0.7\textwidth]{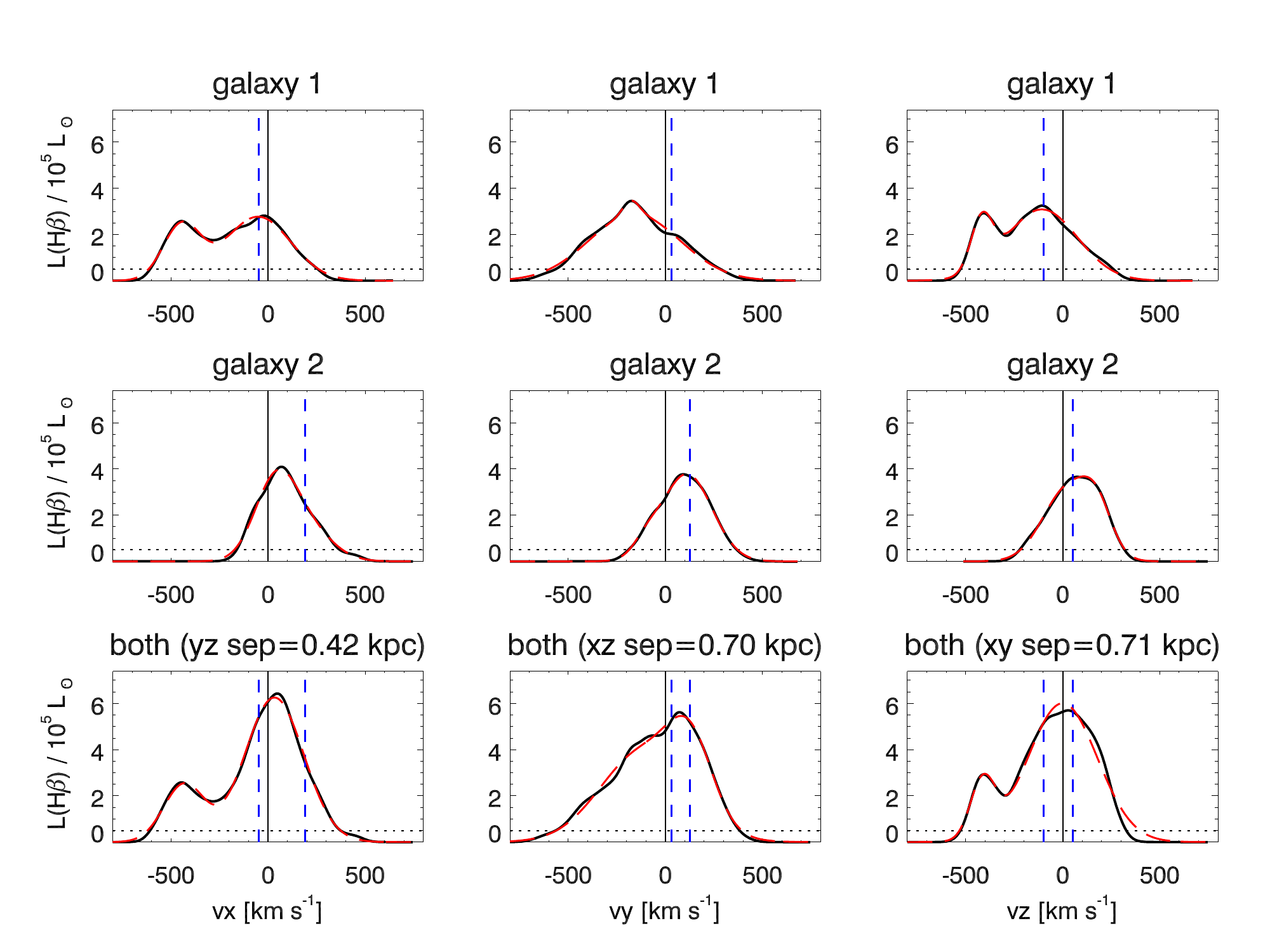}}
\hspace{32pt}
\subfloat[]{\includegraphics[width=0.1983\textwidth]{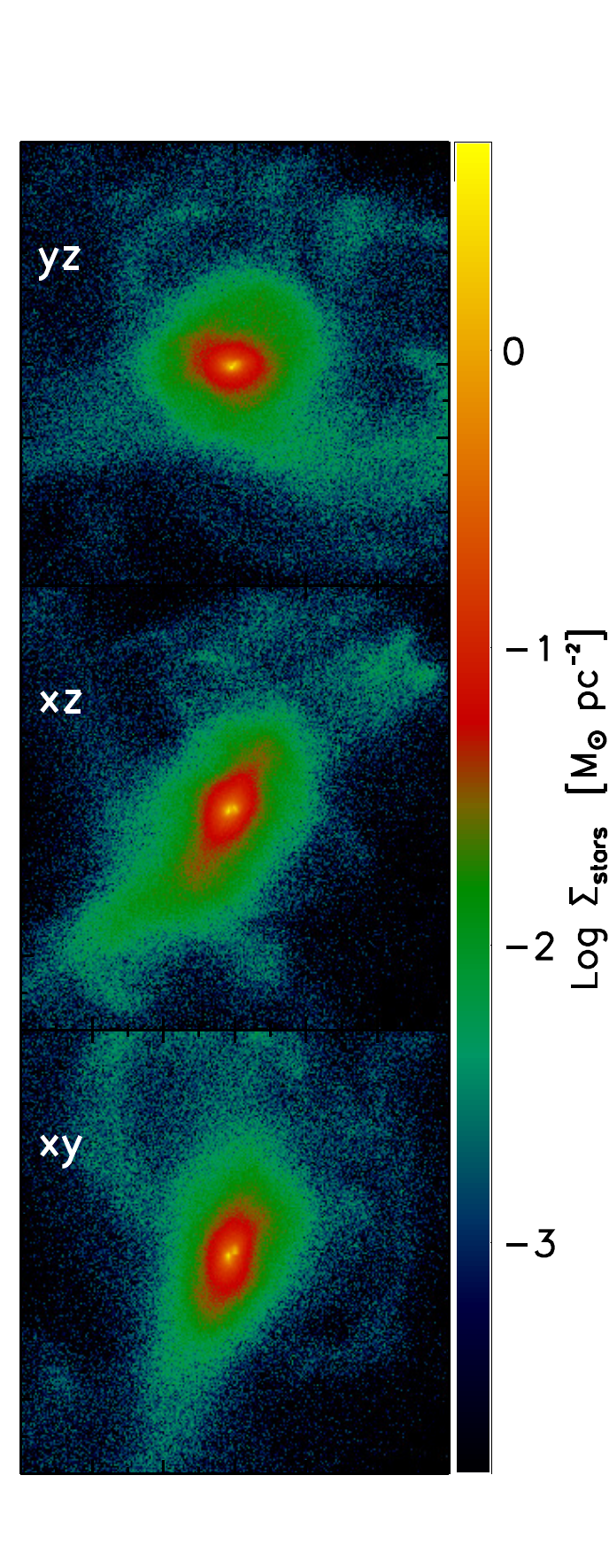}}
\caption[]{NLR velocity maps, \hbeta\ luminosity maps, and 1-D velocity profiles, as well as stellar density maps, are shown in the same manner as Fig.~\ref{fig:dir_dnl_wide}, for a Phase II snapshot in {\rev $q=0.5$, \fgas\ $= 0.3$ simulation.} This snapshot occurs 20 Myr prior to SMBH merger, when the SMBH separation is 1.1 kpc. \label{fig:kin_dnl_uneven}}
\end{figure*}

In Figs.~\ref{fig:dir_dnl_wide} - \ref{fig:kin_dnl_uneven}, we illustrate some examples of kiloparsec-scale dual AGN with double-peaked NLs, looking in detail at their kinematic features. Foremost, these examples demonstrate that a variety of velocity structures can give rise to dNL AGN during the kiloparsec-scale phase. Figure~\ref{fig:dir_dnl_wide} (the $x y$ projection) is an example of the ``standard" picture that motivates the association of dNL AGN with dual SMBHs. In Fig.~\ref{fig:dir_dnl_wide}a, we see two distinct NLRs with LOS velocities corresponding to the SMBH velocities. The velocity profile in Fig.~\ref{fig:dir_dnl_wide}d ($v_z$; last column) shows a peak with virtually no offset arising from the NLR associated with ``galaxy 1", and a blueshifted peak arising from the other NLR. (Note that the asymmetry of the latter owes to the asymmetric distribution of the NLR gas, as seen in Fig.~\ref{fig:dir_dnl_wide}b.) The combined profile, which is what an observer would see, is double-peaked with a velocity splitting of $\sim 300$ \kms. This is an example of a dNL AGN resulting directly from SMBH motion, at a time when the NLRs are non-overlapping (the SMBH separation is 2.3 kpc). 

Figure~\ref{fig:dir_dnl_tight} also shows a dNL AGN that results directly from SMBH motion, but in a very different regime. Here, the SMBH separation is comparable to the size of the NLR, so we would not expect two stellar density peaks to be resolvable. The luminosities of the two SMBHs are quite different; we see in the middle panels of Fig.~\ref{fig:dir_dnl_tight}d that the NLR emission arises almost entirely from ionization by the SMBH in ``galaxy 1". However, the velocity map (Fig.~\ref{fig:dir_dnl_tight}a) demonstrates clearly that the second SMBH influences the NLR kinematics; as described in \S\ \ref{ssec:nlmorph}, the SMBH carries a cusp of bound gas and stars, which in this instance plows through the edge of the first NLR. The resulting double-peaked velocity profile is apparent in all three orthogonal projections. We emphasize that in this example, a dNL AGN results directly from SMBH motion even though the SMBHs do not have comparable luminosities and their separation is less than the size of the NLR. Such features will be very short-lived, of course, but should be hard to avoid once the SMBH separation falls below the size of the NLR.

In addition to dNL AGN produced directly by SMBH motion, we find many examples where double-peaked profiles are produced by gas kinematics but are still {\em influenced} by the SMBH motion. In Figures \ref{fig:kin_dnl_uneven}a \& d,  the combined velocity profile is a redshifted, uneven double peak in which the double peak arises from NLR1, but is influenced by the blueshifted peak in NLR2. Thus, the SMBH motion affects both the centroid and the peak ratio of the combined NLR profile. 

In many other cases we find that the double-peaked features arising from gas kinematics are not significantly influenced by the SMBH motion but are simply coincident with the kpc-scale phase, because the SMBHs are simultaneously active more frequently during the late merger stages. We note that examples of this scenario have in fact been found in real systems \citep{fu12}. 

It is of great interest, both for future observations and for our theoretical understanding of dual SMBHs, to estimate the fraction of Phase II dNL AGN in our simulations that are affected either directly or indirectly by the SMBH motion. Accordingly, we have undertaken a visual analysis of the velocity profiles at each Phase II snapshot, from three orthogonal viewing angles, in two of our simulations (mergers with $q=1$, \fgas $=0.1$ and with $q=0.5$, \fgas $=0.3$). We have visually classified each observable double-peaked profile in Phase II  (for both $a_{\rm max} = 5.5$ \& 21 kpc) as {\bf (i)} directly induced by SMBH motion, {\bf (ii)} indirectly influenced by SMBH motion, {\bf (iii)} merely coincident with Phase II, or {\bf (iv)} a complex or highly asymmetric profile. 

The fraction of snapshots in each category varies substantially between the three sight lines and two simulations, owing to small number statistics and to intrinsic variability with viewing angle and galaxy merger model. Nevertheless, we can draw some general conclusions from this analysis. The directly-induced dNLs occur in only a few snapshots, corresponding to lifetimes of 3-6 Myr. These primarily occur for SMBH separations $< 5.5$ kpc, and thus they account for up to 60\% (33\%) of the Phase II dNL lifetime $a_{\rm max} = 5.5$ (21) kpc. The total fraction of Phase II for which the SMBH motion has {\em some} influence on the velocity profiles, directly or indirectly, is between 30 and 80\% depending on $a_{\rm max}$ and the merger model. Most of the remaining double-peaked profiles arise from gas rotation in a single, disk-like NLR. However, between 0 and 30\% of the Phase II profiles classified as double peaks are actually complex (multi-peaked) or highly asymmetric profiles. These are important signatures as well, because they are usually also associated with relative SMBH motion. The complex profiles, in particular, typically arise either from a highly-disturbed NLR gas distribution following a close passage or from the superposition of two double-peaked profiles with different velocity centroids.

\subsubsection{Stellar Structure}
\label{sssec:stars}

In addition to the NLR kinematics, subfigures {\bf (c)} in Figs.~\ref{fig:dir_dnl_wide} - \ref{fig:kin_dnl_uneven} show stellar density maps for each of these examples. In both Fig.~\ref{fig:dir_dnl_wide} and Fig.~\ref{fig:kin_dnl_uneven}, two density peaks are present in all projections of the stellar density maps. In Fig.~\ref{fig:kin_dnl_uneven}, very high resolution imaging would be required to resolve the two peaks; the $y z$ projection in particular has the smallest projected separation but the largest relative LOS SMBH motion. This is not an uncommon feature of kpc-scale SMBH pairs; owing to their rapid inspiral on these scales, their orbits are generally {\em not} circular. Thus, the largest LOS velocity separation may occur when the LOS is oriented along the long axis of the eccentric, plunging orbit and the SMBHs are at pericenter, which corresponds to the smallest projected spatial separation. This has unfortunate implications for attempts to confirm dual SMBH candidates via spatially-resolved stellar cusps, but it also means that some apparently single-core galaxies with dNL AGN may in fact be hiding dual SMBHs at small projected separations. 

In such cases, other clues may hint at the object's true nature. For example, the $y z$ and $x z$ projections in Fig.~\ref{fig:kin_dnl_uneven}c show disturbed morphology that indicates the galaxy's ongoing merger state. The subfigures {\bf (e)} in Figs.~\ref{fig:dir_dnl_wide}-\ref{fig:kin_dnl_uneven} have a larger field of view and a lower minimum density by a factor of 20 than the subfigures {\bf (c)}. This reveals diffuse tidal features indicating the highly disturbed state of the galaxy. While disturbed morphology in a dNL AGN host galaxy does not not necessarily indicate a SMBH pair, as the SMBHs may have already merged, this signature would establish the system as a late-stage merger, thus removing the possibility of an isolated galaxy or a galaxy with a close, but still widely separated companion. Therefore, sensitive imaging of dNL AGN hosts, for example with {\em HST}, could capture faint tidal features and aid in distinguishing dual AGN from those with double peaks arising from gas kinematics.

\citet{comerf12} note that the objects in their dNL AGN sample are preferentially aligned with the major axis of the host galaxy and propose this as an additional diagnostic for the presence of a SMBH pair. In Figs.~\ref{fig:dir_dnl_wide}c we see that there is indeed a strong alignment between the orientation of the SMBH pair and the apparent ellipticity of the merger remnant. In this case, when the SMBHs are separated by $> 2$ kpc, this ``ellipticity" of the remnant results directly from the unmerged galaxy cores. The lower-density tidal features shown in Fig.~\ref{fig:dir_dnl_wide}e make clear that the overall morphology is highly disturbed rather than ellipsoidal. A similarly disturbed morphology is apparent in Fig.~\ref{fig:kin_dnl_uneven}e; here, however, the SMBHs have a separation of $\sim 1$ kpc, and the progenitor galaxy cores have essentially merged. In this case the SMBH pair is less closely aligned with the apparent ellipticity of the remnant, as the BH orbits are now decoupled from their progenitor host galaxies. We therefore predict an inverse correlation between projected SMBH pair separation and the degree of alignment with the apparent major axis of the merger remnant, though with large scatter owing to variation in the SMBH orbits, the viewing angle, and the structure of the unrelaxed merger remnant. 

\begin{figure*}
\resizebox{0.82\hsize}{!}{\includegraphics{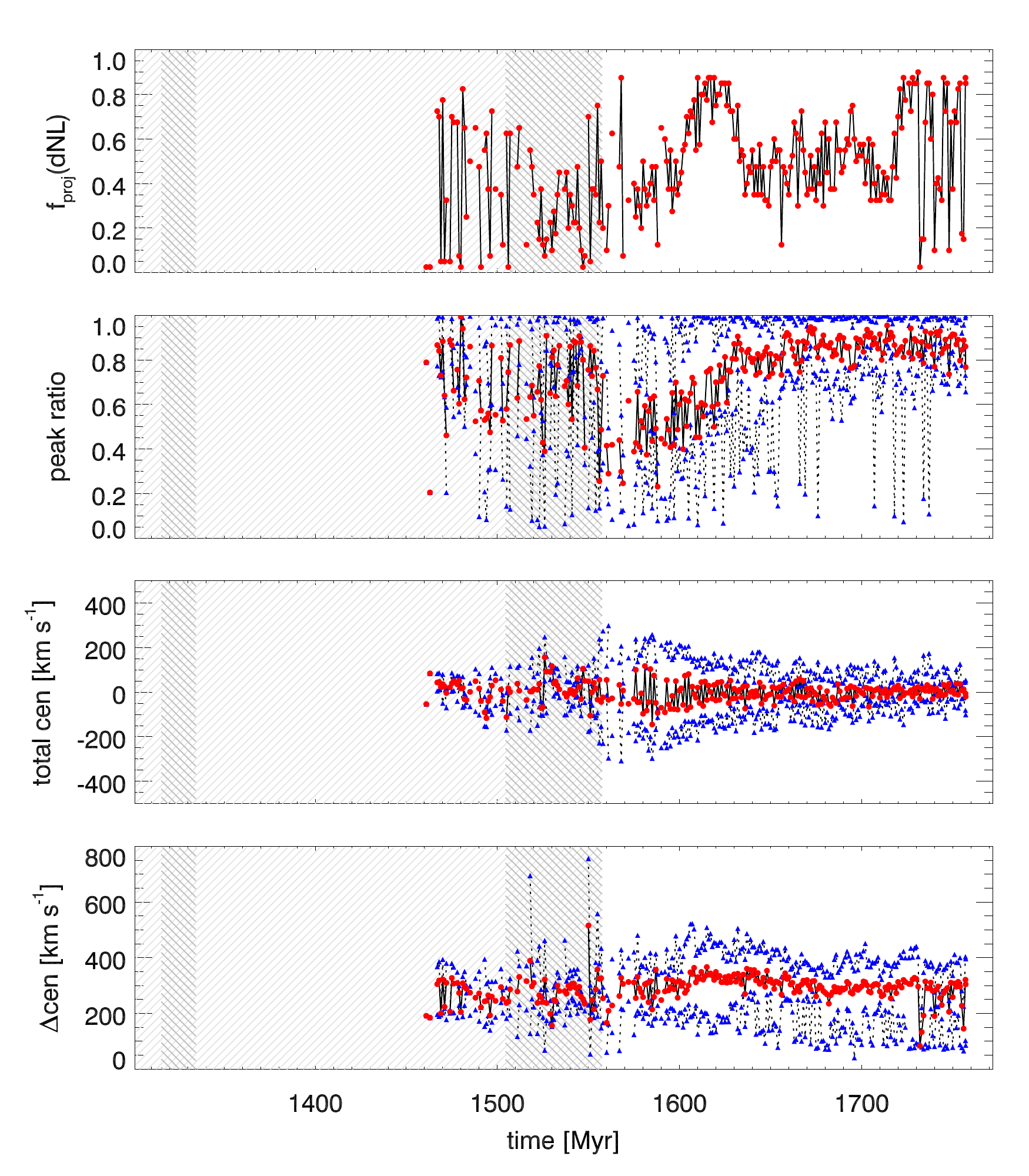}}
\caption[]{Line-of-sight-averaged properties of \hbeta\ velocity profiles versus time. For greater clarity, only the late-merger portion of the simulation is shown, starting 15 Myr before the onset of Phase II. Phase II snapshots are shaded with gray cross-hatching for $a_{\rm max} = 5.5$ kpc (light gray) and 21 kpc (dark gray). The simulation shown is an equal-mass merger with \fgas\ $= 0.1$ initially. The top panel shows, as a function of time, the fraction of the 40 random sight lines sampled from which the velocity profile has an observable double peak. ``Observable" double peaks are those with a velocity splitting larger than FWHM$_{\rm a}$ (corresponding to the brightest peak), a peak ratio larger than 0.05, and a peak \lhbeta\ $> 5\times 10^4$ \lsun. Only snapshots in which these criteria are met for at least one viewing angle are shown. The data shown represent the combined velocity profiles (NLR1 + NLR2). In the other panels, various quantities from the profile fits are shown, averaged over {\em only} the sight lines for which the profile is double-peaked. The mean values {\rev are plotted as red circles, and the maximum and minimum values are shown with blue triangles.} The extremal values are connected with dotted lines, while solid lines connect the mean values. Starting with the second panel from the top, the panels show: the peak ratio of the double peaks, the offset of the entire profile, and the velocity splitting of the two peaks ($\Delta_{\rm cen}$). \label{fig:fitdata1}}
\end{figure*}

Our results indicate that the absence of two resolvable stellar components does not rule out the presence of a dual AGN, even for spatial resolutions of $\sim 1$ kpc. Although a candidate dual AGN has recently been found with a separation of 150 pc \citep{fabbia11}, the recent work by \citet{comerf12}---in which all 81 objects observed with slit spectroscopy showed spatially-distinct emission components with separations of at least 200 pc---suggests that sub-kpc-scale dual AGN may be relatively rare. If so, this likely owes to their  short lifetimes, which are typically a few Myr, or 1-2\% of Phase II. Thus, for most of our simulations, $\sim 10-40$\% of  of the dNL AGN in Phase II contain sub-kpc SMBH pairs. However, for gas-poor mergers (\fgas\ $= 0.04$ initially), we find that $\ga 90$\% of the Phase II dNL AGN may be associated with sub-kpc SMBH pairs. Because these mergers are less dissipative, the SMBH inspiral is more gradual, and the SMBHs spend more time at small separations prior to merger. Also, the low-level AGN activity in gas-poor mergers means that the NL AGN may be under-luminous until the peak of SMBH accretion, near the time of SMBH merger. Therefore, we argue that some systems with a single stellar component but with disturbed morphology may contain dual AGN with separations of a few hundred parsec (see Fig.~\ref{fig:dir_dnl_tight}); these may be good candidates for high-resolution {\em Chandra} \citep[c.f.][]{fabbia11} or EVLA \citep[c.f.][]{fu11b} observations. 

\subsubsection{NL Velocity Profile Diagnostics}
\label{sssec:diagnostics}

We can compare the peak luminosities of the simulated \hbeta\ profiles to observations, similar to the comparison of the {\em total} \hbeta\ luminosities in \S~\ref{ssec:general}. The observed dNL AGN profiles for which such data is published have peak \hbeta\ fluxes ranging from non-detections to nearly $10^{-15}$ erg cm$^{-2}$ s\inv, with typical peak fluxes of a few $\times 10^{-16}$ erg cm$^{-2}$ s\inv\ \citep{liu10a,liu10b,comerf09a,comerf12}. At an average sample redshift of $z \sim 0.1$, these upper and typical values correspond to log \lsun $\sim$ 6.7 and 6.2, respectively. Similarly, in our simulations, the \hbeta\ profile peak luminosities range from below the detectable limit (log \lsun\ $= 4.7$, corresponding to $10^{-17}$ erg cm$^{-2}$ s\inv\ at $z = 0.1$, as described in \S~\ref{sssec:hbeta_profs}) to as high as log \lsun $= 6.8$ in our most gas-rich, equal-mass merger simulation. In most of our simulations, the maximum peak luminosity (for 40 random sight lines throughout the merger) is log \lsun $\sim$ 5.7 - 6.2. Thus, the range of simulated \hbeta\ profile peak luminosities is consistent with observations.

Figure~\ref{fig:fitdata1} shows the evolution of several line-of-sight-averaged quantities for the \hbeta\ velocity profiles in the late-merger stage of an example simulation.  For each snapshot, the values shown are averages resulting from fitting profiles for 40 random sight lines, and the definitions of merger phases are based on the projected SMBH separation for each sight line. This equal-mass merger with initial \fgas\ $= 0.1$ was chosen for illustration because it has sufficient NLR activity to demonstrate all the salient features of the \hbeta\ profiles, but still has a moderate gas content and thus should not be strongly affected by a rapid post-merger increase in \vesc\ or by possible dust obscuration.  For each snapshot, the data shown are averaged over the sight lines for which the profile has an observable double peak. 

Several trends are apparent in Fig.~\ref{fig:fitdata1}. The top panel shows the fraction of sight lines from which the profile appears double-peaked, and the second panel shows the luminosity ratio of the two peaks. Both quantities are generally lower and more variable during Phase II than during Phase III. This is because double peaks arising from a single NLR are generally caused by rotating gas disks in our simulations, whereas some of the double peaks in Phase II are caused or influenced by SMBH motion. In principle, the 1-D velocity profile for two NL AGN in a circular orbit could appear quite similar to that of a rotating disk of the same size. However, the dual SMBHs in our simulations generally have non-circular orbits; thus, they appear as double-peaked for a smaller fraction of viewing angles than do the rotating disks. 

Additionally, we see a tendency for double peaks in Phase III (those caused exclusively by rotating gas) to have peak ratios closer to unity than those in Phase II. This is the diagnostic proposed by \citet{smith11} to differentiate between dNL AGN arising from rotating disks versus dual SMBHs. While we see that this distinction does exist in an average sense, the peak ratios in Phase III still have substantial variation with viewing angle. In the best cases, the peak ratios for a single snapshot range from $\sim 0.6 - 1.0$ depending on the sight line, while at other times the {\em mean} peak ratios are $\la 0.5$. Furthermore, in almost every snapshot with an observable double peak, including those in Phase II, the maximum peak ratio is $> 0.8 - 0.9$; i.e., there is usually at least one viewing angle from which the profile has nearly even peaks. We therefore conclude that for individual systems, an uneven-peaked dNL AGN has at most a modestly higher probability of containing a dual SMBH than one with an even-peaked profile, even for peak ratios as low as $\sim 0.1 - 0.2$.  

The velocity splitting of the peaks in each double-peaked profile ($\Delta_{\rm cen}$, bottom panel of Fig.~\ref{fig:fitdata1}) also shows distinct behavior in different merger phases. The spikes in $\Delta_{\rm cen}$ during Phase II correspond to pericentric passages of the SMBHs shortly before their merger, revealing a brief but direct effect of the SMBH motion on the \hbeta\ velocity profiles. Generally, we find that values of $\Delta_{\rm cen} \ga 500$ \kms\ occur only during pericentric passages in Phase II, and only for SMBH pair separations $< 1$ kpc. They are also short lived, with a cumulative lifetime of at most a few Myr in a given merger. In our unequal-mass, gas-poor simulations with shallower central potentials, NL velocity splittings $> 500$ \kms\ are never achieved. Nonetheless, the association between large $\Delta_{\rm cen}$ and sub-kpc SMBH pairs suggests that spatially-unresolved dNL AGN with large $\Delta_{\rm cen}$ may be good candidates for high-resolution follow-up imaging studies.

Excluding these brief, large spikes in $\Delta_{\rm cen}$, we note that the maximum $\Delta_{\rm cen}$ for each snapshot is typically a bit larger in Phase III than in Phase II, owing primarily to the higher gas velocities that reflect a (modest) increase in the depth of the central potential during this time. In the $q=1$, \fgas\ $= 0.3$ merger model (the most extreme model used in terms of central gas density and \vesc), the maximal $\Delta_{\rm cen}$ may be as high as 800 \kms\ in Phase III, but obscuration is likely important in such an environment and may reduce the probability of observing such large velocity splitting in a single NLR. 

We also calculate the velocity centroid of the entire double-peaked profile (the median velocity of the two peak centroids) and examine phases in which it differs from zero (second panel from bottom in Fig.~\ref{fig:fitdata1}). In principle, a double-peaked profile with an offset centroid could arise from one orbiting SMBH in a pair, as in Fig.~\ref{fig:kin_dnl_uneven}. The total offsets are generally less than 150 \kms, but can briefly reach values of $\sim 300 - 400$ \kms. Again, there is a correspondence between larger offsets and pericentric passages of the SMBHs, but such offsets also occur in Phase III owing to some ``sloshing" of the recently-merged SMBH settling into the steepening central potential. (In the simulations, some of this post-merger BH motion may also be numerical in origin; the BH particle is much larger than the surrounding baryon and DM particles but is not perfectly pinned to the center of the galaxy.)However, offset profiles are expected to occur after the SMBH merger as a result of SMBH motion. Comparable-mass BH mergers can easily result in a GW recoil kick of at least 100 - 200 \kms, even if the BH spins are partially aligned \citep[e.g.,][]{vanmet10,louzlo11,lousto12}. At such kick speeds, the SMBH could retain much of its central cusp and NLR, such that a double-peaked NL profile with an overall offset could result. Finally, in our ``extreme", $q=1$, \fgas\ $= 0.3$ merger, the total offsets during Phase III owing to SMBH sloshing are $< 300$ \kms\ despite the significantly deeper central potential, indicating that this signature is less sensitive to the underlying density profile than is the velocity splitting.

\subsection{Lifetimes of Double-NL AGN}
\label{ssec:tdnl}

\begin{figure*}
\resizebox{0.33\hsize}{!}{\includegraphics{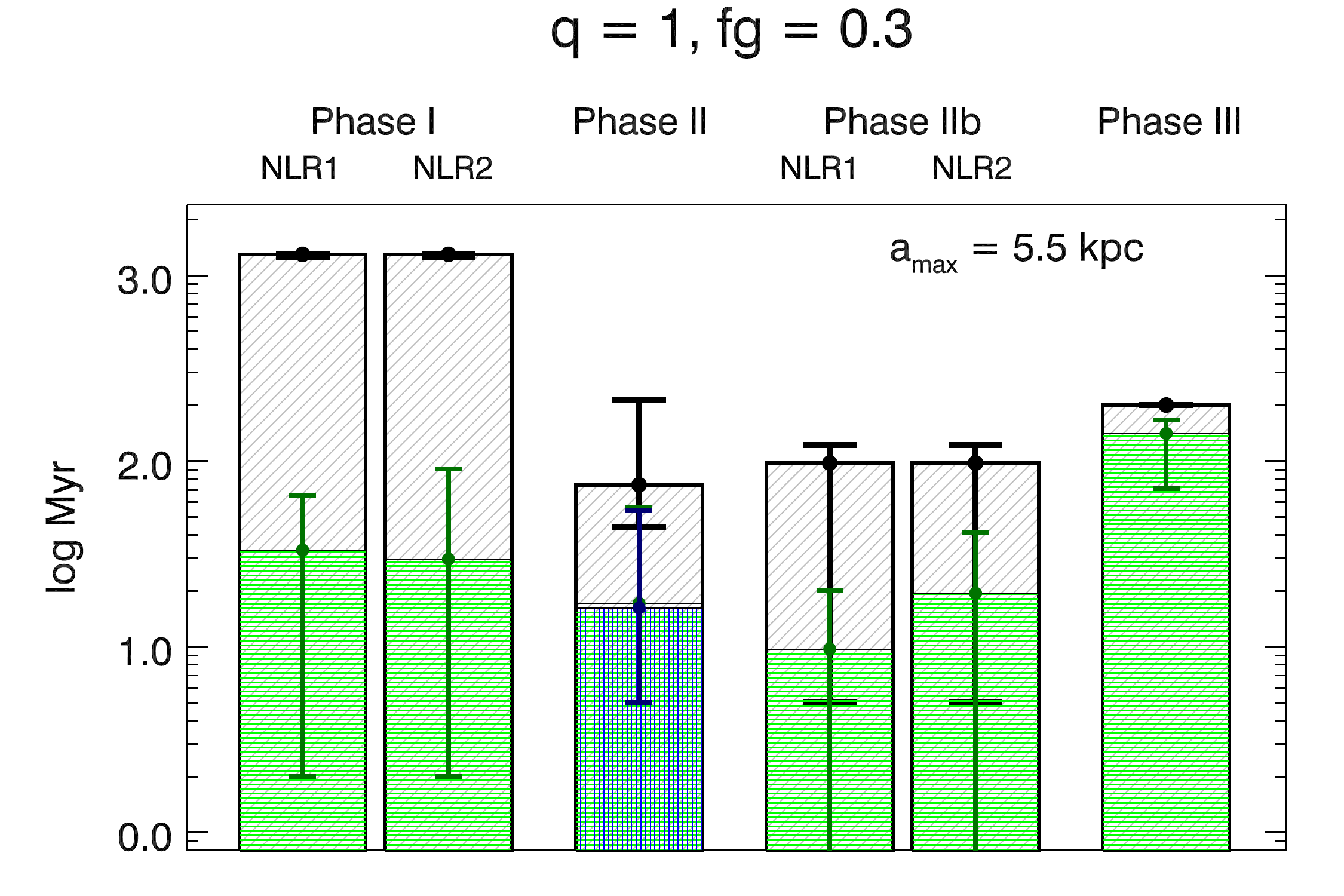}}
\resizebox{0.33\hsize}{!}{\includegraphics{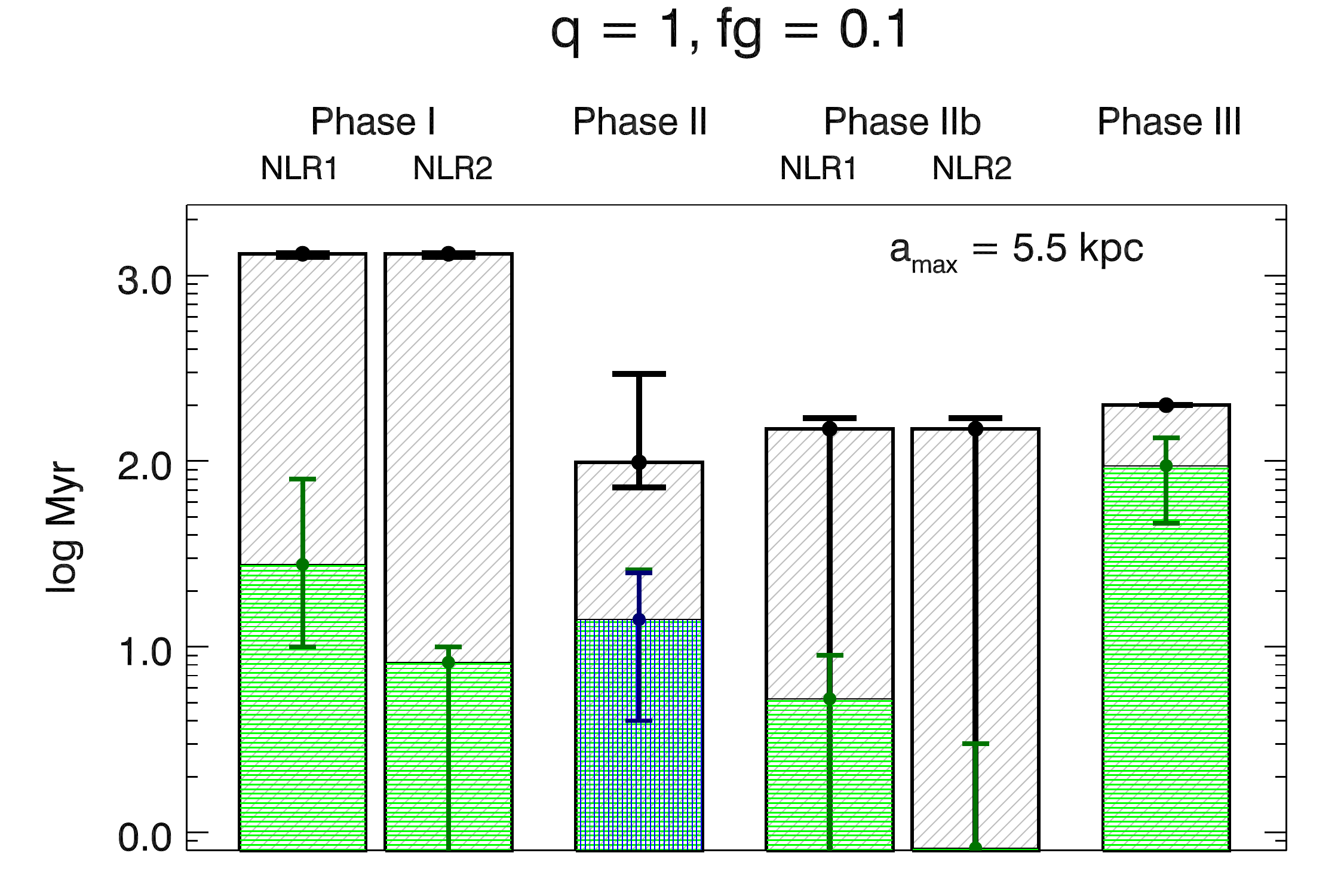}}
\resizebox{0.33\hsize}{!}{\includegraphics{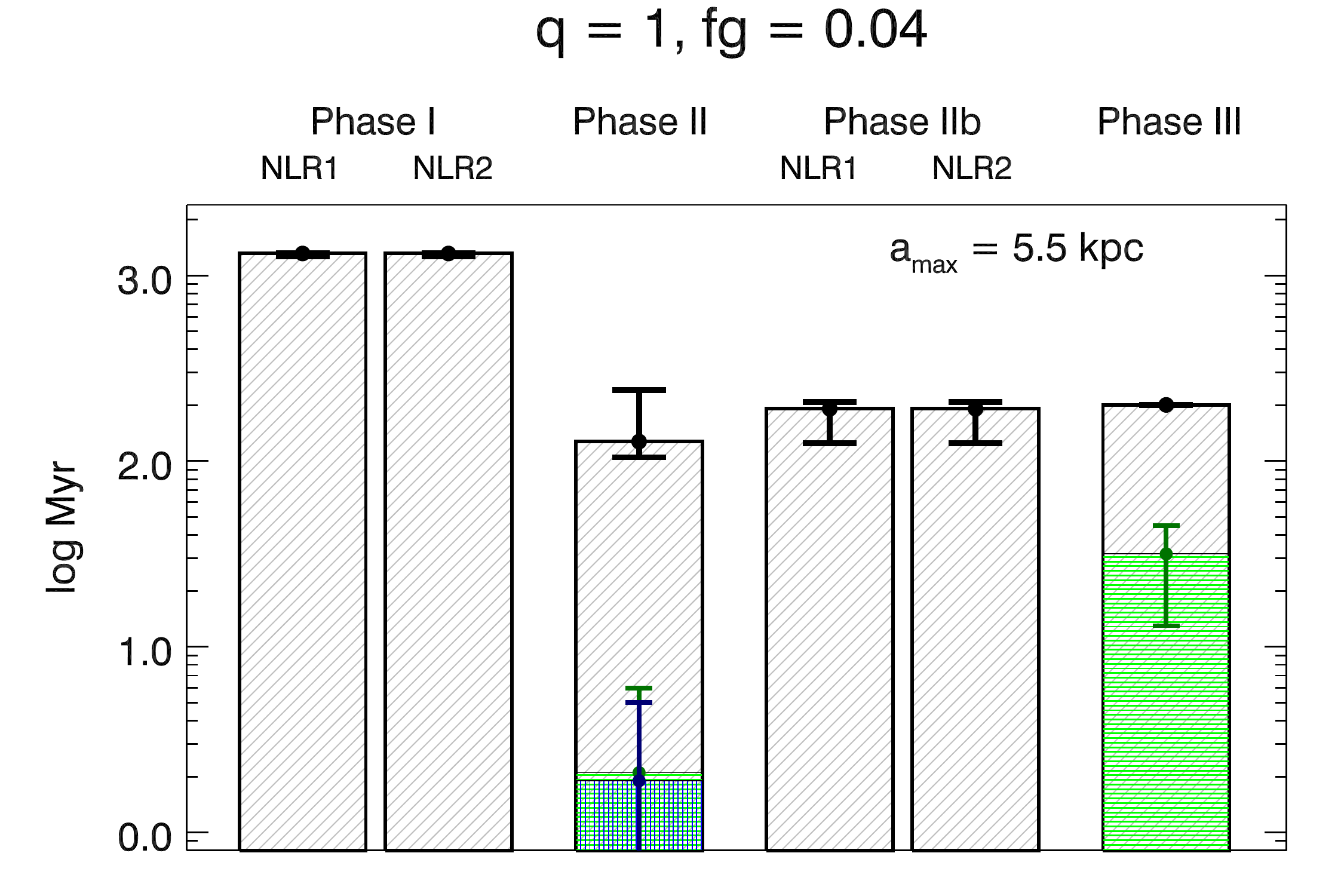}} \\
\vspace{12pt}
\resizebox{0.33\hsize}{!}{\includegraphics{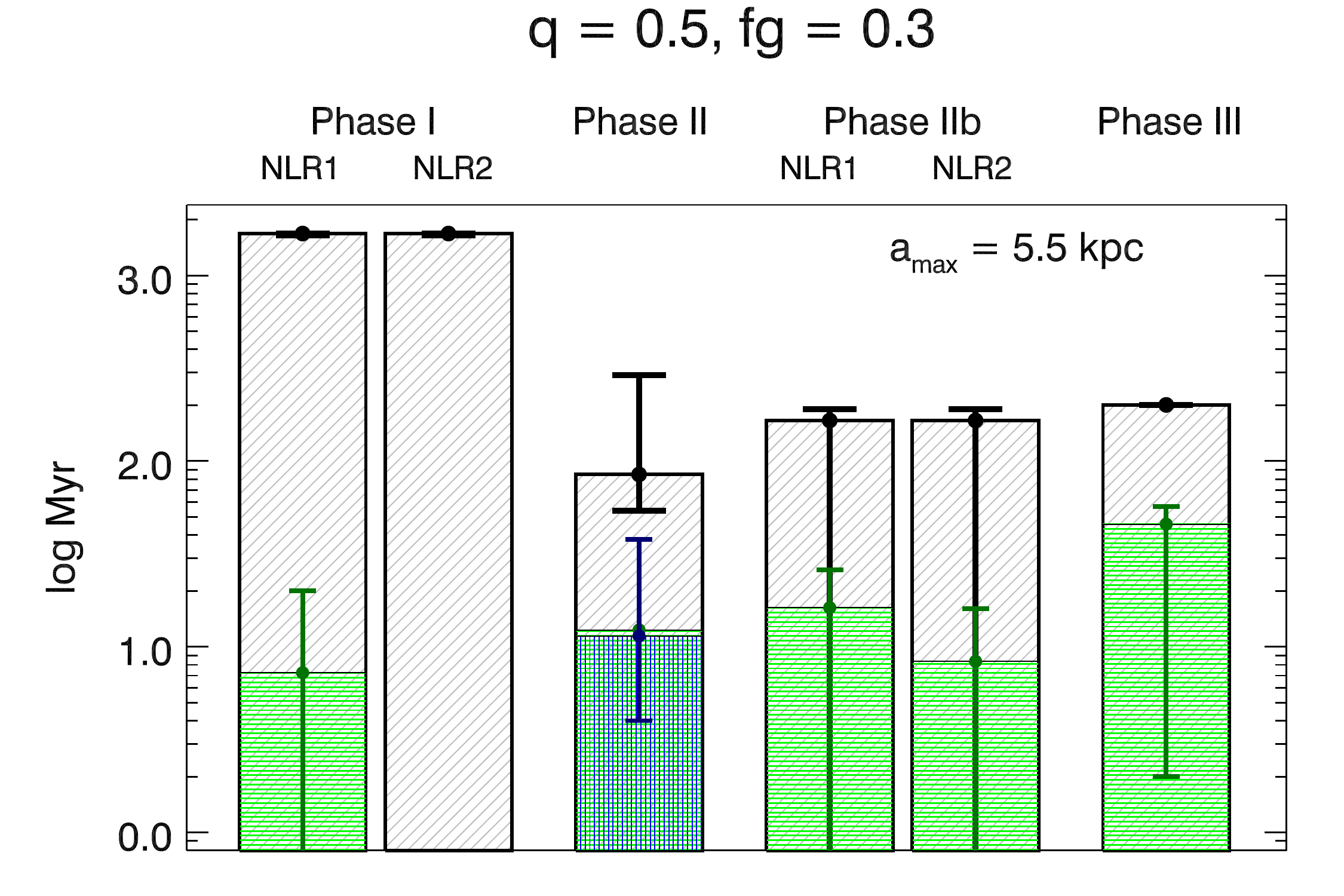}}
\resizebox{0.33\hsize}{!}{\includegraphics{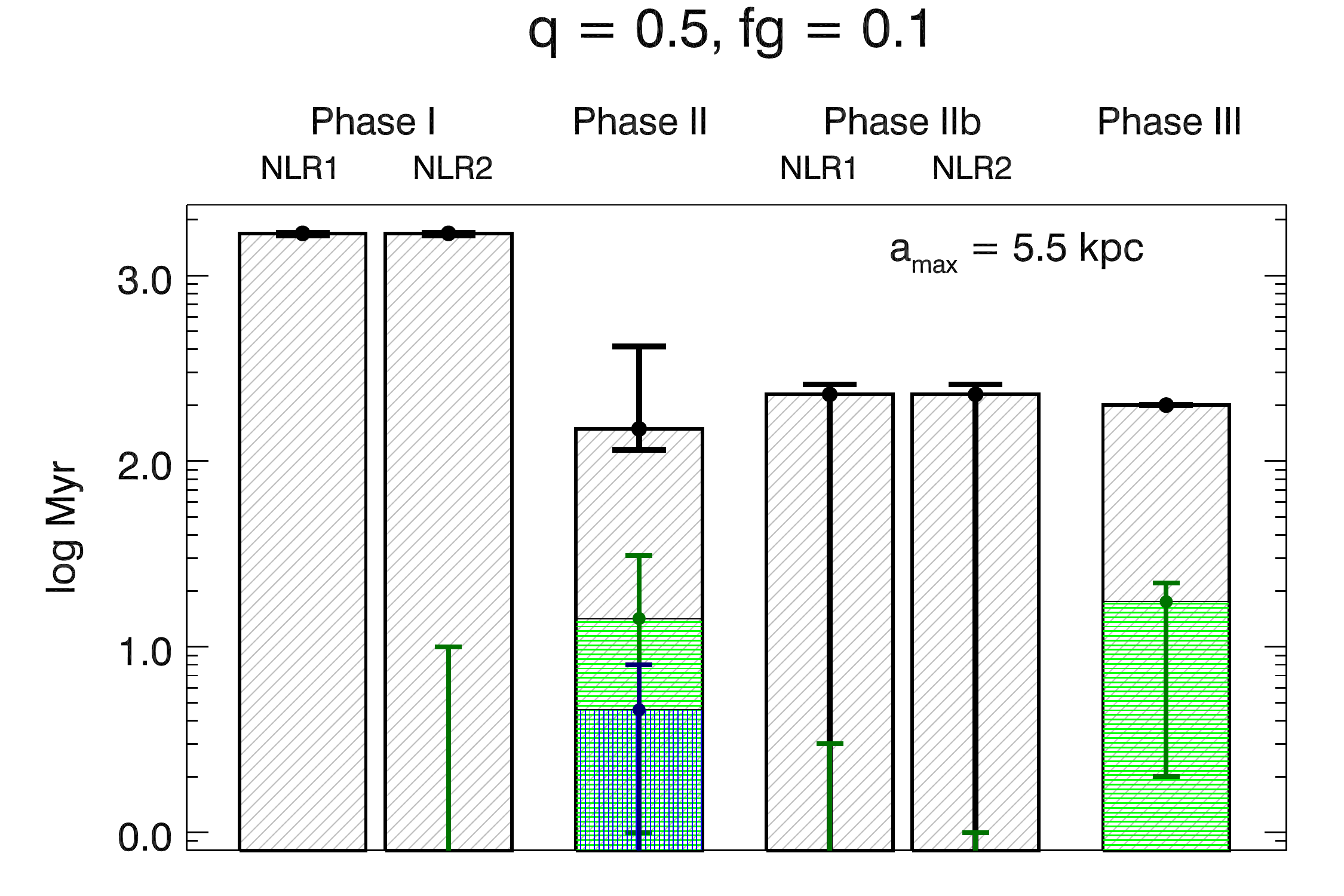}}
\resizebox{0.33\hsize}{!}{\includegraphics{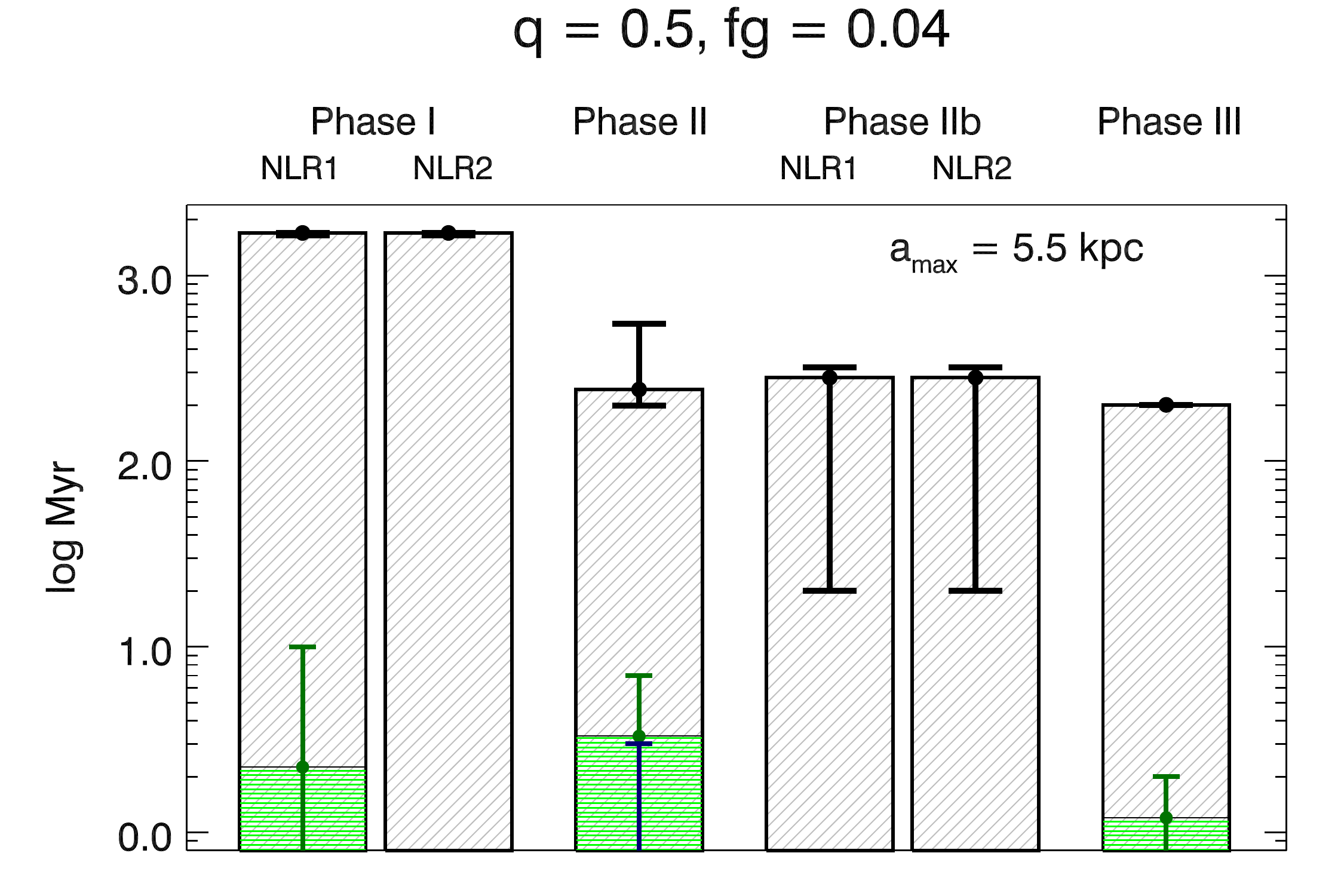}} \\
\vspace{12pt}
\resizebox{0.33\hsize}{!}{\includegraphics{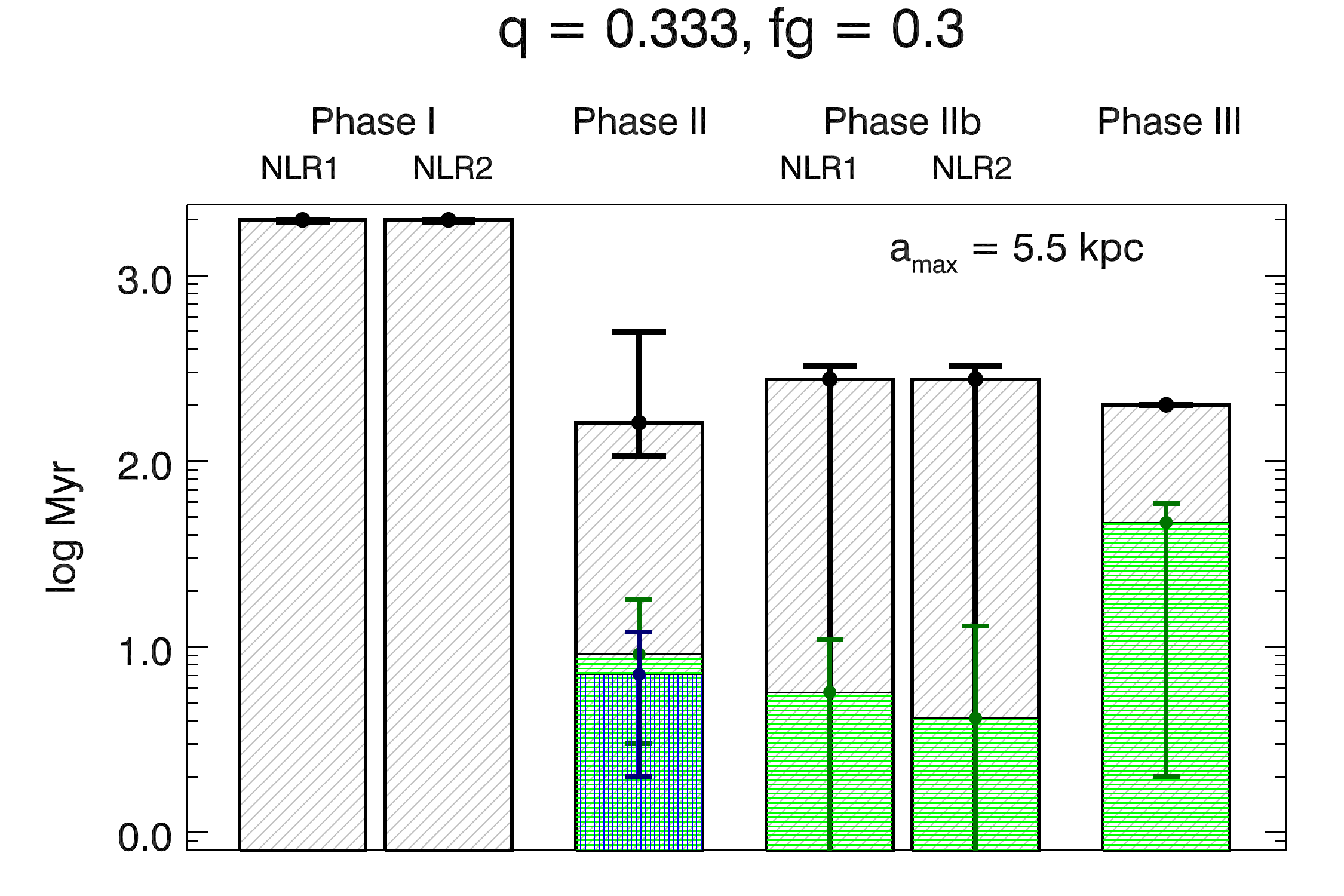}}
\resizebox{0.33\hsize}{!}{\includegraphics{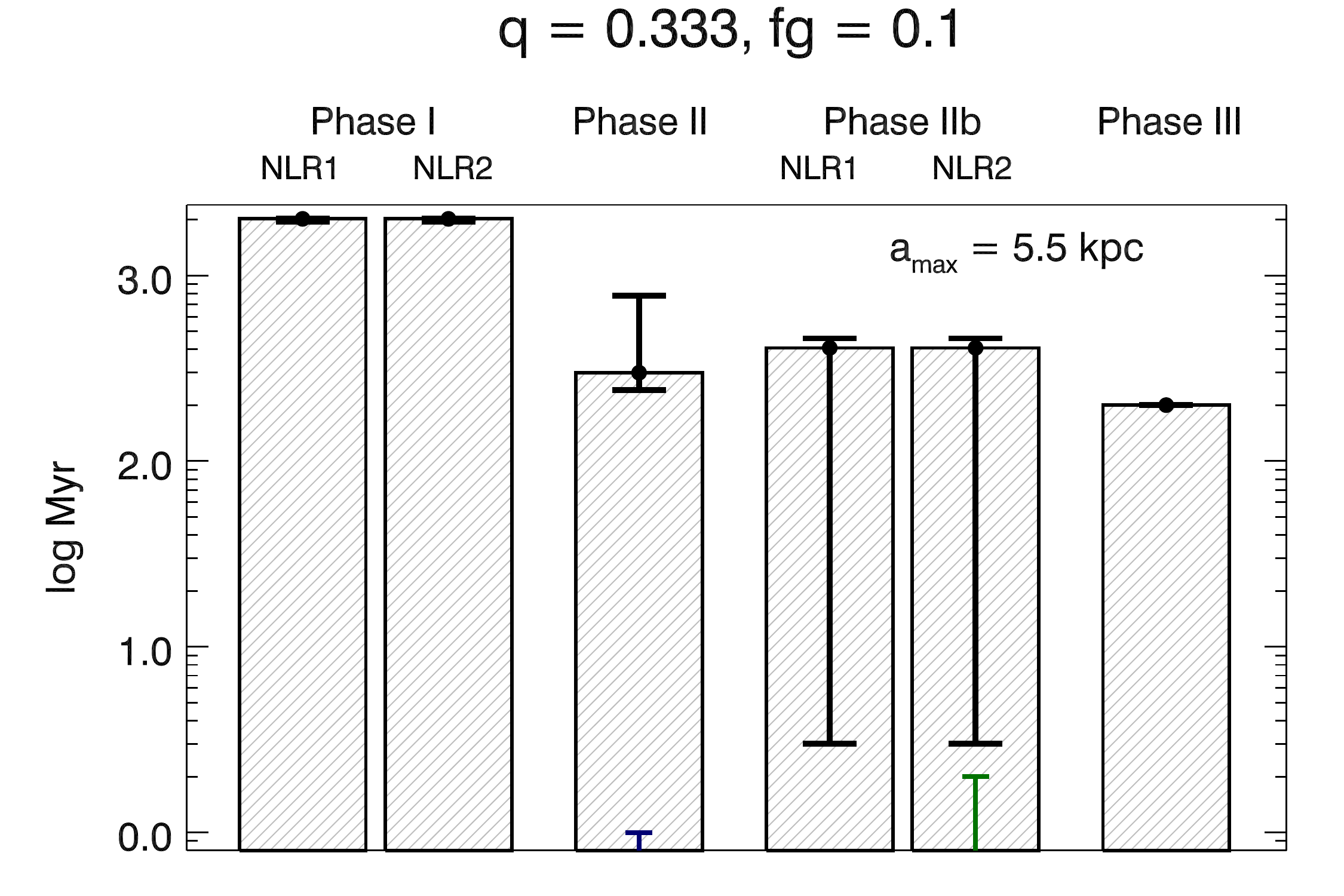}} \\
\vspace{12pt}
\resizebox{0.33\hsize}{!}{\includegraphics{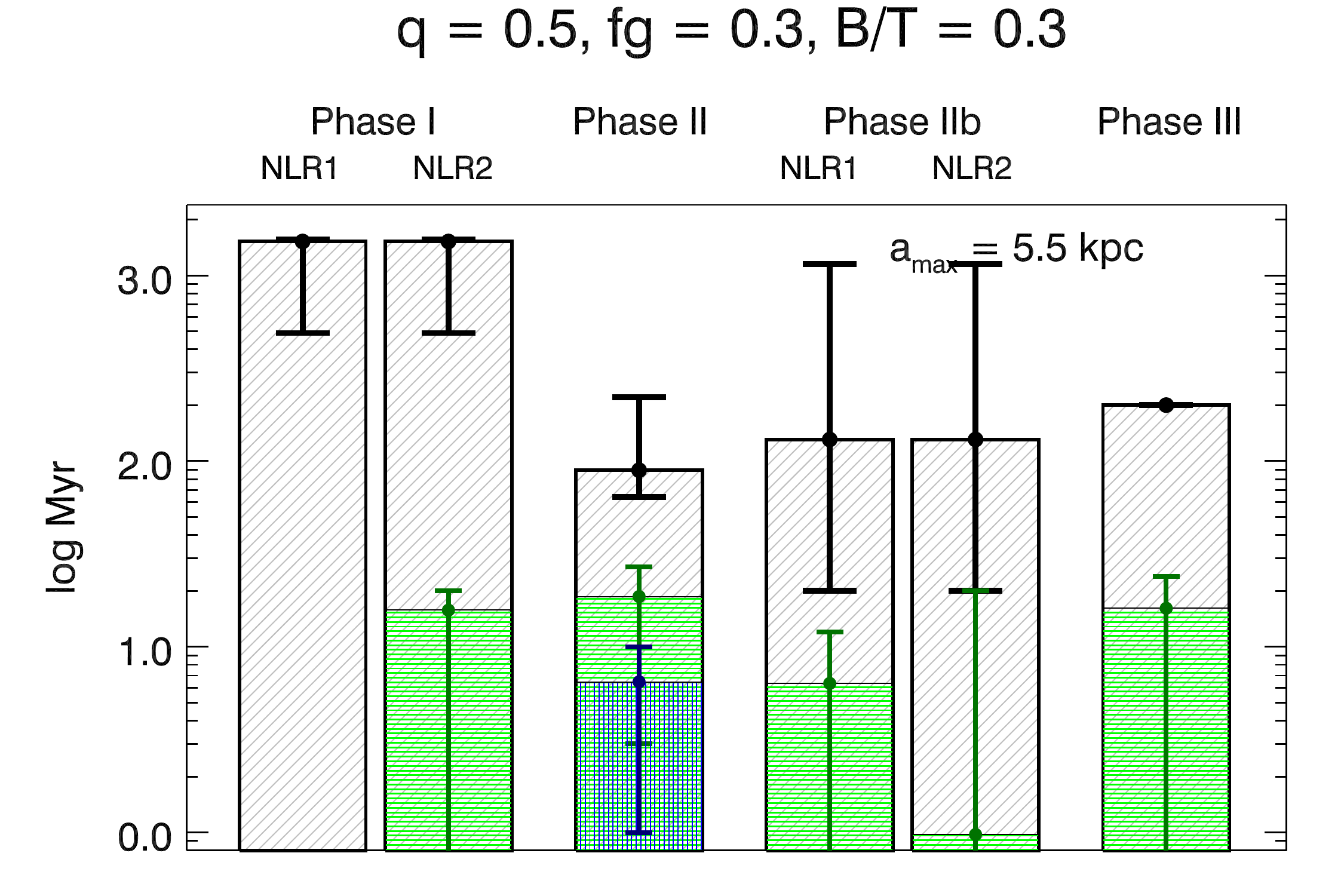}}
\resizebox{0.33\hsize}{!}{\includegraphics{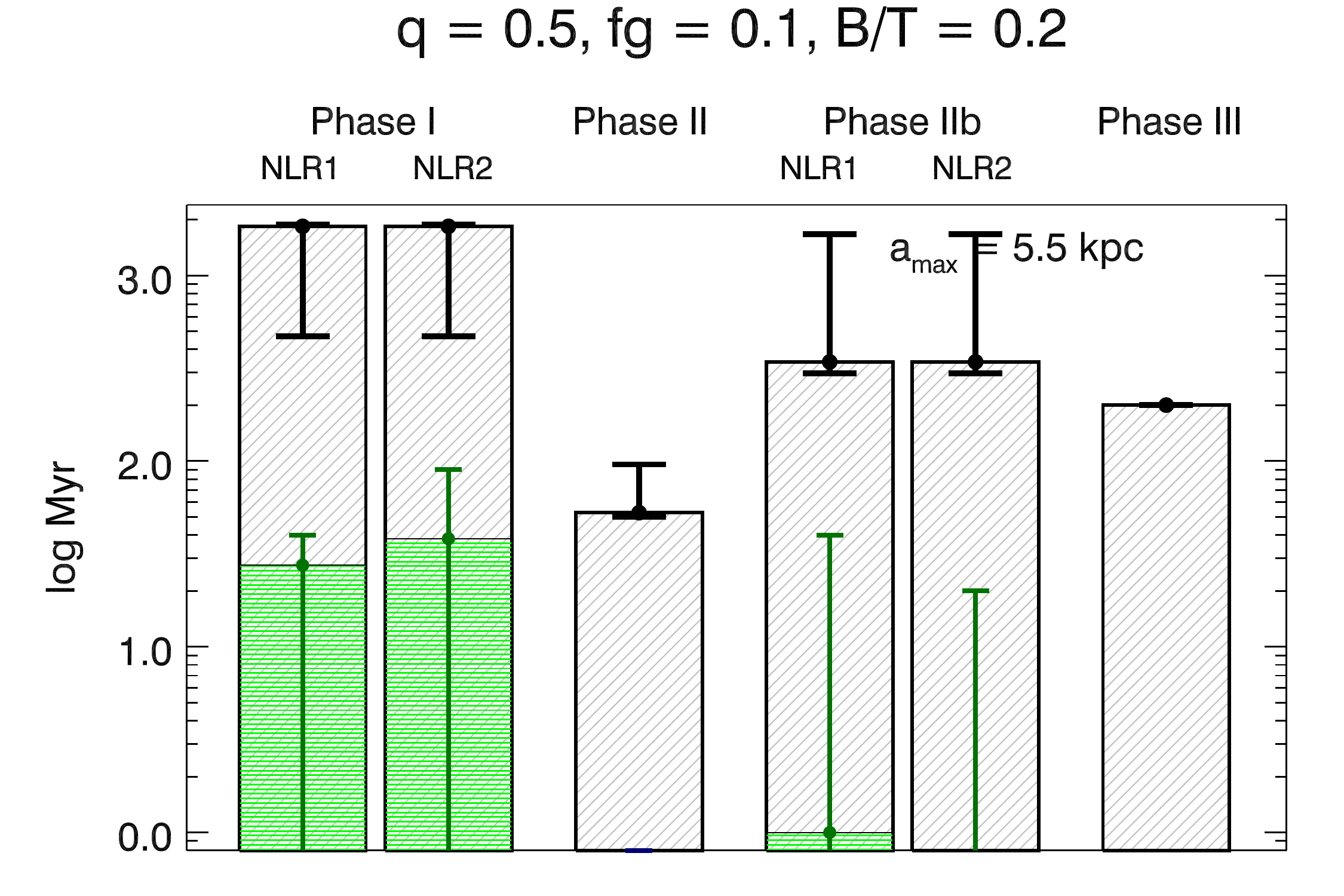}}
\caption[]{The total time during the merger, separated by merger phase, for which the NL AGN have an observable double-peaked profile (\tdnl). Data are shown for ten simulations with different galaxy models, with $q$, \fgas, and, in the bottom row, B/T values as indicated on the plots. Each plot shows the lifetimes for a kpc-scale phase definition of $a_{\rm max} = 5.5$ kpc; for clarity, \tdnl\ for $a_{\rm max} = 21$ kpc are not shown. Within each panel, Phase I is shown in the first two gray bars {\rev (diagonal hatching)}; Phase I is separated into NLR1 and NLR2 because during these phases, the NLR from each galaxy would not be observable in the same spectrum. These are followed by Phase II (the kpc-scale phase, as defined in the text), then Phase IIb (NLR1 \& NLR2) and finally, Phase III. The height of each gray bar represents the mean total duration of each phase averaged over 40 random sight lines (as the phase definitions depend on {\em projected} SMBH separation). The error bars give the range of values sampled. Within each phase, the green bars {\rev (horizontal hatching)} denote the mean lifetime for which the system has an observable double-peaked NL profile {\em and} $L_{\rm bol} > 3\% L_{\rm Edd}$ for one BH. Similarly, in Phase II, the {\rev blue bar (vertical hatching)} denotes the double-peaked NL AGN lifetime for which {\em both} BHs have $L_{\rm bol} > 3\% L_{\rm Edd}$. 
\label{fig:tdnl}}
\end{figure*}

Figure \ref{fig:tdnl} shows the time for which the NLRs have an observable double-peaked profile (\tdnl) and are active as AGN ($L_{\rm bol} >$ 3\% $L_{\rm Edd}$) in each merger phase, for ten different galaxy models. As in Fig.~\ref{fig:fitdata1}, the values shown are averaged over 40 random sight lines. When the \hbeta\ line flux is above the minimum observable criterion, the continuum luminosity is almost always $> 3\% L_{\rm Edd}$; thus, we do not make a distinction between these observability criteria in Fig.~\ref{fig:tdnl}.  In Phase II, lifetimes are shown for when one (green) or both (cyan) SMBHs meet the AGN criterion. In the equal-mass mergers, the SMBHs have similar Eddington ratios during Phases II and III, as both SMBHs are near their peak luminosities. 

For clarity, we do not plot \tdnl\ for the larger definition of Phase II ($a_{\rm max} = 21$ kpc). However, we find that the definition of $a_{\rm max}$ does not affect our results qualitatively, with the exception of Phase IIb, which depends strongly on both $a_{\rm max}$ and the LOS. More importantly, the variation in merger phase duration and in \tdnl\ for different values of $a_{\rm max}$ is much less than the variation for different sight lines. 

As discussed in \S~\ref{sssec:U_ndens}, we use the narrow \hbeta\ line for our calculations instead of the more luminous \oiii\ line because the latter is more sensitive to the exact conditions in the ISM. Accordingly, we note that these lifetimes can be considered lower limits on the lifetime of NL emission above the observability threshold, because at times when \lhbeta\ is slightly below this threshold, the stronger \oiii\ line may be detectable.
 
Comparing \tdnl\ in each phase, we see from Fig.~\ref{fig:tdnl} that in many simulations, Phase III (the post-BH-merger phase) has the longest average \tdnl. These lifetimes are typically tens of Myr, but range from zero to nearly 200 Myr (the total duration of Phase III, at which point we stop the simulation) for the merger models and sight lines shown. Phase II generally has somewhat shorter dNL lifetimes, typically a few Myr to a few tens of Myr, but ranging from 0 - 110 Myr. \tdnl\ for Phase I is somewhat more variable, but in most cases it is shorter than \tdnl\ for Phase II or III.

It is clear that \tdnl\ varies strongly with the merger parameters; moreover, some systematic trends can be identified. As discussed in \S~\ref{ssec:general} \& \S~\ref{sssec:morphology}, \lbol\ and \lhbeta\ typically peak during Phase I (after the first pericentric passage of the galaxies) and again during Phases II \& III (during final coalescence). \S~\ref{sssec:morphology} describs how the relative strength of these peaks depends on galaxy mass ratio, gas content, bulge-to-total mass ratio, and initial SMBH mass. In essence, most of the fiducial (bulgeless) simulations with initially small SMBHs have more dNL AGN activity during Phases II and III than during Phase I, despite the fact that the {\em total} duration of Phase I is much greater than the combined duration of Phase II and Phase III. Mergers with relatively large bulges (B/T = 0.3) tend to exhibit the same behavior, as the bulge stabilizes the stellar and gas disk to perturbations until final coalescence (see Fig.~\ref{fig:tdnl}, bottom left).
 
In contrast, mergers with unequal mass and low gas fractions have little NL AGN activity overall, and may have less AGN activity at coalescence than in Phase I. The merger with $q=0.333$ and initial \fgas\ $= 0.1$, which is still a major merger, has almost no double NL AGN in any merger phase, and the $q=0.5$, \fgas\ $= 0.04$ simulation has \tdnl\ of at most a few Myr. The simulation with $q=0.5$, \fgas $= 0.1$, and B/T $= 0.2$ similarly has little dNL AGN activity, except in Phase I, where the LOS-averaged \tdnl\ is tens of Myr. All of these simulations suffer from the fact that after the initial burst of AGN and SF activity following a close passage, little cold gas remains for the SMBH to accrete at final coalescence. The simulation with a stellar bulge and low \fgas\ additionally has a much larger initial SMBH, so it does have dNL AGN activity in Phase I lasting tens of Myr, with very little activity thereafter. As discussed in \S~\ref{sssec:morphology}, this particular simulation is the most extreme such example we find, in that most mergers with initial bulges have AGN that peak in Phase II or III.

\section{Discussion}
\label{sec:disc}

Motivated by recent identifications of candidate dual SMBHs via double-peaked narrow emission line signatures, we have developed the first model for narrow-line emission from AGN in full hydrodynamic simulations of evolving, interacting galaxies. Specifically, our model was applied to the output of SPH/N-body (\gadthree) simulations of merging galaxies with central, accreting SMBHs, allowing us to obtain NL velocity profiles for arbitrary sight lines throughout the galaxy merger. While our work represents only a first attempt to understand the complex dynamical and radiative processes in the centers of merging galaxies, this approach highlights several general properties of the kinematics and luminosity of photoionized gas around AGN in mergers. When combined with further information regarding, for example, the underlying stellar distribution and the AGN continuum luminosity, these results inform the prospects for confirming candidate dual SMBHs via future follow-up observations of double-peaked NL AGN.

\subsection{Evolution and Kinematics of Double-Peaked Narrow-Line AGN}

The kpc-scale phase of SMBH evolution in a major galaxy merger (Phase II in the above discussion) lasts up to a few hundred Myr, while the merger itself takes a few Gyr from first infall to final coalescence. The SMBHs have observable double-peaked NL AGN profiles for only a fraction of  Phase II, with lifetimes ranging from $< 1$ Myr to a few tens of Myr. Thus, dNL AGN seem to be a generic but relatively short-lived feature of major, gaseous mergers. 

The lifetimes of merger-triggered dNL AGN activity depend heavily on the underlying SMBH accretion rates, which vary strongly with conditions in the merging galaxies. For example, more NL AGN activity occurs in mergers with higher mass ratio (closer to equal mass) and gas content, i.e. those that drive the strongest inflow of gas to the central galactic regions. Second-order effects steepen this trend somewhat, as discussed in \S~\ref{ssec:params}.
 
We find that NL AGN often have relatively low luminosities during the early merger phase (Phase I in the above discussion). Therefore, much of the observable NL emission triggered by mergers should occur in the late stages of merging. This is consistent with a recent study by \citet{vanwas12}, which also found that merger-triggered AGN pairs are most active in the late merger phase. Owing to this coincidence in timing, dNL AGN triggered by major mergers are most likely to be associated with kpc- or sub-kpc-scale SMBH pairs, or with recent SMBH mergers (Phases II and III). However, we note that unequal mass mergers of galaxies that are relatively gas poor (\fgas\ $\la 0.1$) and disk-dominated (B/T $\la 0.2$) may follow the reverse trend, provided their SMBHs lie close to the $M_{\rm BH} - M_{\rm bulge}$ relation prior to merger. As outlined in \S~\ref{sssec:morphology}, empirical constraints from \citet{liu10a} and \citet{ge12} limit the fraction of dNL AGN in early-merger phases to $\la 2/3$, and this fraction may be lower. Thus, mergers in which NL AGN activity is dominated by Phase I should make at most a moderate contribution to the observed population of dNL AGN.

Follow-up observations of dNL AGN indicate that a minority of these objects show evidence for dual SMBHs; the rest of the double-peaked features are either ambiguous or are presumed to result from gas kinematics. If a substantial fraction of dNL AGN were {\em directly} associated with kpc-scale SMBH pairs, our results might be in tension with this finding. However, we find that only a minority of dNL AGN result directly from SMBH motion, {\em even within the kpc-scale phase of SMBH evolution}. The canonical picture in which double peaks arise from two distinct NLRs orbiting about a common central potential is only one of several possible mechanisms, and it is not the most common. 

As the example in Fig.~\ref{fig:dir_dnl_tight} shows, if the SMBH separation is comparable to the size of the NLRs, then double peaks may arise from the direct interaction of the SMBHs with the NL gas. In other cases, a double-peaked feature may arise from gas kinematics in a single NLR, but still may be {\em influenced} by the relative SMBH motion. For example, the SMBH motion may alter the peak ratio or impart an overall velocity shift to the observed dNL profile. The fraction of double-peaked profiles in Phase II that are influenced by SMBH motion either directly or indirectly can be up to $\sim 80$\%, depending on the merger model and the definition of Phase II. 

We also see many NL profiles in which the dNL AGN arising from gas kinematics are not significantly affected by the SMBH motion and are simply {\em coincident} with the kpc-scale phase.  This is consistent with the empirical finding that a minority of dNL AGN show clear association with dual SMBHs. It also indicates that ``serendipitous" discoveries of dual AGN with double-peaked profiles, but in which the double peak results from gas kinematics \citep[cf.][]{fu12}, may in fact be a relatively common occurrence. The remainder of double-peaked NLs mostly arise from rotating gas disks, though in some cases almost 30\% of profiles in Phase II classified as double peaks have complex (multi-peaked) or highly-asymmetric profiles. These are generally also indicate by relative SMBH motion via disturbed or superimposed NL emission regions.

We note that our thermal feedback model for AGN does not allow for AGN outflows or jets, which may be additional mechanisms for producing double-peaked NLs. This could mean that the prevalence of dNLs induced by gas rotation is lower in reality than in our simulations. However, dNL AGN produced by outflows or jets may be distinguishable from dNLs  produced by dual SMBHs. \citet{rosari10} demonstrate that jets in some dNL AGN may be resolved via radio imaging, and \citet{comerf12} suggest that spatially-extended NL emission components may be caused by outflows, though the possibility of a dual AGN with outflows cannot be excluded. However, a key result of this study is that dNL AGN induced by SMBH motion are generic to gaseous major mergers; variations in the gas kinematics alter only the importance of dual SMBH-induced dNL AGN relative to dNL AGN produced by other means.

Another important consideration is that the observed kinematic features of NL AGN are strongly dependent on viewing angle. To account for this, we fit velocity profiles and calculate dNL AGN lifetimes for many lines of sight in each simulation snapshot. The LOS variation in dNL AGN lifetimes is typically at least an order of magnitude. Therefore, efforts to infer statistics of dual SMBHs from observations of dNL AGN are limited by this fundamental uncertainty.

\subsection{Additional Signatures of Dual AGN}

dNL AGN with two resolved peaks in stellar surface brightness that are spatially coincident with the NL emission components can be considered strong dual SMBH candidates. We find that after the first close passage of the galaxies triggers a burst of star formation, each SMBH is surrounded by a dense cusp of stars and gas. Regardless of the AGN luminosity, these cusps should appear as two brightness peaks with sufficient spatial resolution. However, in most of our simulations, $\sim 10 - 40$\% of dNL AGN induced by SMBH motion have projected SMBH separations less than a kpc and may be difficult to resolve. Gas-poor mergers have an even higher fraction of dNL AGN associated with sub-kpc-scale SMBH pairs, $> 90$\% in some cases. This is further evidence that a significant population of dNL AGN may not show clear association with a SMBH pair, even though a pair is present. Other means must be employed to confirm the presence of a dual SMBH in such systems.

Some additional information may be obtained from sensitive, high-resolution (i.e., {\em HST}) imaging that can capture diffuse tidal features in the merger remnant. If the galaxy appears undisturbed, or if it shows obvious signs of being in an early-merger phase (for example, two widely-separated galaxies, perhaps with characteristic tidal bridge and tail features), a kpc-scale dual AGN would be disfavored. If a kpc-scale SMBH pair is present, the galaxy should appear as a late-stage merger remnant---a single galaxy, possibly with a double core, that is morphologically disturbed. However, because the morphology should appear much the same after the SMBHs have merged, this diagnostic cannot rule out a post-BH-merger (Phase III) dNL AGN.

We also consider the degree of alignment between the projected SMBH orbital plane and the apparent ellipticity of the host galaxy, as \citet{comerf12} find a correlation between these quantities for dNL AGN. When the merging galaxies are physically overlapping (in projection) but their cores have not quite merged, the system may appear to have a significant ellipticity that is strongly aligned with the SMBH orbit (as in Fig.~\ref{fig:dir_dnl_wide}). However, in this stage the merger remnant is still highly disturbed rather than a true ellipsoid, and deeper imaging in such cases could reveal fainter tidal features. Once the cores of the progenitor galaxies have merged, the SMBHs begin to decouple from this orbital plane as they continue to inspiral via dynamical friction. Thus, SMBH pairs with projected separations $\la 1$ kpc should be more weakly correlated with the ellipticity of the host galaxy.

\citet{smith11} have suggested that dNL AGN with even-peaked profiles are less likely to be dNL AGN, on the grounds that rotating gas disks should have double-peaked profiles with roughly even peaks, and that dual AGN are unlikely to have very similar luminosities. We do see a trend toward higher peak ratios (closer to unity) for double-peaked NLs arising from rotating gas disks around single SMBHs than for those resulting from SMBH pairs. However, in the single-NLR case the peak ratios still vary substantially, and in the dual-BH case there is almost always at least one viewing angle for which the double-peaked profile has nearly even peaks. Thus, dNL AGN with uneven peak ratios ($\la 0.5$) should have a slightly higher probability of containing a dual SMBH, but this appears to be a fairly weak correlation.

In addition, we find that large velocity splittings in double-peaked profiles ($\Delta_{\rm cen} \ga 500$ \kms) are often associated with relative motion of sub-kpc SMBH pairs, because $\Delta_{\rm cen}$ increases during pericentric passages of the SMBHs, especially just before the SMBHs merge. Large dNL velocity splittings may also occur in AGN with high-velocity outflows, though these cases may be more easily distinguishable as spatially-extended emission components. Thus, dNL AGN with spatially-compact emission components separated by $\ga 500$ \kms\ may be good candidates for high-resolution imaging that could detect a sub-kpc dual AGN.

Finally, double-peaked profiles with overall velocity offsets may indicate the presence of an inspiraling SMBH pair. Again, this signature could also arise from a recently-merged SMBH if it sloshes though the central region before settling down, especially as SMBHs may often receive GW recoil kicks $\ga 100 - 200$ \kms\ \citep[e.g.,][]{lousto12}. In either case, our results indicate that double-peaked profiles with discernible velocity offsets should be associated with SMBH motion of some kind, and that these systems have a somewhat higher probability of containing dual SMBHs.

\subsection{Model Assumptions and Prospects for Future Work}
In this initial study of NL AGN in merging galaxies, the qualitative conclusions are insensitive to the choice of parameters for our simulations and NL model. However, the quantitative results do have some dependence on these model parameters. The variation in our results with these parameters is in many cases smaller than the intrinsic variability of NL activity with viewing angle, and we have chosen parameters such that our simulated NLRs broadly reproduce the size and luminosity of observed NL AGN. Here we discuss this model dependence in more detail, as well as possibly relevant physics to be included in future work. 

As discussed in \S~\ref{ssec:params}, our results have some quantitative dependence on the gas EOS used; softer EOS models result in more stable gas disks, and thus lower levels of AGN and NL activity. We find that a nearly-isothermal EOS (softened EOS parameter $q_{\rm EOS} = 0.05$) results in higher sustained accretion rates following the initial close passage of the galaxies and significantly more NL activity in the early merger phase. However, this variation in dNL AGN activity is almost entirely attributable to the underlying variation in NL and AGN luminosites; i.e., the choice of $q_{\rm EOS}$ does not have a measurable effect on the {\em kinematics} of the NL gas.

Our results are generally independent of the choice of mass and spatial resolution, though we do require that the NLRs be at least marginally resolved in order to impart physical meaning to these results. By conducting a small resolution study we do find a weak trend toward higher central densities and accretion rates for higher resolution, but this has little effect on our results. In particular, despite small variations in the AGN lightcurves and peak luminosities, the {\em total} amount of SMBH accretion remains constant regardless of resolution.

Without arbitrarily high resolution, we cannot exclude the possibility that gas kinematics on sub-resolution scales could shape the resulting velocity profiles in an unpredictable manner. However, this will only affect double-peaked profiles arising from gas kinematics, and it is possible that turbulent motion in rotating NLRs would smear out some of these double peaks, thereby increasing the fraction of dNL AGN caused by SMBH motion. We can robustly predict that for at least a small fraction of the kpc-scale (and sub-kpc scale) phase, the NL gas kinematics will be dominated by the SMBH motion.

A related issue is the extent to which our NL models are limited by the multi-phase ISM treatment used in the simulations \citep{sprher03}. While this model has shown success in reproducing the global properties of merging galaxies \citep[e.g.,][]{cox06,hopkin06}, its treatment of gas physics on small scales is necessarily highly simplified. For example, if star formation feedback drives strong turbulence in the ISM, then the assumed pressure equilibrium between hot and cold gas phases would be violated. Future work that relies on a turbulent-pressure-driven ISM model will produce more realistic distributions of gas density and temperature, allowing a more accurate identification of the NL-emitting gas.

Perhaps the most important caveat to the work presented here is the lack of obscuring dust in our models. While we do account for self-shielding of NL clouds in an average sense, such that AGN continuum photons are not double-counted, our models do not include radiative transfer calculations and thus are unable to account for the effects of obscuration and reprocessing of emission by dust. We accordingly restrict our analysis to galaxies with relatively low initial gas fractions, as higher gas fractions generally drive stronger starbursts and produce much more dust. Starbursts can additionally produce substantial NL emission from stellar photoionization, which is also not accounted for in our model. The results from models with initial gas fractions of 30\%, the highest included in this study, may be considered less robust in this sense. However, we note that in cases where the dust and gas have a clumpy distribution, such that the central galactic region is only partially obscured, it is possible that double-peaked features arising from kinematics in a single NLR would be preferentially smeared out by dust reprocessing (relative to two well-separated NLRs). This could potentially strengthen the correlation between dNL AGN and dual SMBHs. The incorporation of full radiative transfer calculations in future work, including models for dust and stellar photoionization, will allow a more direct comparison of our simulated results to observed AGN.

\subsection{Broader Implications}

It is important to note that all of the analysis presented here applies only to AGN triggered by major galaxy mergers. We therefore cannot comment on the contribution of AGN triggered by minor mergers or secular processes to the population of double-peaked NL objects. More generally, a strong consensus on the fraction of AGN triggered by mergers does not currently exist, as surveys with different selection criteria have obtained widely disparate results \citep[e.g.,][]{cister11,elliso11,koss11b}. 

{\rev Our results indicate that merger-triggered dNL AGN may often be observed during the late stages of the merger, either soon before or soon after the SMBHs themselves merge. Thus, many dNL AGN that do not contain dual SMBHs may be associated with recent SMBH mergers, which is interesting in its own right. As mentioned above, BH mergers result in GW recoil kicks, which in many cases may be $\ga 100 - 200$ \kms. At these velocities, the SMBH may retain much of its cusp of gas and stars, such that the SMBH motion, while short-lived, might be detectable as a {\em narrow-line} offset in the AGN spectrum. At larger velocities that could displace the SMBH significantly from the center of the galaxy ($\ga 500 - 1000$ \kms), only the inner region would be remain bound to the SMBH. This would result in the offset {\em broad line} signature that is more commonly considered for recoiling SMBHs \citep[e.g.,][]{loeb07,blecha11}, and in such cases the post-BH-merger phase clearly would not contribute to the dNL AGN lifetime.}

We have shown that the scenario conceived to motivate searches for dual SMBHs in double-peaked NL AGN---i.e. two well-separated, observable NLRs in relative motion---occurs for only a small fraction of the kpc-scale phase. Observations indicate that a minority of dNL AGN show clear evidence of association with dual brightness peaks in imaging. Therefore, we may ask the question of whether double-peaked NL signatures are a good tool for identifying dual SMBHs. In relative terms, studies of dNL AGN have certainly proved to be an {\em effective} means of finding good dual SMBH candidates; the number of strong candidates has greatly increased via follow-up observations. Less obvious is whether observing dNL AGN is an {\em efficient} means of identifying dual SMBHs. We have demonstrated that even within the kpc-scale phase, a wide variety of gas and SMBH configurations may give rise to double-peaked features. 

{\rev The relatively short lifetimes of observable dNL AGN associated with dual SMBHs indicate that the converse situation should occur as well; i.e., kpc-scale AGN pairs should be found that do not have double-peaked emission lines. Indeed, as noted in \S~\ref{sec:intro}, several known (X-ray detected) AGN pairs are not associated with unusual narrow-line kinematics \citep{komoss03, bianchi08, green10, koss11a, fabbia11}. Hard X-ray selection of AGN can reveal nuclear sources that are highly obscured at optical wavelengths, and thus it is to some extent complementary to optical emission line selection techniques. This is an important consideration for AGN in merging galaxies; almost all ultra-luminous infrared galaxies (ULIRGs) are undergoing mergers, and many contain buried AGN \citep[e.g.,][and references therein]{sanmir96,lonsda06}. \citet{koss11b} find that ultra-hard X-ray selected AGN are significantly more likely than optically-selected AGN to be associated with merging galaxies, and the dual AGN in NGC 6240 \citep[e.g.,][]{komoss03} and Mrk 739 \citep{koss11a} show little to no evidence of AGN-like optical emission lines. 

More interesting in the present context are those less-obscured AGN pairs that are identifiable as such in both optical and X-ray (or radio) bands. The detection of dual compact X-ray or radio sources in an active galaxy provides the most unambiguous evidence that a system hosts an SMBH pair. Such systems in which optical AGN spectra are also present allow a uniquely robust comparison of optical emission-line kinematics in dual AGN. X-ray \citep{comerf11} and radio \citep{fu11b} follow-up of dNL AGN has already revealed two such confirmed AGN pairs, but based on our results, one would expect at least as many similar sources {\em without} double-peaked emission lines.

One such example is the binary quasar SDSS J1254+0846; it has resolved (optical and X-ray) dual nuclei, and the $\sim 21$ kpc separation allows for spectroscopy of the individual components \citep{green10}. The optical spectra are in fact offset by $\sim 200$ \kms, which is consistent with dual AGN motion on $\sim$ 1 - 10 kpc scales but could also arise from AGN outflows. Owing partly to the large luminosity ratio ($> 10$), this velocity offset would not produce double-peaked NLs in a combined spectrum. This is perfectly consistent with our results, which indicate that in addition to the relatively short-lived nature of simultaneously-active, dNL AGN in mergers, the NL radial velocity offsets should vary widely with viewing angle. In this instance, the NL equivalent widths are quite different between the two quasars; in other cases, NLs may not be observable at all in one of the AGN. Such systems may contribute to the large population of offset-NL AGN \citep[e.g.,][]{comerf09a}. Future targeted follow-up of X-ray selected AGN in merging galaxies, as well as high-resolution X-ray and radio imaging of dNL AGN, should reveal more such confirmed AGN pairs with optical counterparts. These objects will provide valuable insight into the NL kinematics of dual AGN. }

{\rev Finally, the fact that dNL AGN are often produced by sub-kpc-scale SMBH pairs, along with the sensitive dependence of NL profiles on viewing angle, indicates that some of the currently ambiguous dNL AGN candidates should contain dual SMBHs.} A critical observation from our results is that regardless of the degree of {\em causation} of dNL AGN by dual SMBHs, a {\em correlation} in the timing of their occurrence does exist.  Furthermore, as described above, some of the physics not accounted for in our current models could potentially increase the fraction of dNL AGN induced by SMBH motion relative to those induced by gas kinematics. {\rev Continued} dedicated, multi-wavelength follow-up observations of dNL AGN should reveal a larger population of strong candidate SMBH pairs. {\rev Some may be directly ``confirmed" by discovery of dual compact X-ray or radio sources. In other cases, multiple} lines of indirect evidence, including the signatures discussed in this work, should be combined to determine the most promising dual SMBH candidates.

\section*{acknowledgements}
We thank the anonymous {\rev referees} for constructive comments that have improved this manuscript. We would also like to thank T.~J. Cox, Lars Hernquist, Volker Springel, and Xin Liu for helpful comments and discussions. This work was supported in part by NSF grant AST-0907890 and NASA grants NNX08AL43G and NNA09DB30A (AL), and by NASA grant NNX11AE16G (RN).

\bibliography{refs_double_NL_v2}

\end{document}